\documentclass[reprint,aps,pre]{revtex4-1}

\usepackage{amsmath,amsfonts,amscd,amssymb,graphicx,subeqnar}

\usepackage{dcolumn}
\usepackage{bm}
\usepackage[percent]{overpic}
\usepackage{color}
\usepackage{adjustbox}
\usepackage{rotating}
\usepackage{float}
\makeatletter
\let\newfloat\newfloat@ltx
\makeatother
\usepackage{algorithm}
\usepackage{algpseudocode}
\algnewcommand\algorithmicinput{\textbf{Input:}}
\algnewcommand\Input{\item[\algorithmicinput]}
\usepackage{soul}
\usepackage{stackengine}

\usepackage{tikz-cd} 

\usepackage[hypertexnames=false,
            colorlinks=true,
            linkcolor=blue,
            citecolor=blue,
            urlcolor=blue]
            {hyperref} 

\usepackage{tikz}
\definecolor{red}{rgb}{0.8500, 0.3250, 0.0980}
\definecolor{green}{rgb}{0.4660, 0.6740, 0.1880}
\definecolor{yellow}{rgb}{0.9290, 0.6940, 0.1250}
\definecolor{blue}{rgb}{0, 0.4470, 0.7410}

\begin{document}

\title{Universal Dynamics of Damped-Driven Systems:\\ The Logistic Map as a Normal Form for Energy Balance}

\author{J. Nathan Kutz$^{*,\dag}$, Aminur Rahman$^*$, Megan R. Ebers$^\S$, James Koch$^{**}$, Jason J. Bramburger$^\ddag$}
 \affiliation{$^*$Department of Applied Mathematics, University of Washington, Seattle, WA 98195-3925} %
 \affiliation{$^\dag$Department of Electrical and Computer Engineering, University of Washington, Seattle, WA 98195}
  \affiliation{$^\S$Department of Mechanical Engineering, University of Washington, Seattle, WA 98195}
  \affiliation{$^{**}$Pacific Northwest National Laboratories, Richland, WA 99354}
  \affiliation{$^\ddag$Department of Mathematics and Statistics, Concordia University, Montr\'eal, QC H3G 1M8}

\begin{abstract}
Damped-driven systems are ubiquitous in engineering and science.  Despite the diversity of physical processes observed in a broad range of applications, the underlying instabilities observed in practice have a universal characterization which is determined by the overall gain and loss curves of a given system.  The universal behavior of damped-driven systems can be understood from a geometrical description of the energy balance with a minimal number of assumptions. The assumptions on the energy dynamics are as follows:  the energy increases monotonically as a function of increasing gain, and the losses become increasingly larger with increasing energy, i.e. there are many routes for dissipation in the system for large input energy.  The intersection of the gain and loss curves define an energy balanced solution.  By constructing an iterative map between the loss and gain curves, the dynamics can be shown to be homeomorphic to the logistic map, which exhibits a period doubling cascade to chaos. Indeed, the loss and gain curves allow for a geometrical description of the dynamics through a simple Verhulst diagram (cobweb plot). Thus irrespective of the physics and its complexities, this simple geometrical description dictates the universal set of logistic map instabilities that arise in complex damped-driven systems.  More broadly, damped-driven systems are a class of non-equilibrium pattern forming systems which have a canonical set of instabilities that are manifest in practice.
\end{abstract}

\maketitle

\section{INTRODUCTION} \label{intro}

Damped-driven systems are ubiquitous in the engineering and physical sciences.    From mode-locked lasers to rotating detonation rocket engines, damped-driven systems exhibit basic and universal instabilities despite the diverse physical processes that dominate a given system.   Common to such systems is a driving mechanism which ultimately raises the total energy (potential, kinetic, etc) in the system.  The additional energy imparted into the system is then diffused in a diverse number of ways which depends upon the particular physical system under consideration. The diffusion of energy becomes a strongly nonlinear loss mechanism if enough energy is driven into the systems and as the system finds potentially a multitude of ways for energy to be damped out of the system.  The interplay of the nonlinear gain and loss dynamics, regardless of the underlying physical mechanisms, are a form of nonequilibrium dynamics which have shown to generate canonical pattern forming instabilities across diverse application areas~\cite{cross1993pattern}.  Like the universal aspects of pattern formation instabilities,  our aim is to identify unifying themes in damped-driven systems and show that despite the inherent complexity and diverse physics of many damped-driven systems, their fundamental energy balance dynamics can be represented by dynamics as simple as that of the logistic map.

Although the physical mechanisms that dominate a given system can vary significantly from one system to another, energy balance is a more universal concept that plays the underlying and fundamental role in dictating the bifurcations and instabilities observed in practice.  Just like the general structure of pattern forming systems~\cite{cross1993pattern}, which allow for a canonical set of dominant balance spatio-temporal patterns to emerge from nonequilibrium dynamics, irrespective of the underlying physics, so too does the consideration of gain-loss dynamics reveal a limited set of canonical instabilities.  Indeed, from the energy balance perspective, the specific form of pattern manifest is unimportant.  Rather, it is the organization of energy as it is continuously imparted and increased that is of central concern.  Thus the overall gain and loss dynamics dominate the universality of the resulting behavior, not the specific physics which organize it.


The simplest model for interaction of gain and nonlinear loss is the logistic map first proposed by the Belgian mathematician Pierre-François Verhulst in 1838~\cite{verhulst1838notice}. Originally proposed as a mathematical model to describe population growth, it has since been rediscovered multiple times~\cite{cramer2004early} and was popularized by Robert May in 1976~\cite{may2004simple}.  Indeed, May's work entitled ``Simple mathematical models with very complicated dynamics," provides the underlying narrative for understanding damped-driven systems.  The popularization of the logistic map coincided with the advent of many modern concepts of dynamical systems, including the coining of the term ``chaos" by Yorke and Li in 1975~\cite{li2004period}.  The logistic map is comprised of two terms:  a linear growth term and a nonlinear loss.  For a given quantity of interest ${\bf x}_n={\bf x}(t_n)$, the linear growth is simply modeled by 
\begin{equation}
    {\bf x}_{n+1} = r {\bf x}_n 
\end{equation}
where we consider discrete time for simplicity and $r$ is a coefficient that specifies the growth rate.  For $r>1$, the quantity ${\bf x}_n$ grows over time.  For $r<1$, the quantity decays to zero.  The simplest nonlinear loss model one can construct is 
\begin{equation}
    {\bf x}_{n+1} = -r {\bf x}_n^2 
\end{equation}
where again the parameter $r$ dictates the loss rate.  Note that the larger the quantity ${\bf x}_n$ the stronger the losses.  The logistic map simply combines these two basic components into a single model
\begin{equation}
    {\bf x}_{n+1} = r {\bf x}_n \left( 1- {\bf x}_n \right) .
    \label{eq:logistic}
\end{equation}
Despite its simplicity, its dynamics as a function of the bifurcation parameter $r$ are quite complicated.  Indeed, as we will show, this simple mapping represents a normal form for the gain-loss dynamics of damped-driven systems, capturing all the salient bifurcation features of diverse and complex physical systems. 

Experimental observations of a diverse number of physical systems, many of which are considered in detail in subsequent sections of this manuscript, are known to generate a period-doubling route to chaos.  The connection of these systems to the logistic map are thought to be anecdotal at best. Specifically, such analogies with the logistic map have been observed in the underlying energy balance physics of mode-locked lasers~\cite{haus2000mode,kutz2006mode}, where global gain dynamics produce a similar cascading bifurcation diagram of mode-locked states~\cite{li2010geometrical,spaulding2002nonlinear,ding2009operating,bale2009transition}.  This has been similarly observed in rotating detonation engines (RDEs)~\cite{anand2015characterization,anand2017amplitude,bennewitz2019modal,koch2020mode,koch2020modeling,koch2021multiscale,koch2021data}
whereby the nonlinear dynamics are governed by the interaction physics of global gain (fuel) depletion and recovery along with local dominant balance physics characterized by the Burgers’ equation~\cite{majda1981qualitative}.
Koch et al provide a theoretical model that captures the experimental cascade of bifurcations and flame-front solutions whose attracting nature is termed mode-locked rotating detonation waves~\cite{koch2020mode,koch2020modeling,koch2021multiscale}.  More recently, the quantum hydrodynamic analog (HQA) system~\cite{CPFB05, CouderFort06} whereby a droplet evolves on a vibrating fluid bath.  The gain-loss dynamics~\cite{rahman2022walking} in this case also display the hallmark features of the instabilities observed in lasers and rockets.

The goal of this work is to show that the connection between damped-driven systems and the logistic map is more than an analogy.  Specifically, we show how a diverse number of damped-driven systems, all subject to significantly different physical processes, map their gain-loss dynamics (energy balance) to the logistic equation and inherit its underlying bifurcation structure.  Thus a quantitative connection can be made to the logistic map as an underlying normal form for the bifurcations in the energy variables (gain-loss dynamics) that occur in damped-driven systems.  Understanding this connection allows one to potentially {\em engineer} their way out of the deleterious effects of the instability mechanisms by understanding the mapping to the bifurcation parameter $r$.  Thus a pathway to circumventing instability exists in many damped-driven systems by simply engineering the overall loss and gain profiles of a system.

\section{The Logistic Map:  A review} \label{sec:Logistic}

\begin{figure}[t]
\includegraphics[width=7cm]{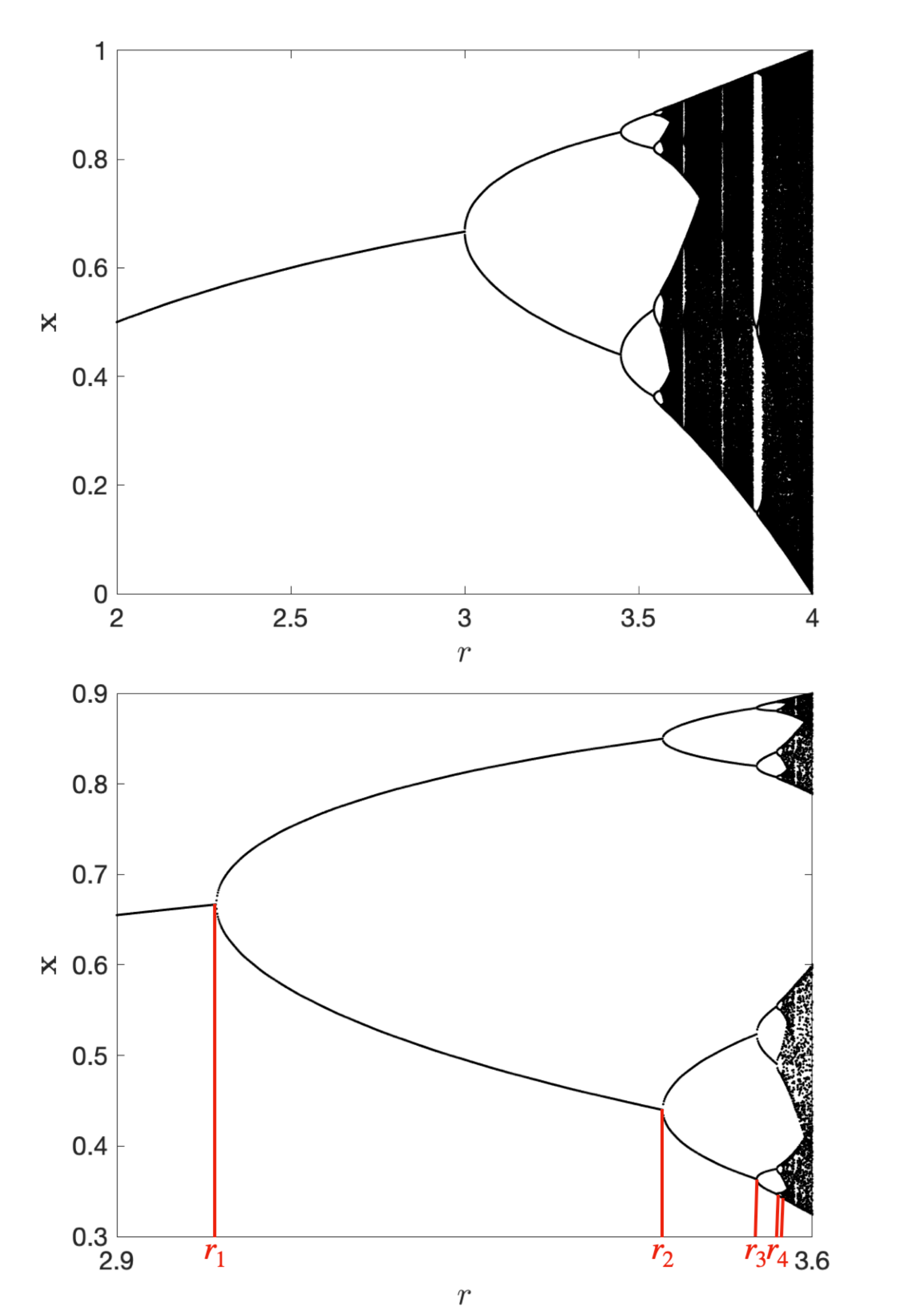}
\caption{Top: the period-doubling bifurcation diagram of the logistic function \eqref{eq:logistic} which plots the attractor versus the bifurcation parameter $r$. Bottom: A zoom of the bifurcation diagram to $2.9 \leq r \leq 3.6$ which shows the first locations of the period-doubling bifurcations, denoted $r_n$, which are used to define the Feigenbaum constant \eqref{Feigenbaum}}
\label{fig:LogisticBifDiag}
\end{figure}

The logistic map \eqref{eq:logistic} has become a simple paradigm for understanding the emergence of chaos through a sequence of period-doubling bifurcations (See Figure~\ref{fig:LogisticBifDiag}). Analysis of the system is well-documented and has become the standard example in papers and textbooks that seek to explain chaos; see for example \cite{Logistic1,Logistic2,Logistic3,Logistic4}. Therefore, in this section we will exclusively highlight the important aspects of the logistic map as they apply to our discussion below. For a more in-depth analysis of the logistic map we direct the reader to the aforementioned references.

The natural setting for the logistic map is to take $r \geq 0$ and $0 \leq x \leq 1$, although similar results to those that follow are known to hold for $r \leq 0$ and $-\frac 1 2 \leq x \leq \frac 3 2$ \cite{Logistic1}. First, the map has exactly two fixed points for all parameter values:
\begin{equation}
    {\bf x} = 0, \frac{r-1}{r},
\end{equation}
which are referred to as the trivial and nontrivial equilibria, respectively. For $0 < r < 1$ every initial condition of the map belonging to the interval $[0,1]$ eventually converges to the trivial state. However, as $r$ crosses 1 a transcritical bifurcation takes place, whereby the trivial and nontrivial equilibria exchange stability. The result is that for all $1 < r < 3$ every initial condition taken from $[0,1]$ asymptotically approaches the nontrivial state $(r-1)/r$. 

At $r = 3$ the complex dynamics of the logistic map begin to emerge. Indeed, at $r = 3$ the nontrivial equilibrium destabilizes in a period-doubling bifurcation, resulting in a stable period 2 orbit which oscillates between the values
\begin{equation}\label{logisticPer2}
    {\bf x}_\pm = \frac{1}{2r}\bigg(r+ 1 \pm \sqrt{(r-3)(r+1)}\bigg).
\end{equation}
This stable period 2 oscillation is not long-lived in the logistic map \eqref{eq:logistic} as it destabilizes at $r = 1 + \sqrt{6}$ via another period-doubling bifurcation. For $r$ values slightly above $1 + \sqrt{6}$ the resulting system dynamics are such that there is a stable period 4 oscillation, while the period 2 oscillation \eqref{logisticPer2} and the nontrivial equilibrium are now unstable. Continuing to increase the value of the parameter $r$ results in successive period-doubling bifurcations whereby the period 4 orbit destabilizes giving way to a period 8 orbit which eventually destabilizes and gives way to a period 16 orbit and so on. These period doubling bifurcations come in rapid succession so that a period $2^n$ orbit destabilizes through a period-doubling bifurcation to a stable period $2^{n+1}$ orbit.   

Let $\{r_n\}_{n = 0}^\infty \subset [0,4]$ be the monotone increasing sequence of points where a period $2^n$ orbit undergoes a period-doubling bifurcation. The sequence $r_n$ forms a convergence sequence with approximate limit point $r_* \approx 3.56995$. Furthermore, the range of parameter values for which a period $2^n$ orbit is stable is given by $r_{n} - r_{n-1}$, referred to as the period $2^n$ window. Feigenbaum \cite{feigenbaum1978quantitative} computed the ratio between successive windows widths to be 
\begin{equation}\label{Feigenbaum}
    \delta := \lim_{n \to \infty} \frac{r_{n -1} - r_{n-2}}{r_n - r_{n-1}} \approx 4.66920,
\end{equation}
which is now referred to as the Feigenbaum constant. Figure~\ref{fig:LogisticBifDiag} provides the first few $r_n$ values for visual reference. What makes the Feigenbaum constant interesting is that it is universal. That is, performing a similar bifurcation analysis on mappings with a single quadratic maximum, so-called unimodal maps, return the same Feigenbaum constant when tracking the ratio of successive window widths. Other examples of unimodal maps that are used throughout the literature include the quadratic map
\begin{equation}\label{eq:Quad}
    {\bf x}_{n+1} = r - {\bf x}_n^2,
\end{equation}
the Ricker function \cite{Ricker}
\begin{equation}\label{eq:Ricker}
    {\bf x}_{n+1} = {\bf x}_n\mathrm{e}^{r(1 - {\bf x}_n)},
\end{equation}
and the hyperbolic tangent map \cite{tanh}
\begin{equation}\label{eq:tanh}
    {\bf x}_{n+1} = r{\bf x}_n(1 - \tanh({\bf x}_n)),
\end{equation}
 all of which lead to the Feigenbaum constant when tracking the stable period $2^n$ window widths. This ``universality'' of the Feigenbaum constant was first articulated through a series of numerical computations. Landford used computer-assisted methods to first prove that all one-dimensional mappings undergo a sequence of period-doubling bifurcations at the same rate \cite{lanford2017computer} - that is according to the Feigenbaum constant. Corrections to Landford's proof were made by Eckmann and Wittwer \cite{eckmann1987complete}, while the first non-numerically assisted proof came in 1999 with the work of Lyubich \cite{lyubich1999feigenbaum}.

Beyond $r_* \approx 3.56995$ the dynamics of the logistic map are chaotic for most parameter values. However, there are {\em islands of stability} where periodic oscillations become stable. Periodic oscillations with any odd period can be found and are destabilized again through sequences of period-doubling bifurcations \cite{may2004simple,LogisticEcon}. Again, the period-doubling windows obey the Feigenbaum constant, giving another testament to its universality. For example, at $r = 1 + \sqrt{8}$ a range of $r$ values begins for which a period 3 orbit is stable giving way to another period doubling cascade into chaos. It was proven in \cite{lyubich1997dynamics,graczyk1997generic} that the set of $r$ for which the logistic map has a periodic attractor with all other periodic orbits hyperbolic, termed regular parameter values, is an open and dense set of $1 \leq r\leq 4$. Parameter values for which the dynamics do not settle into a periodic attractor are referred to as stochastic values and constitute almost every value of $r$ that is not regular \cite{lyubich2002almost}.    

Our fixation in this section on the logistic map is not only for exhibition as nearly every detail of it extends to general parameter-dependent unimodal maps \cite{kozlovski2003axiom,avila2003regular}. In particular, pioneering work by Milnor and Thurston demonstrated the qualitative universality of the logistic map \cite{milnor1988iterated} and under some restrictions every unimodal map generates equivalent dynamics to the logistic map for some value of $r$ \cite{guckenheimer1979sensitive}. Equivalent dynamics in this case are quantified by the existence of a topological conjugacy that provides an invertible change of coordinates to recast the iterative scheme as the logistic map. Thus, unimodal maps have much more than just the Feigenbaum constant in common with the logistic map. 

Beyond just unimodal maps, period-doubling cascades into chaos have been observed in a multitude of chaotic models. This includes simple toy models, such as the R\"ossler system \cite{BramUPO,wilczak2009period}, partial differential equations, such as the Kuramoto--Sivashinsky equation \cite{KSE1,KSE2}, and delay differential equations, such as the Mackey--Glass equation \cite{mackey1977oscillation,junges2012intricate}. Moreover, at least in the case of the Kuramoto--Sivashinksy equation there is numerical evidence showing that the period-doubling follows the Feigenbaum constant. Thus, one is tempted to conjecture that there exists a precise manner by which one can recast the dynamics of these high-dimensional systems as the simple logistic map \eqref{eq:logistic} that appears to emerge as a normal form for destabilizing cascades into chaos.



\section{Energy Balance and Dynamics}

The dynamics of damped-driven systems are characterized by the energy balance that occurs in the system.  As will be shown in our diverse examples, irrespective of the underlying physics of the system, the energy balance dictates the resulting observed instability.  Thus as energy is pumped into the system, energy must be re-organized and dissipated to accommodate the energy gain.  The re-organization and dissipation of the energy can happen in a myriad of ways which largely depends upon the particular system under consideration.   An energy balance is achieved when the driving input energy is equal to the energy losses.  A rather simple geometrical conception of this energy exchange can be constructed, showing how the instability cascade becomes universal in damped-driven systems.

\begin{figure}[t]
\centering
\begin{overpic}[width = 9cm]{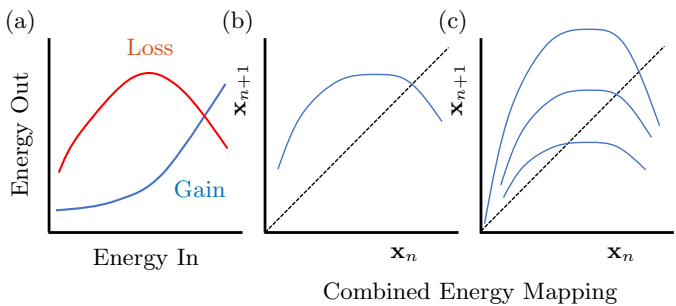}
\put(11,22){Energy In}
\put(-1,32){\rotatebox{90}{Energy Out}}
\put(16,53){\color{red}{Loss}}
\put(23,32){\color{blue}{Gain}}
\put(55,23){${\bf x}_n$}
\put(32,44){\rotatebox{90}{${\bf x}_{n+1}$}}
\put(87,23){${\bf x}_n$}
\put(64,44){\rotatebox{90}{${\bf x}_{n+1}$}}
\put(45,17){Combined Energy Mapping}
\put(-2,57){(a)}
\put(30,57){(b)}
\put(62,57){(c)}
\end{overpic}
\vspace*{-.60in}
\caption{Geometrical description of damped-driven systems.  Panel (a) shows the basic and underlying assumptions of the energy dynamics:  the gain (blue curve) increases monotonically and the losses (red curve) become increasingly nonlinear and significant with increasing energy, i.e. there are many routes for dissipation in the system for large energy inputs.  The intersection of the loss and gain represents the energy balanced solution. Panel (b) shows the combined gain and loss iterative dynamics (denoted by ${\bf x}_n$) in a single curve which is like the mapping (\ref{eq:logistic}).  Panel (c) shows that different gain-loss curves can result in a different iterative maps which are equivalent to modifying the parameter $r$ in (\ref{eq:logistic}).  }
\label{fig:balance}
\end{figure}

We can construct a geometrical viewpoint of the gain-loss dynamics by mapping gain and loss as input-to-outputs over time.  Thus a discrete map is constructed to describe the input-to-output relationship of the gain and a corresponding map for the loss. There are two generic assumptions required for understanding the bifurcation sequence resulting in damped-driven system.  The first assumption concerns the gain, which assumes that there exists some mapping between input to output gain.  Figure~\ref{fig:balance}(a) shows a generic gain curve (blue line) that relates the input to output of the gain (driving force) dynamics.  In many continuous systems, the simplest gain system is modeled by the differential equation $dE/dt=g E$ whose solution is an exponential gain $E(t)=E(0) \exp(gt)$.  Here $E(t)$ is the energy and $g$ parametrizes the strength of the gain.  The second assumption concerns the loss, which assumes that as the energy in the system increases, there are more and more physical processes that dissipate or eject energy from the system.  Thus the diffusion of energy becomes a strongly nonlinear loss mechanism if enough energy is driven into the systems and as the system finds potentially a multitude of ways for energy to be damped out of the system, increasing losses significantly for higher energy input.  Figure~\ref{fig:balance}(a) shows a generic loss curve (red line) where the loss mechanisms create a fold in the input-to-ouput relation.  

Alternatively, the gain and loss curves can be combined into a single input-to-output relationship as shown in Fig.~\ref{fig:balance}(b) (blue curve).  In this figure the diagonal dotted line represents a steady-state solution where the input matches the output.  This point is where the gain and loss are equal in Fig.~\ref{fig:balance}(a) when the blue and red curves intersect.  Figure~\ref{fig:balance}(b) is the equivalent representation of the iterative dynamics of (\ref{eq:logistic}), but with a more complicated right hand side.  As will be shown later, the generic curve shown in Fig.~\ref{fig:balance}(b) can be transformed to the logistic map using deep learning algorithms (See Sec.~\ref{sec:deep_logistic}).  Finally, Fig.~\ref{fig:balance}(c) shows the combined curve for various gain-loss dynamics, showing that the resulting iterative mapping depends greatly on where and when the gain and loss dynamics intersect.  In the logistic map, this dependence is captured by the parameter $r$ in (\ref{eq:logistic}).  Thus with minimal assumptions, that the gain increases monotonically and the losses are higher with increased energy, a generic mapping is created in the energy landscape capable of producing the underlying dynamics of instability captured by the logistic map.

\section{Deep Learning of the Logistic Map}\label{sec:deep_logistic}

The previous section outlined the basic energy balance dynamics of damped-driven systems.  Figure~\ref{fig:balance} is a generic representation of the loss and gain curves that are present in physical systems, as will be shown in detail in the forthcoming section.  With the advent of deep learning methods for dynamical systems, we are able to directly link these loss-gain curves to the logistic map ${\bf x}_{n+1} = r{\bf x}_n(1-{\bf x}_n)$. That is, as discussed in Section~\ref{sec:Logistic}, when a loss-gain curve leads to a unimodal map there exists an invertible change of variable (a topological conjugacy) that transforms the dynamics to the logistic map with some parameter $r$. With the widespread availability of powerful methods from machine learning, we are now able to simultaneously learn this invertible change of variable and the corresponding logistic parameter $r$ using only data gathered from the damped-driven system.  

First, recall that two functions $f:X\to X$ and $g:Y\to Y$ are said to be conjugate if there exists a homeomorphism (or conjugacy) $h:X\to Y$ so that $g \circ h = h \circ f$. The invertibility of $h$ guarantees that we can equivalently write $f = h^{-1}\circ g \circ h$, resulting in the commutative diagram
\begin{equation}\label{ComDiag}
	\begin{tikzcd}
        X \arrow{r}{f} \arrow[swap]{d}{h} &  X\arrow{d}{h} \\
        Y\arrow{u} \arrow{r}{g} & Y\arrow{u}
    \end{tikzcd}
\end{equation} 
which illustrates the dynamical equivalence between orbits generated by $f$ and $g$. In \cite{bramburger2021deep} a method of using deep autoencoder neural networks is proposed for indirect data-driven discovery of $f$ based upon the commutative diagram \eqref{ComDiag}. In what follows we will describe this method only as it pertains to teasing out the logistic dynamics from a damped-driven system.

Our goal is to identify the logistic map dynamics only from data gathered from a damped-driven system. To this end, we will fix $g(x) = rx(1-x)$, where the value of $r$ will be identified through the training process. We begin by gathering input $\{x_1,x_2,x_3,\dots,x_n\}$ and output $\{y_1,y_2,y_3,\dots,y_n\}$ data from a loss-gain curve that corresponds to an unknown function $f$ by $y_n = f(x_n)$. Suppose further that the loss-gain dynamics are unimodal, and therefore based on the work in Section~\ref{sec:Logistic} we wish to identify its conjugate logistic map. To do this, we employ two separate neural networks to identify the conjugacy $h$ and its inverse $h^{-1}$ while treating $r$ as a trainable network parameter.  

\begin{figure}[t]
\includegraphics[width=8cm]{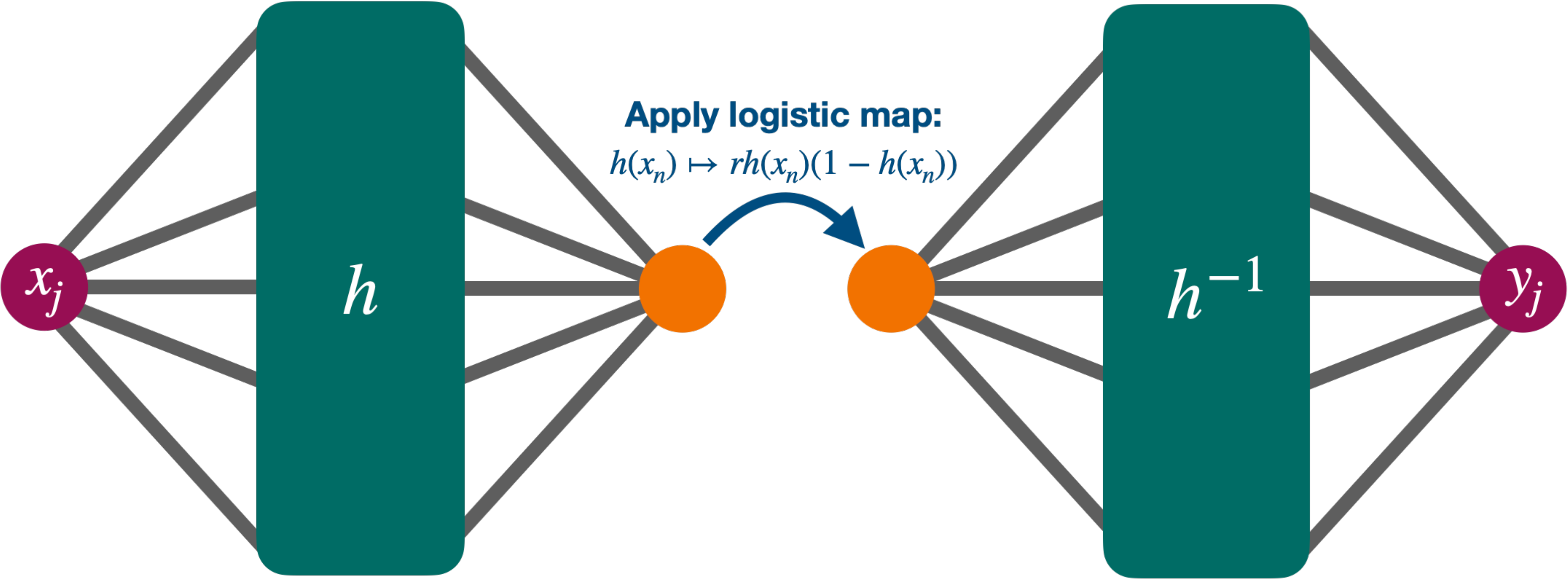}
\begin{overpic}[width=8cm]{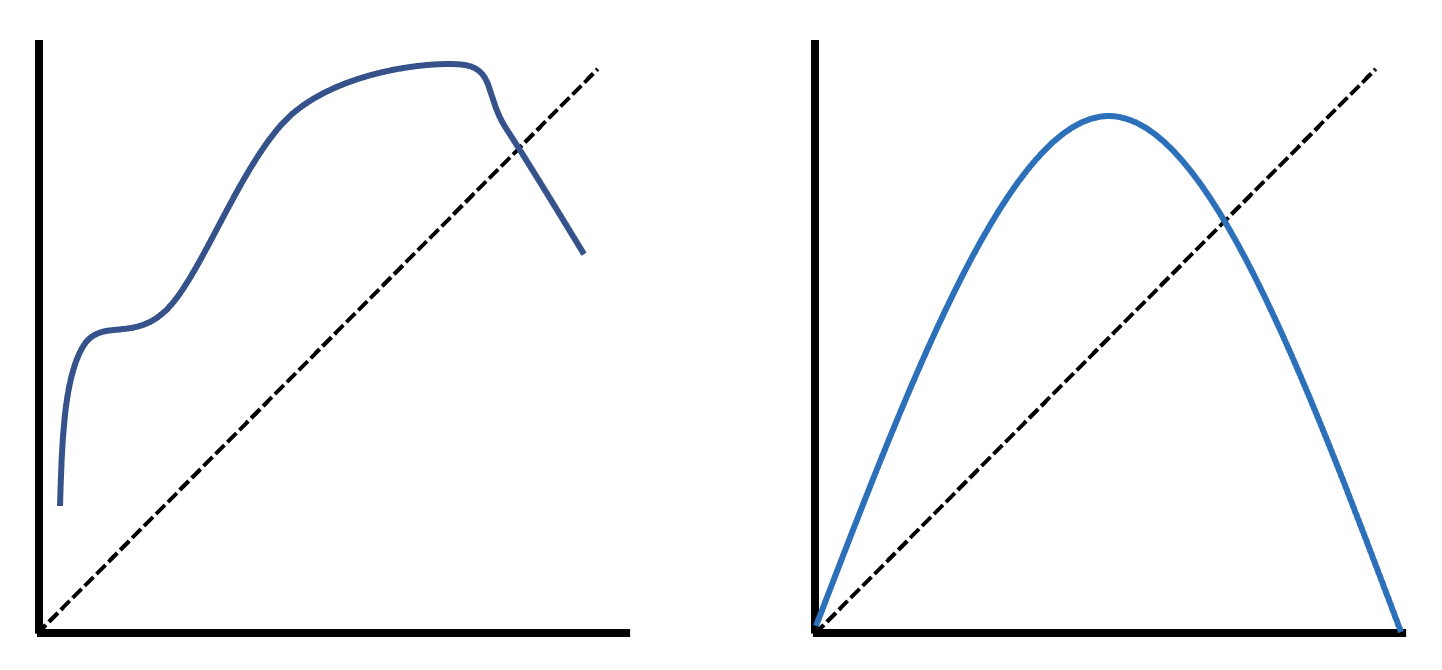}
\put(-5,80){(a)}
\put(-5,35){(b)}
\put(44,25){$h({\bf x}_n)$}
\put(46,20){$\longrightarrow$}
\end{overpic}
\caption{\label{fig:NN_Conj} Using neural networks to identify topological conjugacies with the logistic function. (a) The conjugacy $h$ and its inverse $h^{-1}$ are approximated with neural networks, while moving left to right in the diagram corresponds to moving through the commutative diagram \eqref{ComDiag} via $f = h^{-1}\circ g\circ h$ with $g$ a logistic function. (b) Depiction of the learned transformation of a generic combined gain-loss curve of Fig.~\ref{fig:balance}(b) to the logistic map $rh({\bf x}_n(1-h({\bf x}_n))$}.
\end{figure}

To achieve the desired numerical conjugacy we train the neural network using a loss function which is the sum of three distinct losses, denoted $\mathcal{L}_1$ through $\mathcal{L}_3$. First we have that {\em conjugacy loss} 
\begin{equation}
    \mathcal{L}_1 = \frac{1}{2n}\sum_{j=1}^n |x_j - h^{-1}(h(x_j))|^2,
\end{equation}
which ensures that $h$ is a homeomorphism by identifying $h$ and $h^{-1}$ conditioned on the fact that $h^{-1}\circ h$ should be the identity. One may further append this loss by including a similar sum over the output training data $y_j$. Then we have the {\em prediction loss}
\begin{equation}
    \mathcal{L}_2 = \frac{1}{2n}\sum_{j = 1}^n |y_j - h^{-1}(g(h(x_j)))|^2,
\end{equation}
which learns the iterative dynamics of $f$ through the conjugacy relationship $f = h^{-1}\circ g\circ h$. Next we have the {\em logistic mapping loss} 
\begin{equation}
    \mathcal{L}_3 = \frac{1}{2n} \sum_{j = 1}^n |h(y_j) - g(h(x_j))|^2,
\end{equation}
which encodes the dynamics of the conjugate logistic function, $g$, in the transformed variables. Indeed, from the commutative diagram \eqref{ComDiag} we see that starting with $x_j$ and $y_j = f(x_j)$ we can follow both down into the domain of the logistic function by applying the conjugacy $h$. Finally, we may include a form of network regularization on the parameters in the networks used to build $h$ and $h^{-1}$ to avoid over-fitting. A sketch of the trained conjugacy network is provided in Figure~\ref{fig:NN_Conj}.   

Summing the losses $\mathcal{L}_1,\mathcal{L}_2$, and $\mathcal{L}_3$ results in loss function for training a conjugacy neural network that reflects all of the transformations demonstrated in the conjugacy diagram \eqref{ComDiag}. Successful applications of this method include relating the logistic mapping to the period-doubling cascade observed in Poincar\'e sections of the R\"ossler system, the Kuramoto--Sivashinksy equation, and the Mackey--Glass delay-differential equation \cite{bramburger2021deep}. These results make further use of the longstanding application of autoencoder neural networks to perform dimensionality-reduction \cite{kramer1991nonlinear}, which in this case can be used to remove redundancies in the input-output data $(x_j,y_j)$ and simply project onto the chaotic attractor. The dimension of this attractor can be estimated with the Lyapunov exponents and the the Kaplan--Yorke dimension \cite{Kaplan}. Full Jupyter notebooks that implement this method can be found in the repository \cite{BramHub}.

\section{Physical Models and Systems}

In what follows, we highlight a diverse set of engineering and physics applications where the canonical damped-driven instabilities are observed.  The underlying physics are significantly different in each system, thus highlighting the universality of the energy balance dynamics irrespective of the particular physics.  Three models are extensively discussed:  Mode-locked lasers, rotating detonation engines and the hydrodynamic quantum analog system.  In addition, we feature several additional models in less detail, but highlight the underlying damped-driven behavior manifest. 

\subsection{Mode-locked lasers}
\label{Sec: Lasers}

Lasers are perhaps one of the most revolutionary technologies of the last century:  capable of reshaping light into coherent ensembles of photons that can be used across diverse technologies and applications.  To begin the analysis of lasers, two key physical effects will be considered:  the saturating gain and the nonlinear losses due to the intensity discrimination element (e.g. a saturable absorber).  Understanding the interplay of these two effects alone allows one to understand how an energy balance and a mode-locked (localized, coherent light) state can be achieved.  Further, this simple model allows one to understand one of the most characteristic bifurcations observed in practice, the multi-pulsing instability.  What follows is an analysis based upon an iteration method and simple first order differential equations.  The advantages of the proposed analytic description is that it relates directly to a fairly simple geometrical description of the cavity dynamics, thus allowing for an intuitive understanding of the mode-locking dynamics.  It further highlights the energy balance and its ability to create interesting dynamical (and relevant) states of the system without anything being driven by additional nonlinearity (self-phase modulation) and dispersion.

\begin{figure}[t]
\includegraphics[width=7cm]{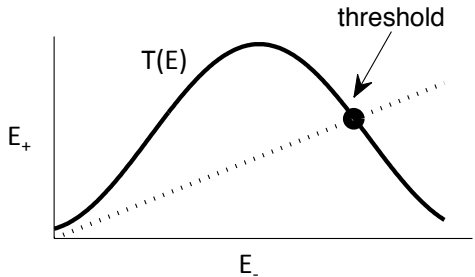}
\vspace*{-0.4in}
\caption{A generic nonlinear loss (or saturable absorption or saturable fluency)
curve (bolded line) showing the standard effect of saturable absorption at high
energies along with a fold over due to higher-order nonlinear loss processes.
The dashed line is a multi-pulsing threshold curve.  Once the input energy
is increased above the threshold point, any perturbation will cause the growth of
an additional pulse so the cavity jumps from $N$ to $N+1$ pulses.
Note that the small and large signal transmission curves (dashed and solid lines respectively) coalesce for low input energies.}
\label{fig:poo1}
\end{figure}
\begin{figure}[t]
\includegraphics[width=8cm]{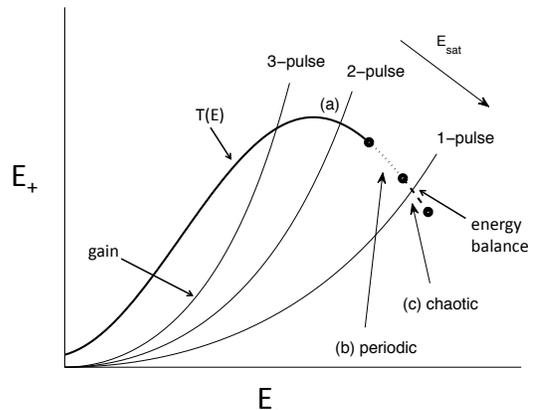}
\vspace*{-0.3in}
\caption{\label{fig:poo2} Nonlinear loss and saturating gain curves for a 1-pulse ($N=1$), 
2-pulse ($N=2$) and 3-pulse ($N=3$) per round trip configuration.  The
intersection of the gain and loss curves represents the mode-locked solution
states of interest.  As the gain parameter $g_0$ is increased, the gain curves
shift to the right.  The 1-pulse solution first becomes periodic (b), then chaotic (c)
before ceasing to exist since it no longer intersects the loss curve.  The solution
then jumps to the next most energetically favorable configuration of 2-pulses
per round trip (a).  This qualitative picture describes the entire $N$ to $N+1$ 
pulse transition.}
\end{figure}

\begin{figure*}[t]
\begin{center}
\hspace*{-0.3in}
\includegraphics[width=16cm]{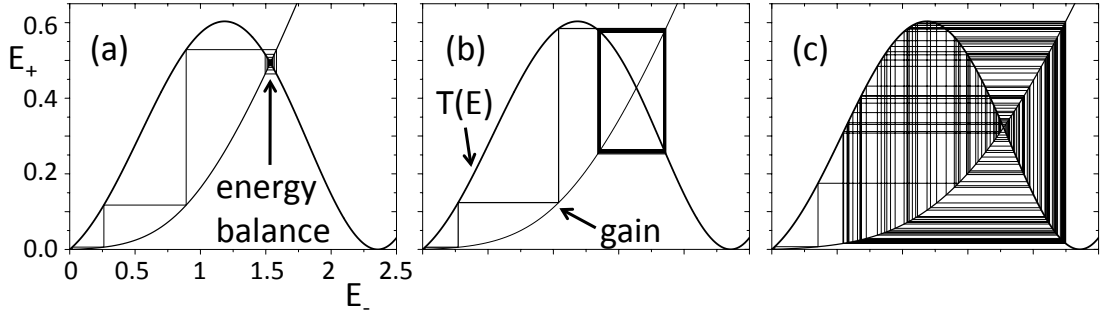}
\end{center} 
\vspace*{-3.0in}
\caption{\label{fig:poo3} Iteration map dynamics for the nonlinear loss and saturating
gain behavior.  Possible iteration behaviors are (a) a steady-state solution,
(b) a periodic solution, and (c) a chaotic dynamics.  }
\end{figure*}
\begin{figure}[t]
\centering
\includegraphics[width = 9cm]{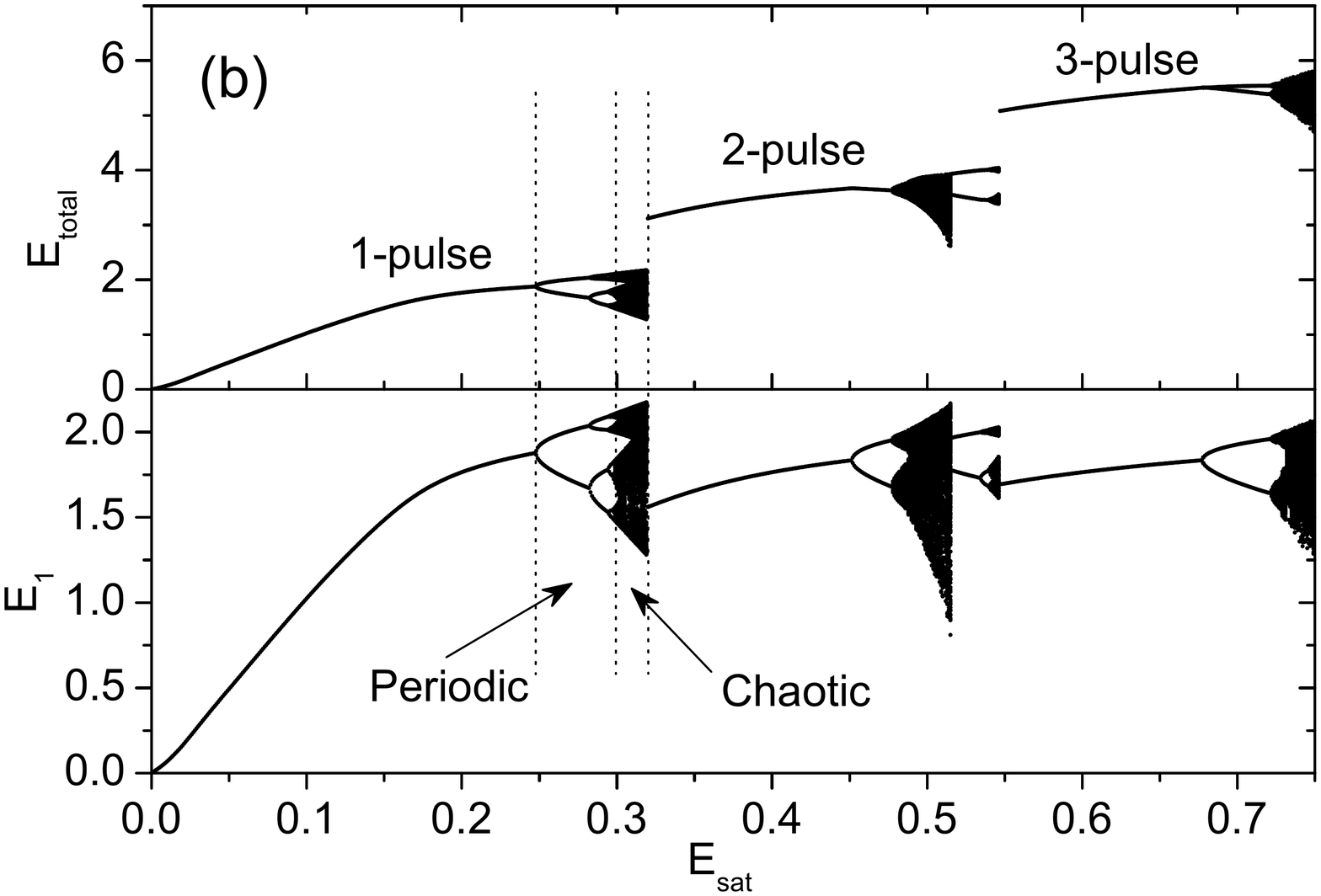}
\vspace*{-.20in}
\caption{Iteration map dynamics 
for the nonlinear loss and saturating gain behavior.  
Shown is the total cavity energy $E_{out}$ (top panel) and the individual pulse
energy $E_1$ (bottom panel) as a function of the cavity saturation energy $E_{sat}$.
The transition dynamics between multi-pulse operation produces a discrete
jump in the cavity energy.  In this case, both periodic and chaotic dynamics are observed
preceding the multi-pulsing transition. }
\label{fig:exam2b}
\end{figure}
The onset of multi-pulsing as a function of increasing laser cavity energy is a well-known physical phenomenon~\cite{haus2000mode,kutz2006mode} that
has been observed in a myriad of theoretical and experimental mode-locking studies in both passive and active laser cavities~\cite{namiki1997energy,kutz2008theory,bale2009transition,xing1999regular,soto2004bifurcations,collings1998stable,fermann1996cladding,grudinin1992energy,guy1994femtosecond,horowitz2000theoretical,davey1991interacting,lederer1999multipulse,lai1997multiple,wang1997pulse,kitano1998stable}.   Aside from two purely theoretical (computational) studies~\cite{namiki1997energy,kutz2008theory}, the
bulk of these observations have been almost exclusively experimental in nature.  Specifically, each of these experiments demonstrates that as the gain pumping is increased, 
the number of mode-locked pulses in the cavity increases in an approximately linear and discrete manner with the cavity saturation energy.  This observation is independent of the specific mode-locking mechanism used, whether it be nonlinear polarization rotation, nonlinear interferometry, quantum saturable absorbers, etc.  Thus the phenomenon is ubiquitous to mode-locked laser cavities.  One of the earliest theoretical descriptions of the multi-pulsing dynamics was by Namiki et al.~\cite{namiki1997energy} in which energy rate equations were derived for the averaged cavity dynamics.  More recently, a full stability analysis of the mode-locking solutions~\cite{kapitula2002stability} was performed showing that the transition dynamics between $N$ and $N+1$ pulses in the cavity exhibited a more complex and subtle behavior than previously suggested~\cite{kutz2008theory}.
Indeed, the theory predicted, and it has been confirmed experimentally since, that near the multi-pulsing transitions, both  
periodic and chaotic behavior could be observed as operating states of the laser cavity for a narrow range of parameter space~\cite{kutz2008theory,bale2009transition,xing1999regular,soto2004bifurcations,kutz1994pulse,hachey2010stability}.  Here, the energy rate equation approach~\cite{namiki1997energy} is generalized which results in an iterative
technique that provides a simple geometrical description of the entire multi-pulsing transition behavior as a function of increasing cavity energy~\cite{li2010geometrical}.  The model captures all the key features observed 
in experiment, including the periodic and chaotic mode-locking regions~\cite{bale2009transition}, and it further provides valuable insight into
laser cavity engineering for maximizing performance, i.e. enhancing the mode-locked pulse energy.

The multi-pulsing instability arises from the competition between the laser cavity's bandwidth constraints and the energy quantization associated with the resulting mode-locked pulses, i.e. the so-called soliton area theorem~\cite{namiki1997energy}.  Specifically, as the cavity energy is increased, the resulting mode-locked pulse has an increasing peak power and spectral bandwidth.  The increase in the mode-locked spectral bandwidth, however, reaches its limit once it is commensurate with the gain bandwidth of the cavity.   Further increasing the cavity energy pushes the mode-locked pulse to an energetically unfavorable situation where the pulse spectrum exceeds the gain bandwidth, thereby incurring a spectral attenuation penalty.  In contrast, by bifurcating to a two-pulse per round trip configuration, the pulse energy is then divided equally among two pulses whose spectral bandwidths are well contained within the gain bandwidth window.

Before proceeding to the analysis, a few comments are made about the 
variable naming conventions used in the manuscript. 
The variable $E$ will represent
the cavity energy.  However, it will rarely appear without superscripts or subscripts
when referring to the laser cavity.   The subscript {\em in} or {\em out} will refer to
the energy at the input or output of a laser cavity element respectively.  The superscript
{\em loss} and {\em gain} will refer to the laser cavity element under consideration.
For convenience, two expressions will be highlighted:   $E_\pm$ will
measure the electric field at the input to the loss element (-) or input to the gain element (+).
More precisely, $E_-=E_{in}^{loss}=E_{out}^{gain}$ and $E_+=E_{out}^{loss}=E_{in}^{gain}$.
Additionally, the expression $E_j$ refers to the energy of the $j$th pulse
at the output coupler and $E_{total}$ is the sum of the
energy of all the pulses.  The final term, $E_{sat}$, is the saturation energy of the
gain medium which is modified is practice as the gain pumping is adjusted in the laser cavity.\\

\noindent {\bf Saturating gain:} The saturating gain dynamics results in the following differential
equation for the gain~\cite{haus2000mode,kutz2006mode}:
\begin{equation}
  \frac{dE_j}{dZ}= \frac{  g_0}{1+ \sum_{j=1}^{N} E_j/E_{sat}} E_j
  \label{eq:gain}
\end{equation}
where $E_j$ is the energy of the $j$th pulse ($j=1, 2, \cdots N$), 
$g_0$ measures the gain pumping strength,
and $E_{sat}$ is the saturation
energy of the cavity.  The total gain in the cavity can be controlled by adjusting
the parameters $g_0$ or $E_{sat}$.  In what follows here, the cavity energy will
be increased by simply increasing the cavity saturation parameter $E_{sat}$.  This
increase in cavity gain can equivalently be controlled by adjusting $g_0$.
These are generic physical parameters that are
common to all laser cavities, but which can vary significantly from one
cavity design to another.   The parameter $N$ is the number of pulses in the 
cavity.  
This parameter, which is critical in the following analysis, 

helps capture the saturation energy received by each individual pulse.
Note that this equation can be solved exactly using
standard methods of differential equations.  However, the resulting solution
is given in implicit form.\\  

\noindent {\bf Nonlinear loss (saturable absorption):} The nonlinear loss in the cavity, i.e. the saturable absorption or saturation fluency curve, 
will be modeled by a simple transmission function:  
\begin{equation}
  E_{out}=T(E_{in}) E_{in} \, .
  \label{eq:trans}
\end{equation}
The actual form of the transmission function $T(E_{in})$ can vary significantly
from experiment to experiment, especially for very high input energies.  For instance,
for mode-locking using nonlinear polarization rotation, the resulting transmission 
curve is known to generate a periodic structure at higher intensities.  Alternatively,
an idealized saturation fluency curve can be modified at high energies due to
higher-order physical effects.  As an example, in mode-locked cavities using waveguide
arrays~\cite{kutz2008theory}, the saturation fluency curve can turn over at high energies due to the
effects of 3-photon absorption, for instance. As a final note, this transmittance function is commonly referred to in the literature
as the cavity's nonlinear loss or saturable absorption~\cite{haus2000mode,kutz2006mode,namiki1997energy,li2010geometrical}.  
In what follows, the terms {\em nonlinear loss} and {\em transmission curves} will be used 
interchangeably.\\

\noindent {\bf Iterative Cavity Dynamics:} The generic loss curve illustrated in Figure~\ref{fig:poo1} along with the
saturable gain as a function of the number of pulses (\ref{eq:gain}) are the
only two elements required to completely characterize the multi-pulsing
transition dynamics and bifurcation. The alternating action of saturating gain and nonlinear
loss produce an iteration map for which only pulses whose loss and gain
balance are stabilized in the cavity.  Specifically, the output of the gain is the input
of the nonlinear loss and vice-versa. This is much like the logistic equation
iterative mapping for which a rich set of dynamics can
be observed with a simple nonlinearity~\cite{drazin1992nonlinear,devaney2018introduction}.  Indeed,
the behavior of the multi-pulsing system is qualitatively similar to the
logistic map with steady-state, periodic and chaotic behavior all
potentially observed in practice.

\begin{figure}[t]
\centering
\includegraphics[width = 8cm]{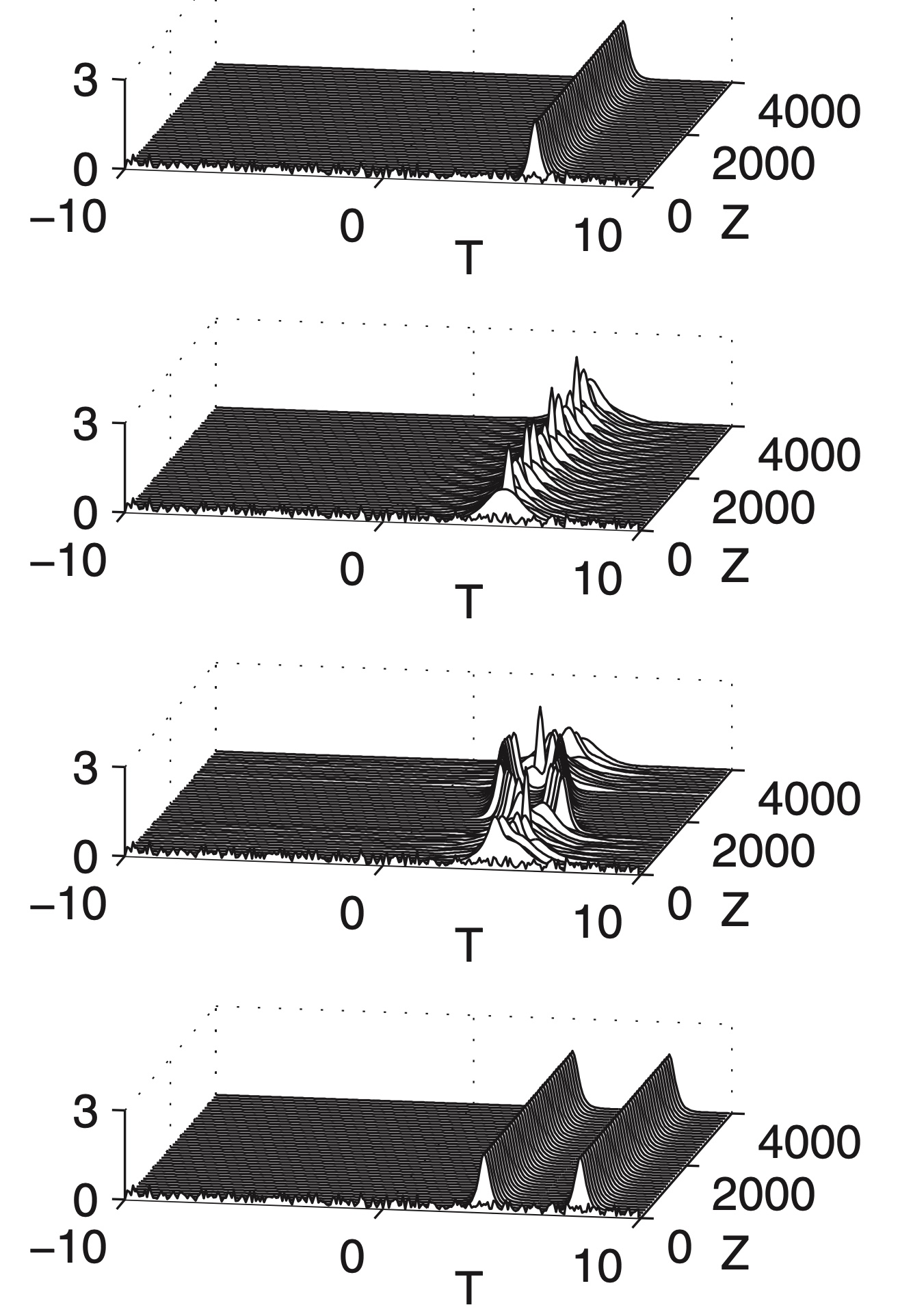}
\vspace*{-.10in}
\caption{Mode-locking cavity simulations where the nonlinear saturable absorption is provided by waveguide arrays. Shown is the intensity of the mode-locked field as a function of normalized propagation distance $z$ and time $t$. As the cavity gain is increased via $g_0$, the stable one-pulse configuration first bifurcates to a periodic solution, and then bifurcates again to a chaotic solution, before finally going to the two-pulse configuration. This cascade of instabilities is a manifestation of the dynamics shown in Figs.~\ref{fig:poo1}-\ref{fig:exam2b}. }
\label{fig:bifurcation}
\end{figure}
%

In addition to the connection with the logistic equation framework,
two additional features are particular to this problem formulation.
First, there are multiple branches of stable solutions, i.e. the 1-pulse,
2-pulse, 3-pulse, etc.  Second the loss curve exceeds the threshold 
energy as shown in Figure~\ref{fig:poo1}.  
{\em Figure~\ref{fig:poo2} is the key figure
of this section and exhibits all the critical features of the energy balance and
multi-pulsing model}.
Exhibited in this model are the input and output relationships
for the gain and loss elements.  Three
gain curves are illustrated for (\ref{eq:gain}) with $N=1$, $N=2$
and $N=3$.  These correspond to the 1-pulse, 2-pulse and 3-pulse per
round trip solutions respectively.   As the gain is increased
through the cavity saturation energy $E_{sat}$, the gain solution
curves move to the right in the graph, producing steady-state (a),
periodic (b) and chaotic (c) behavior 
Figure~\ref{fig:poo3} demonstrates the iterative process for
the generic gain and loss curves illustrated in Fig.~\ref{fig:poo2}.
Specifically, the production of a steady-state (a), periodic (b)
and chaotic (c) behavior in the system is illustrated.  
In the current scenario, the 1-pulse solution branch has reached a point in the solution space where
the iteration between the gain and loss dynamics produces chaotic
energy fluctuations in the laser cavity.  If the gain is further increased, 
the 1-pulse solution branch moves past the threshold point and no
1-pulse per round trip solutions are stable any longer.  Instead, the
mode-locking moves to a multi-pulsing configuration with a higher
number of pulses.  

Figure~\ref{fig:exam2b} shows the quantitative versions of the curves qualitatively represented in Fig.~\ref{fig:poo2}.
Specifically, the nonlinear loss curve along with the gain curves of the 1-pulse, 2-pulse 
and 3-pulse mode-locked solutions are given along with the threshold point as before.
As the cavity energy is increased through an increasing value of $E_{sat}$. The $1$-pulse
solution becomes unstable to the $2$-pulse solution as expected.  In this case,
the computed threshold value does extend down the loss curve to where the
periodic and chaotic branches of solutions occur, thus allowing for the
observation of periodic and chaotic dynamics.  
The multi-pulsing bifurcation occurs as depicted
in Fig.~\ref{fig:exam2b}.  The total cavity energy along with
the single pulse's energy is depicted as
a function of increasing gain.  This process repeats
itself with the transition from $N$ to $N+1$ pulses generating periodic and then chaotic
behavior before the transition is complete.  
This curve is in complete agreement with recent experimental and theoretical
findings~\cite{kutz2008theory,bale2009transition}, thus validating the predicted dynamics.


Generically, this process of increasing the gain shows explicitly how
the mode-locked laser jumps from $N$ to $N+1$ pulses per round trip.
It is simply a consequence of the $N$ solution branch exceeding
the threshold point of the nonlinear loss curve where that particular solution no longer is stable.  
This forces the dynamics to a higher number of pulses per round trip.   
Moreover, depending upon the curvature of the nonlinear loss
curve for high-energy, the transition dynamics can exhibit periodic (b) and
chaotic dynamics (c) before the onset of steady-state multi-pulsing (a).

\noindent {\bf Spatio-Temporal Models of Mode-Locking:}  There are a diversity of models that have been developed to capture the loss and gain mechanisms within a laser cavity~\cite{haus2000mode,kutz2006mode}.  These models capture the full spatio-temporal evolution of the electromagnetic field and can be derived directly from Maxwell's equations.  We highlight one specific laser cavity whose loss mechanism is generated by nonlinear mode-coupling~\cite{kutz2008theory,bale2009transition}. The resulting approximate evolution dynamics describing the
waveguide array mode-locking is given by
\begin{subeqnarray}
&& \hspace*{-.47in}   i\frac{\partial u}{\partial z} \! + \! \frac{D}{2}\, 
  \frac{\partial^2 u}{\partial t^2} \!\!+\!\! \beta |u|^2 u \!\!+\!\!  {C}  v
  \!\!+\!\! i\gamma_0 u 
  \!\!-\!\! ig(z) \! \left(\!\! 1\!\!+\! \!\tau\frac{\partial^2}{\partial t^2} \!\! \right)\!u\!=\!0    \\
&& \hspace*{-.47in}   i\frac{\partial v}{\partial z} \!+\! {C}\! \left(w\!+\! u\right)
  \!+\! i\gamma_1 v =0  \\
&&  \hspace*{-.47in}  i\frac{\partial w}{\partial z} \!+\! {C} v
  \!+\! i\gamma_2 w = 0  \, , 
\label{eq:oneside2}
\end{subeqnarray}
where
\begin{equation} 
g(Z) = \frac{2g_0}{1+\|Q\|^2/e_0} \, ,
\end{equation}
is the saturating cavity gain.  The variables are all normalized and represent the electric field envelopes in the central ($u(z,t)$) and neighboring waveguides ($v(z,t)$ and $w(z,t)$).  The variable $t$ represents time
in the rest frame of the mode-locked pulse, while $z$ is the propagation distance in the laser cavity.   The bandwidth limited gain in the fiber is incorporated through the dimensionless parameters 
$g$ and $\tau$.  The parameter $\tau$ controls the spectral gain bandwidth of the
mode-locking process, limiting the pulse width.    
 
Simulations of the governing equations (\ref{eq:oneside2}) generically produce the multi-pulsing instability shown in Fig.~\ref{fig:bifurcation}.  As noted, this bifurcation sequence is observed in a myriad of models as well as experiments~\cite{namiki1997energy,kutz2008theory,bale2009transition,xing1999regular,soto2004bifurcations,collings1998stable,fermann1996cladding,grudinin1992energy,guy1994femtosecond,horowitz2000theoretical,davey1991interacting,lederer1999multipulse,lai1997multiple,wang1997pulse,kitano1998stable}.  This illustrates how the dynamics, instabilities and bifurcations of a complex spatio-temporal PDE can be reduced to a simple energy balance description.

\subsection{Rotating detonation engines}

\begin{figure}[]
\begin{center}
\begin{overpic}[width = 8cm]{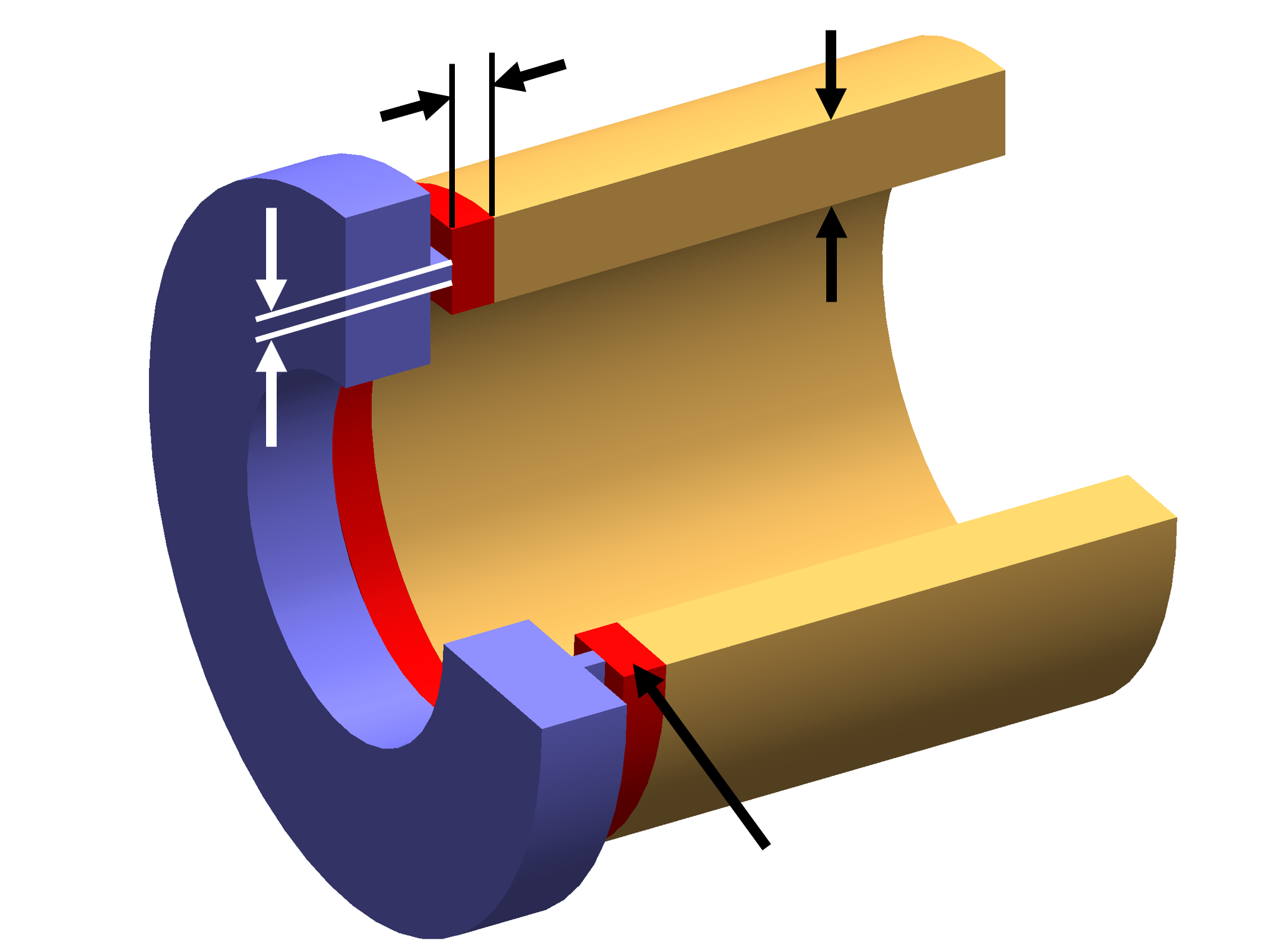}
\put(23,38){\color{white} Injection Area}
\put(62,45){\color{black} Combustor Area}
\put(20,71){\color{black} Reactant fill depth}
\put(55,3){\color{black} Reaction zone}
\end{overpic}
\end{center} 

\caption{\label{fig:rde_01} Schematic of notional annular-type combustor for housing rotating detonations. In this configuration, gaseous fuel and oxidizer are fed into the annular chamber through an injection scheme at or near the head-end of the device. After ignition (via an automotive spark plug, for example), the flowfield can self-organize into a number of rotating combustion fronts that travel about the annular chamber so long as new fuel and oxidizer are supplied. High velocity exhaust exits the device at the aft end. In propulsion applications, this produces usable thrust. Combustion dynamics are largely influenced by modifying time scales through combustor geometry (e.g. injection-to-combustor area ratio) or injection schemes.}
\end{figure}
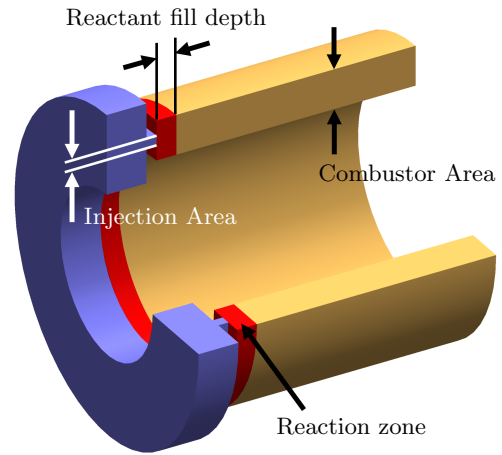

A chemically reacting fluid flow -- such as those found in continuous-flow combustion devices -- can develop \textit{thermoacoustic instabilities} when fluid flow fluctuations couple with heat release fluctuations \cite{lieuwen2021unsteady}. In many settings, this coupling leads to positive feedback whereby a perturbation in heat release induces large gradients in pressure and density, further accelerating chemical kinetics. This constitutes the satisfaction of the so-called Rayleigh criterion \cite{rayleigh1878explanation,durox2009rayleigh}, where the local energy gain in a fluid (e.g. through combustion) exceeds dissipative losses (e.g. turbulence or unconstrained expansion). The manifestation of these instabilities can vary from the benign, such as an oscillating flame (as in Huang et al. \cite{huang2004bifurcation}), to catastrophic, such as the destructive instabilities that plagued the Apollo Program's F-1 rocket engine development \cite{zinn1968theoretical,oefelein1993comprehensive}. 

Designing combustion devices to avoid these instabilities is nontrivial, with modern designs utilizing advanced combustor geometries~\cite{schadow1992combustion,gutmark1995suppression}, flow control~\cite{mcmanus1993review,annaswamy2002active}, or fuel injection and mixing schemes~\cite{miller2007combustion}, as examples. Around the same time as the development of the F-1 rocket engine, engineers investigated an alternative approach of engine design: to \textit{saturate} thermoacoustic instabilities as opposed to their suppression~\cite{clayton1965experimental}. The \textit{Rotating Detonation Engine} (RDE) is a combustor that follows this design philosophy. The standard RDE combustion chamber, a notional depiction of which is shown in Fig. \ref{fig:rde_01}, features an annular combustion chamber along which fuel and oxidizer are continuously injected and mixed~\cite{rankin2017overview}. Upon ignition of the fuel and oxidizer mixture (via an automotive spark plug, for example), the RDE's geometry provides the necessary confinement (or acoustic focusing) for the local accumulation of energy to self-accelerate. Gradients in pressure and density self-steepen and eventually form a shock wave coupled with a region of intense combustion. These shock-reaction structures, or detonations, travel around the combustion chamber so long as there is continued supply of fuel and oxidizer. The combustion products rapidly expand and exit the combustor at high speeds, producing usable thrust or otherwise high enthalpy flow. 

\begin{figure}[]
\begin{center}
\begin{overpic}[width=8cm]{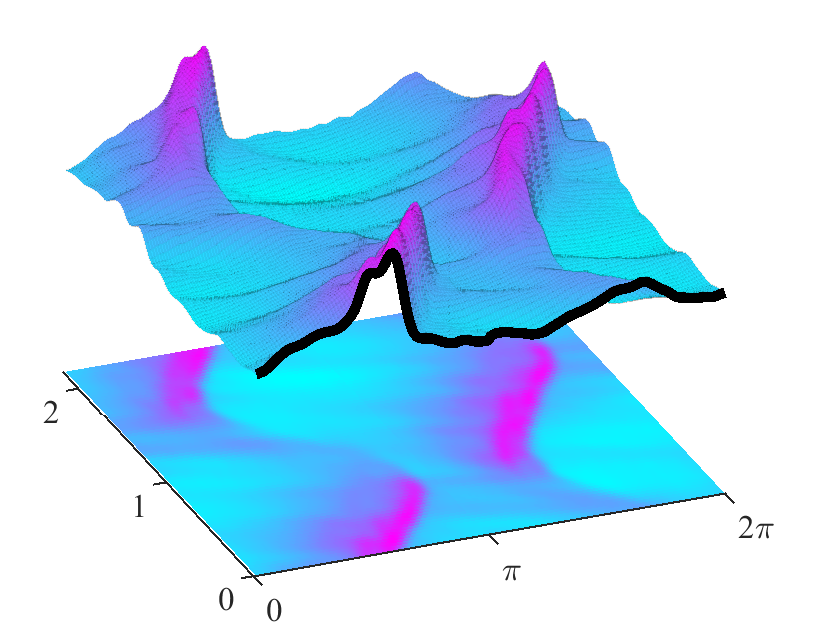}
\end{overpic}
\end{center} 
\caption{\label{fig:rde_02} High-speed camera imagery enables capturing detailed space-time histories of the combustion waves present in the combustor of the engine. For annular combustors, the high-speed imagery can be recast as 1-D spatial snapshots by `unwrapping' the annulus into a linear domain with periodic boundaries. Shown here is example high-speed camera footage stacked frame-by-frame to create a space-time view of the wave dynamics. While this particular experiment exhibits large-amplitude wave oscillations, stable wave propagation is desirable from an engineering standpoint.}
\end{figure}

For a given RDE geometry and propellant, the collection of detonation waves contained within the RDE have been observed to assume characteristic operating modes depending on one or more experimental parameters: the chemical time scale, the convective time scale, and the mixing and injection time scale~\cite{koch2019operating,koch2020mode,koch2020modeling,koch2021multiscale}. These modes include a number of co- or counter-rotating waves (mode-locked), modulated waves, plane waves, and chaotic wave propagation. By imaging the aft-end of an RDE with a high-speed camera, detonation wave positions can be tracked through time to produce a complete space-time history of combustor. An example space-time history for an experiment with modulated wave behavior is shown in Fig. \ref{fig:rde_02}. In this experiment, each of the two waves periodically exchange strength (speed and amplitude) as each interacts with the tail of the preceding wave.

\begin{figure}[]
\begin{center}
\begin{overpic}[width=9cm]{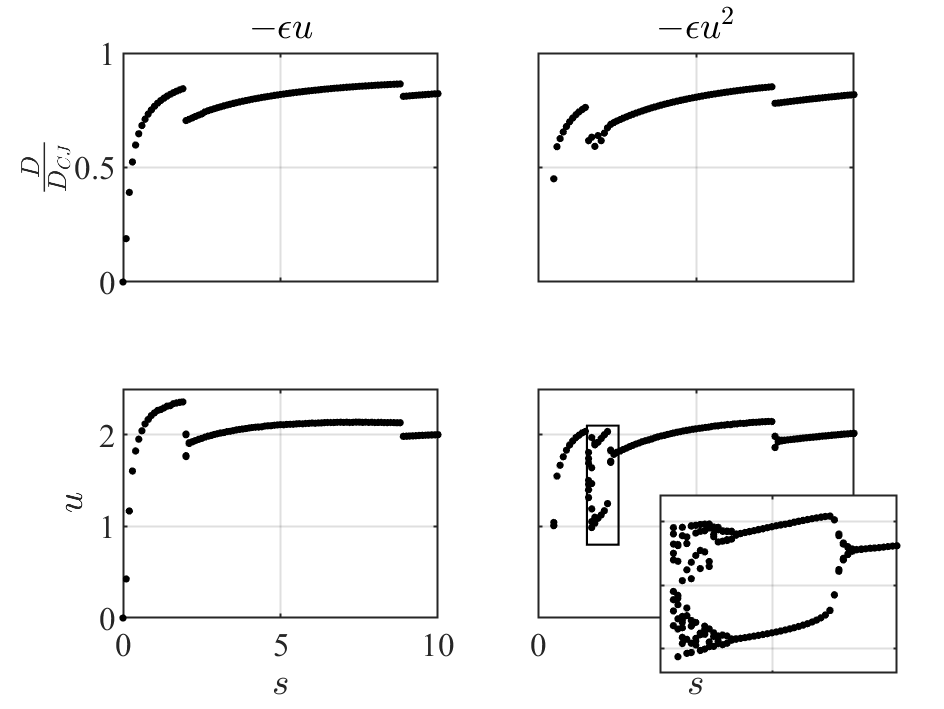}
\put(14,71){(a)}
\put(58,71){(b)}
\put(14,36){(c)}
\put(58,36){(d)}
\put(89,6){(e)}

\put(20,54){\vector(-1,1){5}}
\put(15,50){$N=1$}

\put(30,59){\vector(0,1){6}}
\put(25,55){$N=2$}

\put(40,54){\vector(1,2){5}}
\put(35,50){$N=3$}

\end{overpic}
\end{center} 
\caption{\label{fig:rde_03} Bifurcation diagrams from direct numerical simulation of the RDE analog system through values of the parameter $s$. The first column corresponds to a loss form that is linear in the state $u$, whereas the second column exhibits behavior for a quadratic loss term. The top row shows variation of wave speed and the bottom row describes the peak wave amplitude. For the quadratic loss term, and for this choice of model parameters, a set of period-halving bifurcations are seen for increasing $s$ through the transition from one to two waves. (e) is a enhanced view of the region of transition of (d).}
\end{figure}
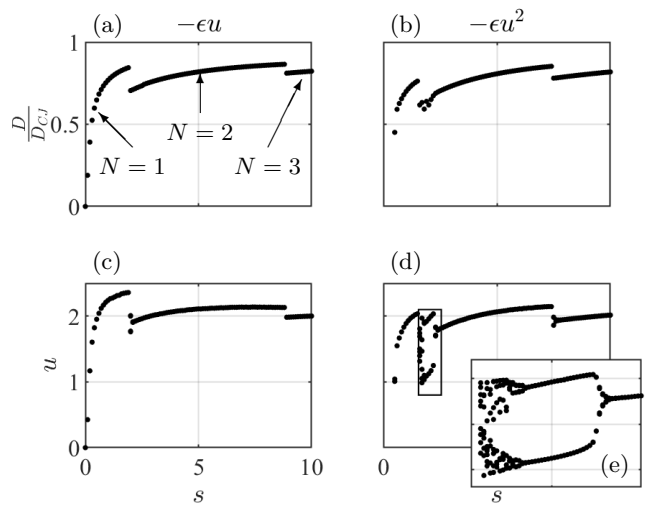

The dynamical properties of rotating detionation waves can be analyzed through a minimal system for shock physics subject to gain depletion (fast combustion), gain recovery (slow injection and mixing) and loss (background exhaustion). The two-component PDE for the minimal system is~\cite{koch2020mode}:
\begin{subequations}
\begin{align}
u_t + u u_x &=  q (1-\lambda) \omega \left(u\right) + \epsilon f\left(u\right) \label{Eq:rde_analog_u} \\
\lambda_t &= (1-\lambda)\omega \left(u\right) - g\left(u\right)\lambda \label{Eq:rde_analog_l} ,
\end{align}
\end{subequations}
where $u(x,t)$ is representative of the internal energy of the system and $\lambda(x,t) \in (0,1)$ is a combustion progress variable. Succinctly stated, the model system relates the shock physics of Burgers' equation to energy release through an analog of Arrhenius kinetics, $\omega(u)$, and losses, $\epsilon f(u)$. The combustion progress variable evolves according to the competition between combustion and regeneration, captured by the term $g(u)\lambda$. Critically, as in the case with real RDEs, the model system can the necessary gain-loss dynamics and feedback loops to reproduce the nonlinear dynamics observed in experiments with appropriate choices for the gain, loss, and gain regeneration terms. In these detonation analogs, the gain term is typically taken to be an exponential:
\begin{equation}
    \omega(u) = \exp{\left(\frac{u-u_c}{\alpha}\right)},
\end{equation}
where $u_c$ and $\alpha$ are parameters describing the activation of the kinetics. Similarly, the gain regeneration term is designed to be \textit{saturable}; that is, consistent with the real-world behavior that propellant injection and mixing can be slowed or halted by high chamber pressures. In this model, these effects are captured by deactivating gain regeneration based on the quantity $u$:
\begin{equation}
    g(u) = \frac{s}{1 + \exp{\left(k(u-u_p)\right)}},
\end{equation}
where $s$ is the time constant for gain regeneration in the absence of perturbations and $k$ and $u_p$ modify the deactivation behavior of the regeneration model. Varying the shape of the loss function -- analogous to crafting a particular engine's geometry or nozzle configuration -- can dramatically alter the qualitative behavior of the solution to the model system in Eqns. \ref{Eq:rde_analog_l}-\ref{Eq:rde_analog_u}. Bifurcation diagrams for a linear loss ($\epsilon u$) and quadratic loss ($\epsilon u^2$) are shown in Fig. \ref{fig:rde_03} with the gain regeneration time constant $s$ as the bifurcation parameter. In both cases, there exists a series of bifurcations incrementing the number of waves from 1 to 2 to 3 (further increases eventually lead to the merger of waves into a single deflagration front). For the quadratic loss case, a series of period-halving bifurcations exist between the stable 1-wave and 2-wave branches. In this transition region, a series of period-halving bifurcations exist that slowly introduce order with increasing $s$.

\begin{figure}[t]
\centering
\begin{tikzpicture}
\color{blue} 
\draw[-stealth, line width = 1mm] (-6,0) -- (-5,0);
\draw (-6 ,0.5) node[right] {\color{black} $E_\text{in}^\text{gain}$};
\node[draw, align = center, fill = green] at (-3.65,0) {\color{white} gravity\\ \color{white}+\\ \color{white}bath acceleration};
\draw[-stealth, line width = 1mm] (-2.35,0) -- (-0.35,0);
\draw (-2.35,0.5) node[right] {\color{black} $E_\text{out}^\text{gain} = E_\text{in}^\text{loss}$};
\node[draw, align = center, fill = yellow] at (0.75,0) {\color{white} hydrodynamic\\ \color{white} damping};
\draw (1.85,0.5) node[right] {\color{black} $E_\text{out}^\text{loss}$};
\draw[line width = 1mm] (1.85,0) -- (2.75,0);
\draw[line width = 1mm] (2.7, 0) -- (2.7, -1);
\draw[line width = 1mm] (2.75, -1) -- (-6,-1);
\draw[line width = 1mm] (-5.95,-1) -- (-5.95,0);
\end{tikzpicture}
\caption{Schematic illustrating the interaction between the driving phase and damping phase as a recurrence relation.}\label{Fig: Schematic}
\end{figure}
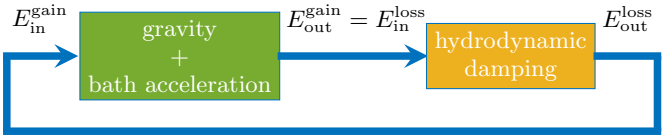

\begin{figure*}[t]
\includegraphics[width = 16cm]{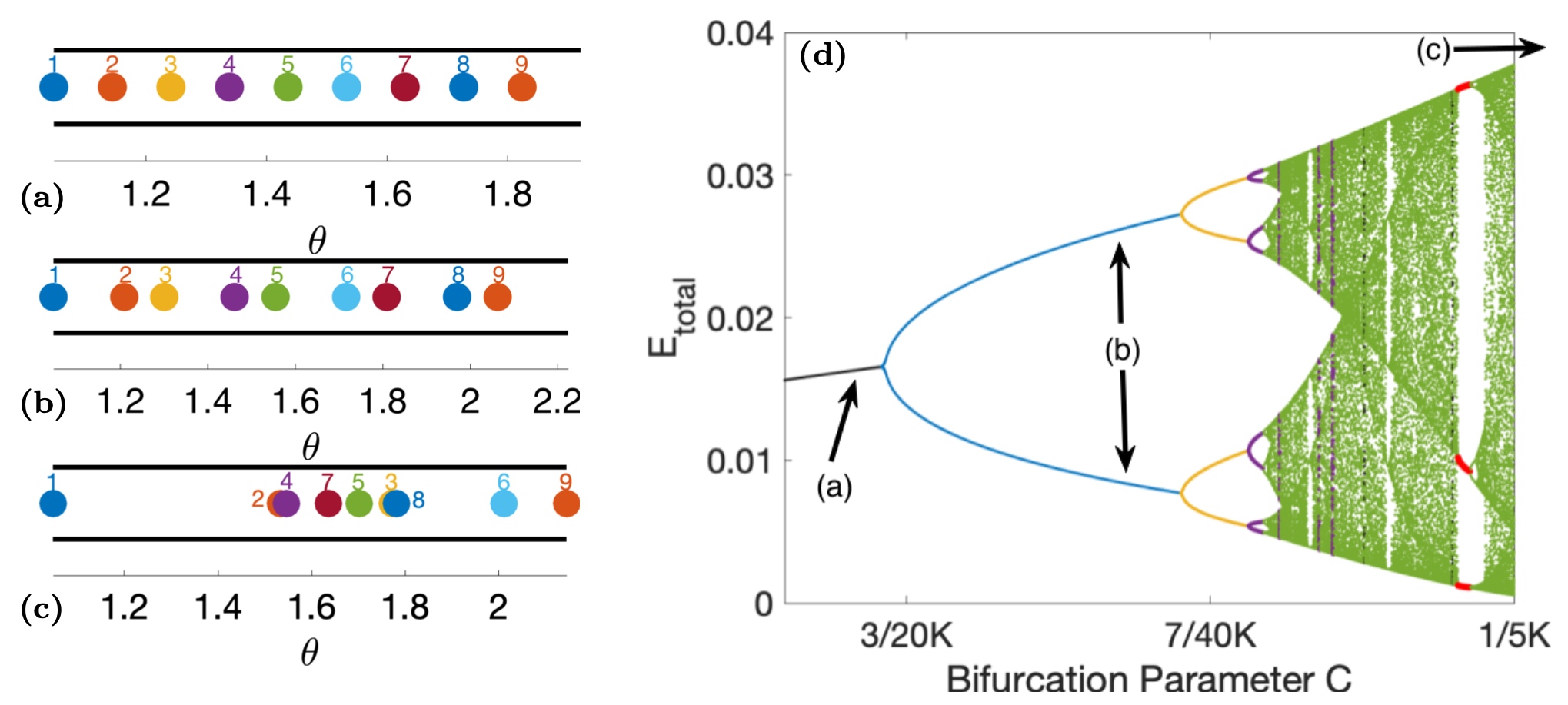}
\caption{\label{Fig: HQA illustration} (a) - (c) Transition from the steady walking phase with $C = 1/10K$ where $\theta$ corresponds to the position around the annulus from the velocity model of Rahman \cite{Rahman18} (a) to the two-step walk with $C = 1/6K$ (b), and finally to chaotic walking with $C = 1/2K$ (c).  (d) Bifurcation diagram, from $C = 7/50K$ to $C = 1/5K$, illustrating period doubling bifurcations that give way to orbits of arbitrarily large periods.  The black curves represent single fixed points, the blue curves represent two cycles, the yellow curves represent four cycles, the purple curves represent eight cycles, and the green curves represent longer periodic orbits.  Finally, the red curves represent three cycles, which in the original 1-D kick map, was sufficient to prove chaos \cite{Rahman18}.  The labels (a) and (b) show the general locations in the bifurcation diagram of each type of walking illustrated in panels (a) - (b).  Note that (c) will occur past the axes of the plot, but will have a similar fractal structure to the chaotic regions in the plot.}
\end{figure*}
\subsection{Hydrodynamic quantum analogs}\label{Sec: Walkers}

Fluid droplets bouncing on a vertically vibrating fluid bath can replicate quantum-like phenomena, as is shown in review articles by Bush and collaborators \cite{Bush10, Bush15a, Bush15b, BushOza20_ROPP, TCB18}.  These droplets produce waves on the bath at each impact that then then propel them horizontally. This was initially observed in the seminal works of Couder and collaborators \cite{CPFB05, CouderFort06}.  Each impact produces a complex hydrodynamic process; studied in great detail by Mola\v{c}ek and Bush \cite{MolBush13a, MolBush13b}.  The dynamics on average can be roughly summarized as follows:  the droplet experiences standard projectile motion in the air, energy gain from the vertical acceleration of the bath, and energy loss from hydrodynamic damping \cite{rahman2022walking}.  This is illustrated as a schematic diagram in Figure~\ref{Fig: Schematic}.

\begin{figure*}[htbp]
\centering
\stackinset{l}{}{t}{}{\textbf{(a)}}{\includegraphics[width = 0.22\textwidth]{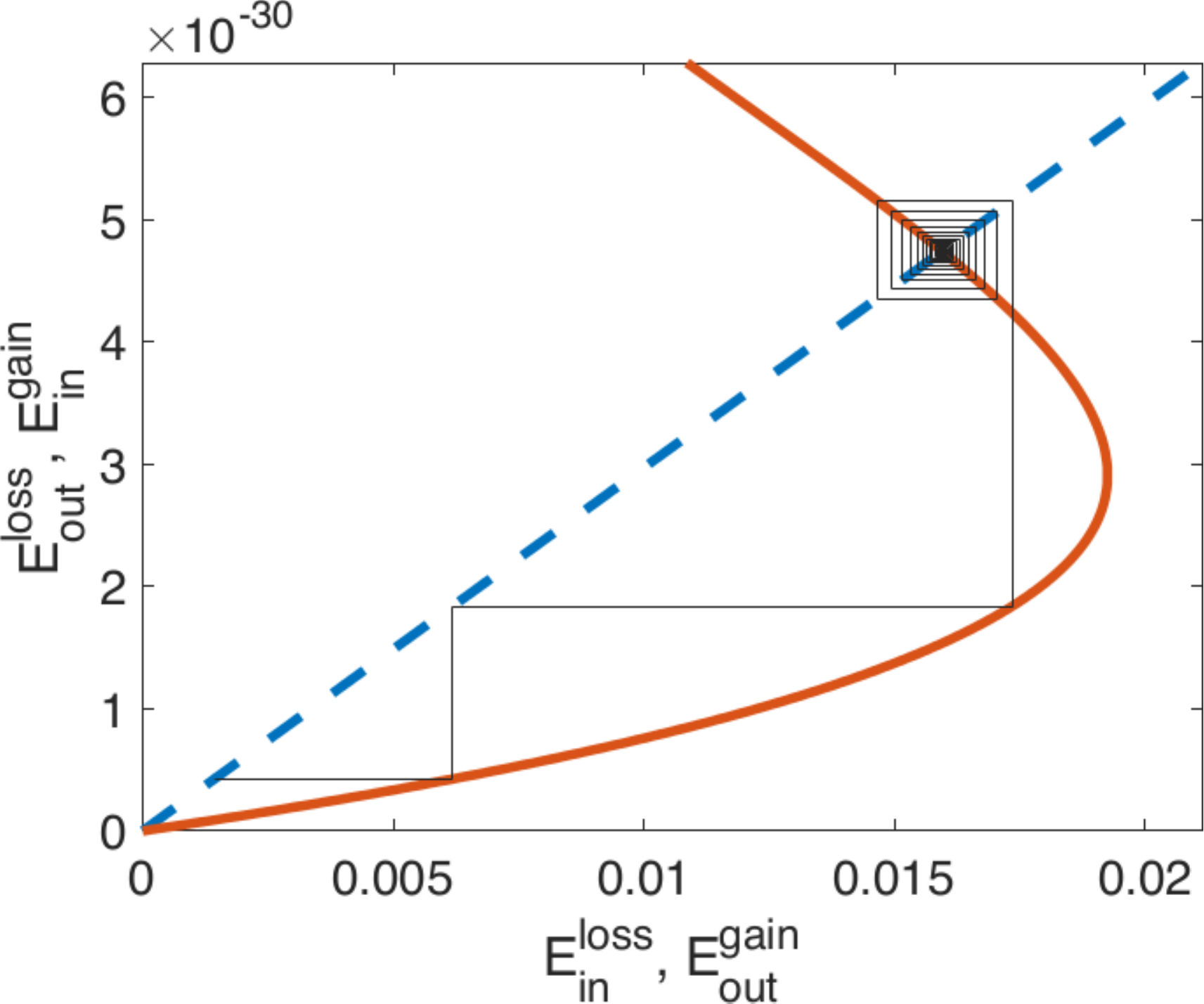}}\qquad
\stackinset{l}{}{t}{}{\textbf{(b)}}{\includegraphics[width = 0.22\textwidth]{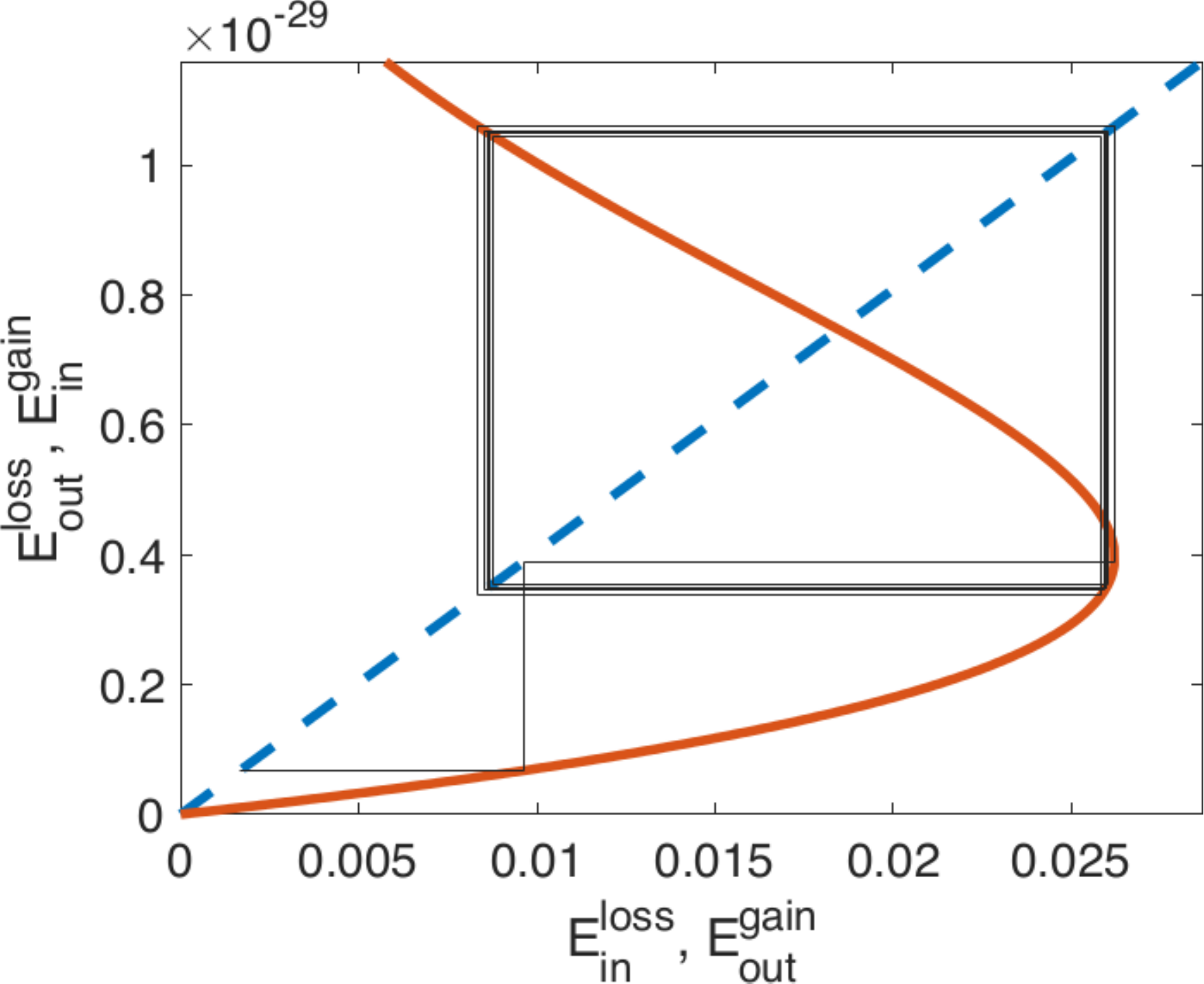}}\qquad
\stackinset{l}{}{t}{}{\textbf{(c)}}{\includegraphics[width = 0.225\textwidth]{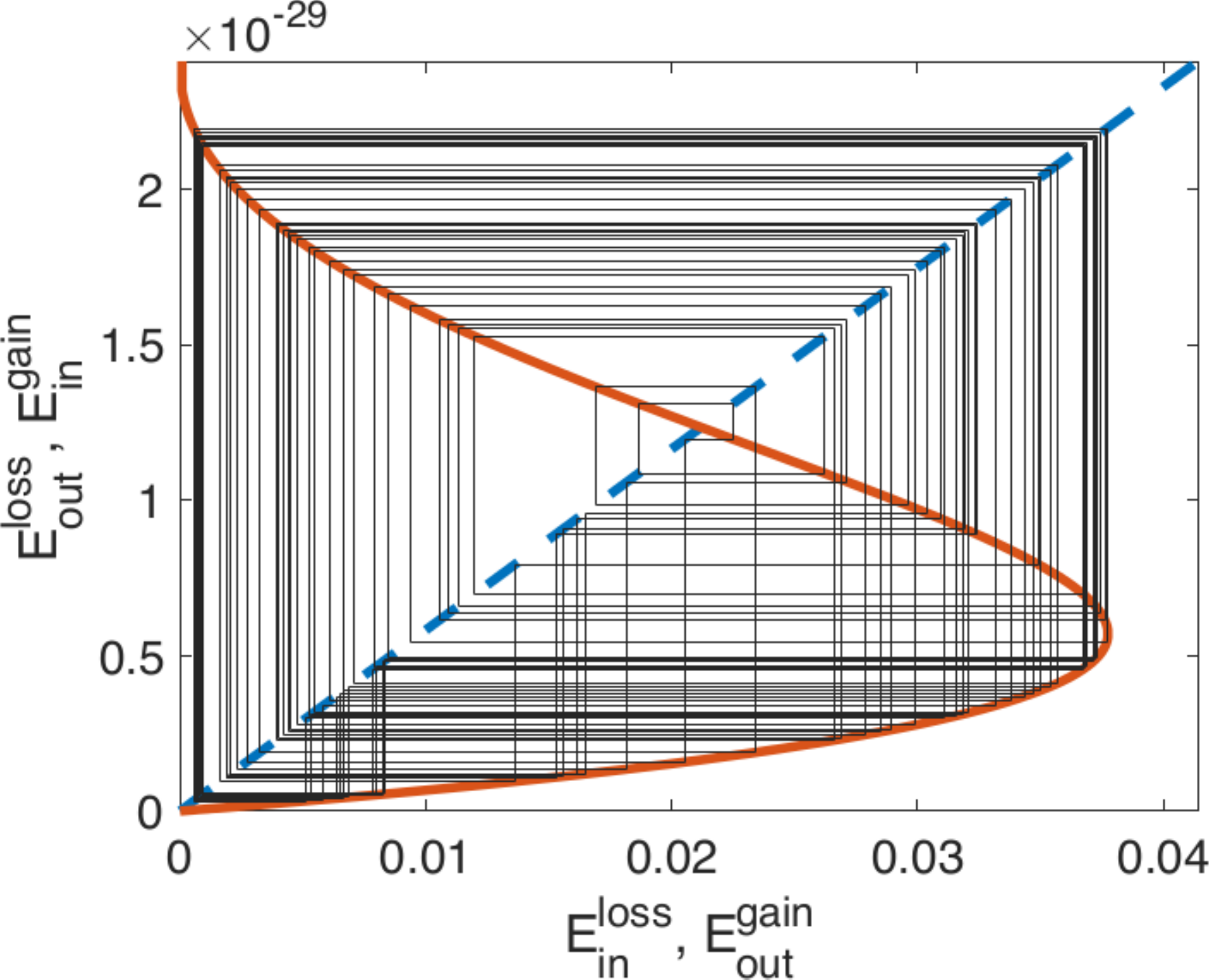}}\qquad
\stackinset{l}{}{t}{}{\textbf{(d)}}{\includegraphics[width = 0.22\textwidth]{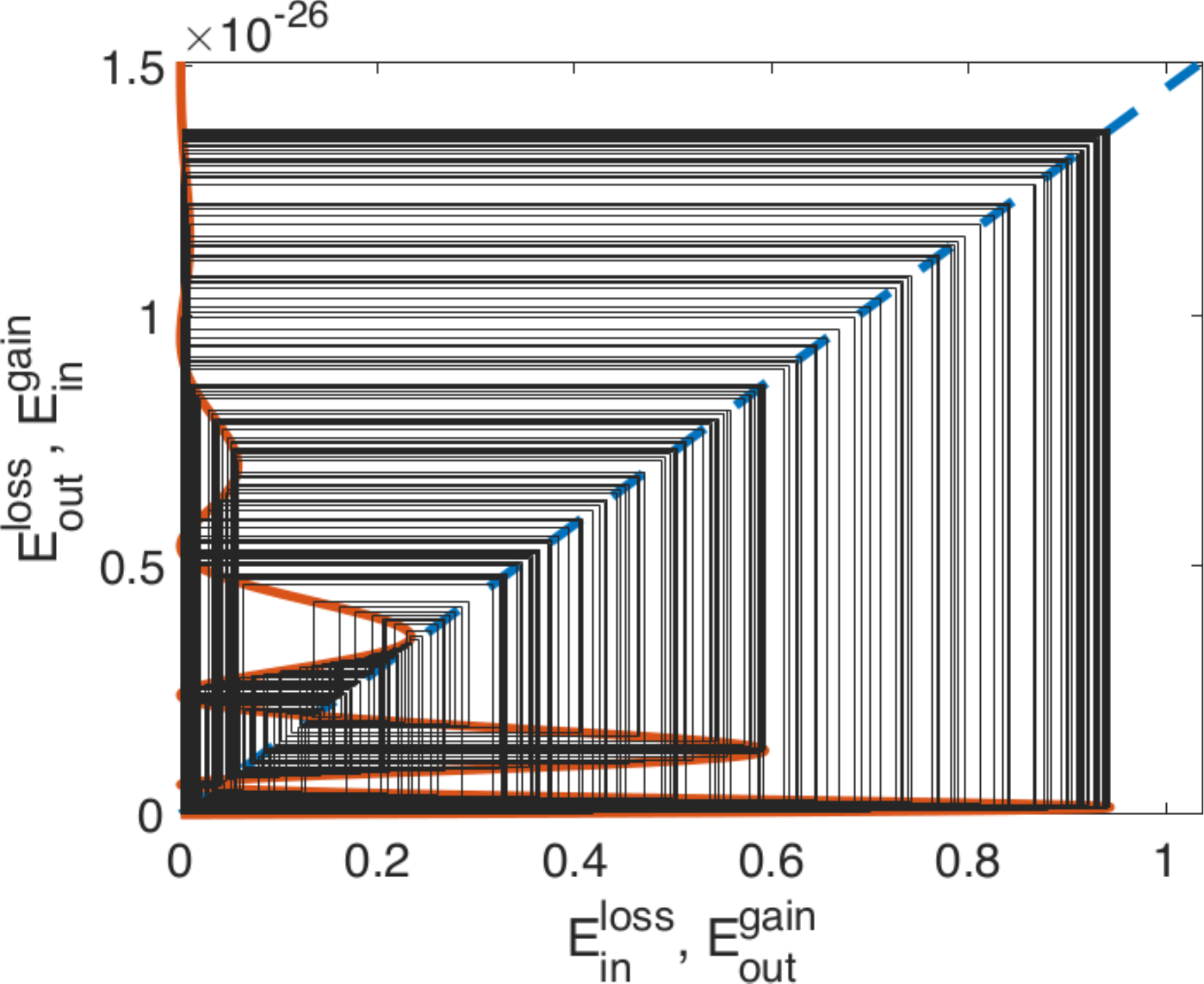}}
\caption{Cobweb plots illustrating the route to chaos.  \textbf{(a)} Convergence to non-trivial fixed point for $C = 1/7K$.  \textbf{(b)} Convergence to a low-period orbit for $C = 1/6K$.  \textbf{(c)} Evidence of seemingly random orbits of arbitrarily large period emerge for $C = 1/5K$.  \textbf{(d)}  Full-scale chaos for $C = 1/K$ as proved in \cite{Rahman18}.}\label{Fig: Kick Transition}
\end{figure*}

From the averaged dynamics of a walking droplet, a velocity model was derived by Rahman \cite{Rahman18}, which was then converted to an energy gain-loss model by Rahman and Kutz \cite{rahman2022walking}.  The model
\begin{subequations}
\begin{align}
E_\text{out}^\text{loss} &= C^2E_\text{in}^\text{loss},\label{Eq: Kick Loss curve}\\
E_\text{out}^\text{gain} &= E_\text{in}^\text{gain} + C^2K^2\sin^2(\omega \sqrt{E_\text{in}^\text{gain}}/C)e^{-2\nu E_\text{in}^\text{gain}/C^2}\nonumber\\
&\qquad + CK\sqrt{E_\text{in}^\text{gain}}\sin(\omega \sqrt{E_\text{in}^\text{gain}}/C)e^{-\nu E_\text{in}^\text{gain}/C^2}\label{Eq: Kick Gain curve}
\end{align}
\label{Eq: Kick Loss-Gain curves}
\end{subequations}
shows how the exchange of energy gain and energy loss, even when averaged out between impacts, leads to period doubling of the droplet velocity, which eventually becomes chaotic.  An illustration of this process and its physical interpretation is shown in Figure \ref{Fig: HQA illustration}.  Just as in Section \ref{Sec: Lasers} we may also plot the cobwebs (Figure \ref{Fig: Kick Transition}) in the gain-loss plane to better understand how the droplet velocity undergoes a sequence of period-doubling bifurcations \cite{Kuznetsov}, eventually destabilizing into chaotic behavior.

\begin{figure}[htbp]
\centering
\includegraphics[width=0.45\textwidth]{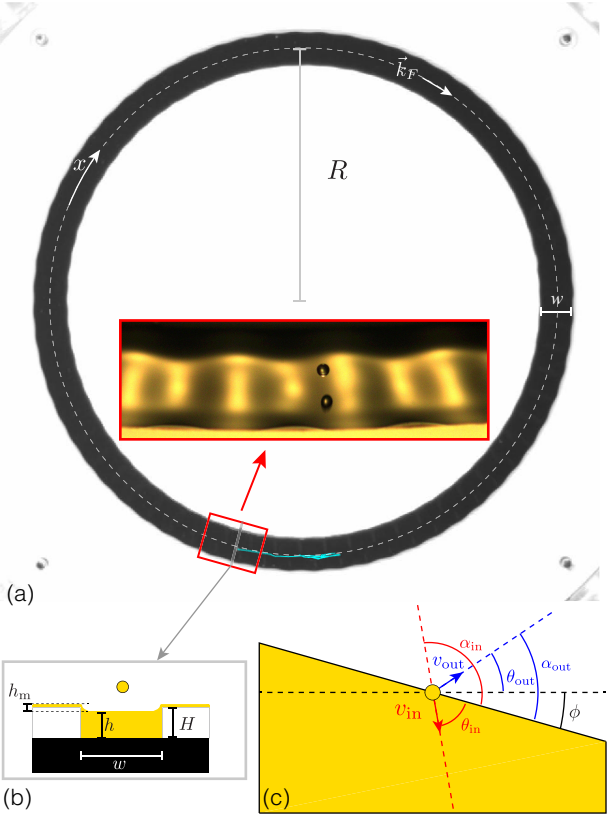}
\caption{A droplet bouncing on a quasi-1D standing wave. (a) Top view of the liquid bath with the annular channel (in black).  We represent $x$ as the circumferential direction, $w$ as the width of the annulus, and $h$ as the height of the fluid. The droplet's recent trajectory is illustrated in cyan. Red-framed inset: snapshot of a droplet with diameter $0.56$~mm while it is bouncing on the standing wave with vertical bath acceleration $\gamma/\gamma_F=1.007$ from \eqref{Eq: HQA Wavefield} \cite{RahmanHQAExperiments2022}. (b) Vertical section of the liquid bath. (c) Schematics of the droplet bouncing for the theoretical model \cite{RahmanHQAAnalysis2022}.}
\label{Fig: HQA Experimental setup and schematic}
\end{figure}

In addition to an averaged energy gain-loss model, we may also conduct similar analyses \cite{RahmanHQAAnalysis2022} more directly from experiments \cite{RahmanHQAExperiments2022}.  Consider a fluid with with a density of $\rho=950$ kg/m$^3$, kinematic viscosity of $\nu=20.9$ cSt, and a surface tension of $\sigma=20.6$ mN/m.  We drop a fluid droplet onto the vertically vibrating fluid bath in an annular cavity. We denote $x$ as the circumferential direction and $y$ as the vertical direction (up into the air), as depicted in Figure \ref{Fig: HQA Experimental setup and schematic}.  We treat the droplet as an undamped projectile when it is in the air, and assume the droplet undergoes inelastic collision with a flat slope at impact that adheres to the model of Mola\v{c}ek and Bush \cite{MolBush13a, MolBush13b}.  The path of the droplet in the air is given by the dimensionless equation
\begin{subequations}
\begin{align}
\xi &= v_n^xt + x_n,\\
\eta &= -\frac{1}{2}Gt^2 + v_n^yt + y_n,
\end{align}
\label{Eq: HQA ProjectilePath}
\end{subequations}
where $t$ is the time between impacts, $x_n$ and $y_n$ are the $x$ and $y$ coordinates at the previous impact, $\xi$ and $\eta$ are the $x$ and $y$ coordinates between impacts, $v_n^x$ and $v_n^y$ is the launch velocity right after the previous impact, and $G$ is the dimensionless effective gravity.  We also need the evolution of the fluid bath to calculate the location of impact, which is given by
\begin{equation}
\Psi = A\cos(2\pi T)\cos(2\pi\xi) + \gamma\cos(2\pi[2T-\varphi]),
\label{Eq: HQA Wavefield}
\end{equation}
where $T$, $\varphi$, $A$, and $\gamma$ are the dimensionless elapsed time (across all impacts), phase, wave amplitude, and bath forcing amplitude, respectively.  Applying the hydrodynamic model of  Mola\v{c}ek and Bush \cite{MolBush13a, MolBush13b} to the approach velocity of the projectile, $v_\text{in}$ (as shown in Fig. \ref{Fig: HQA Experimental setup and schematic}), at impacts give us the launch velocities
\begin{subequations}
\begin{align*}
v_n^x &= \sqrt{\left(v_n^T\right)^2 + \left(v_n^N\right)^2}\cos\theta_n,\\
v_n^y &= \sqrt{\left(v_n^T\right)^2 + \left(v_n^N\right)^2}\sin\theta_n,
\end{align*}
\end{subequations}
where the tangential and normal components, $v_n^T$ and $v_n^N$, and the launch angle $\theta_n$, are found through trigonometric calculations arising from the hydrodynamic analysis.

\begin{figure*}[t]
\centering
\stackinset{l}{1mm}{b}{}{\text{(a)}}{\includegraphics[width = 0.42\textwidth]{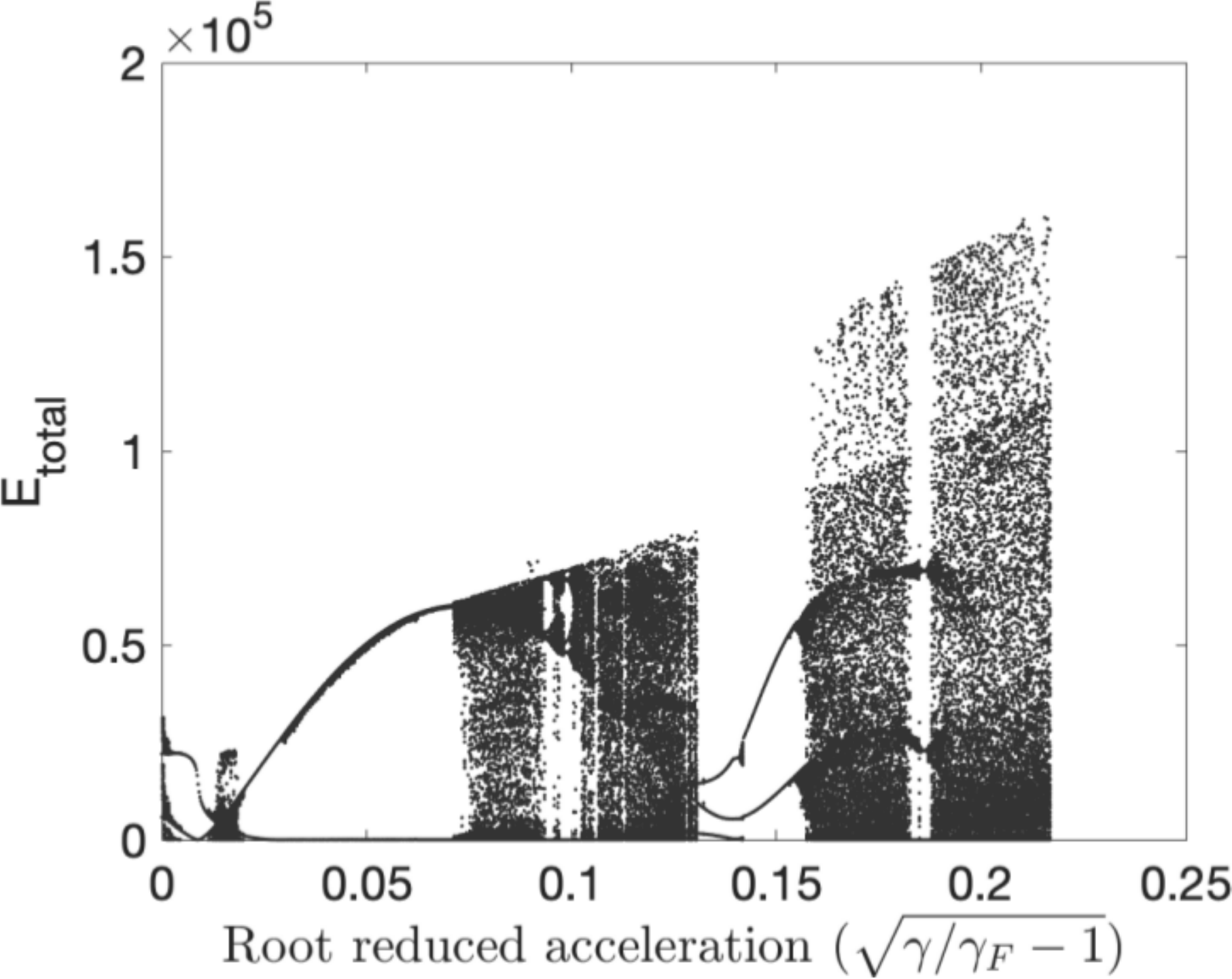}}
\stackinset{l}{1mm}{b}{}{\text{(b)}}{\includegraphics[width = 0.42\textwidth]{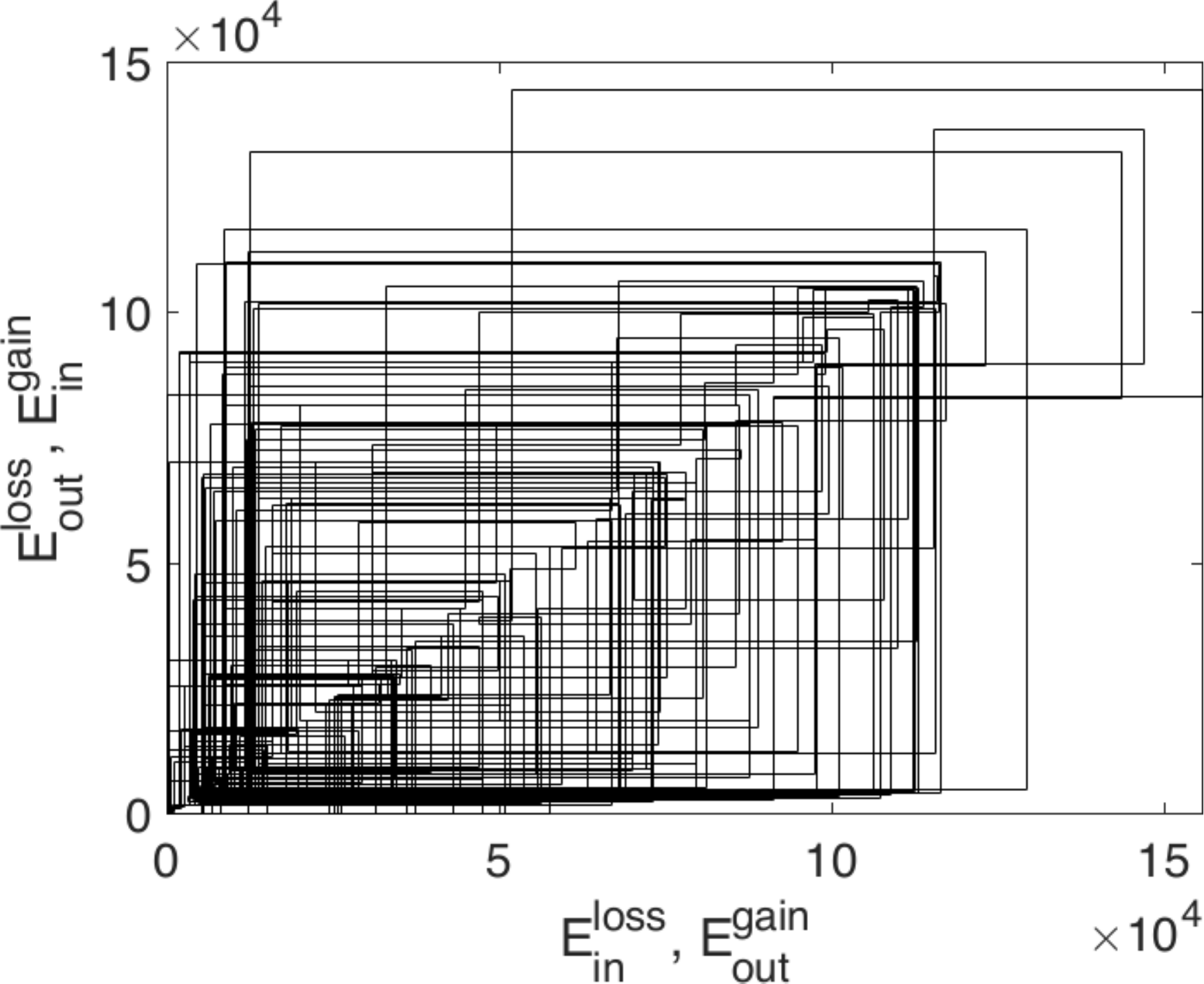}}
\caption{Indication of the route to chaos for the energy gain-loss dynamics.  (a) Evidence of a potential flip bifurcation as the bath acceleration $\gamma$ is increased \cite{RahmanHQAExperiments2022, RahmanHQAAnalysis2022}.  We observe striking similarities to that of other damped-driven systems.  (b) Cobweb plot of \eqref{Eq: Energy loss H-K} and \eqref{Eq: Energy gain H-K} in the chaotic regime.}
\label{Fig: HQA Period Doubling}
\end{figure*}

In such a setup we may segregate the damping mechanism,
\begin{equation}
    E^\text{loss} = (v_n^x)^2 + (v_n^y)^2,
    \label{Eq: Energy loss H-K}
\end{equation}
corresponding to the hydrodynamic losses at impact, from the driving mechanism, 
\begin{equation}
    E^\text{gain} = v_\text{in}^2,
    \label{Eq: Energy gain H-K}
\end{equation}
corresponding to the gain in kinematic energy from the projectile motion aided by the bath acceleration \cite{RahmanHQAExperiments2022, RahmanHQAAnalysis2022}.  Unlike the averaged dynamics of \eqref{Eq: Kick Loss-Gain curves}, the coupling of the gain and loss mechanisms is too convoluted to write in closed form.  However, we can still apply the same-type of analysis through numerical means, described in detail in \cite{RahmanHQAAnalysis2022}.  Similar to other damped-driven systems and the averaged model \eqref{Eq: Kick Loss-Gain curves}, we can conduct a bifurcation analysis accompanied by cobweb plots as shown in Figure \ref{Fig: HQA Period Doubling}.

\begin{figure*}[t]
\centering
\stackinset{l}{}{t}{}{\textbf{(a)}}{\includegraphics[width = 0.42\textwidth]{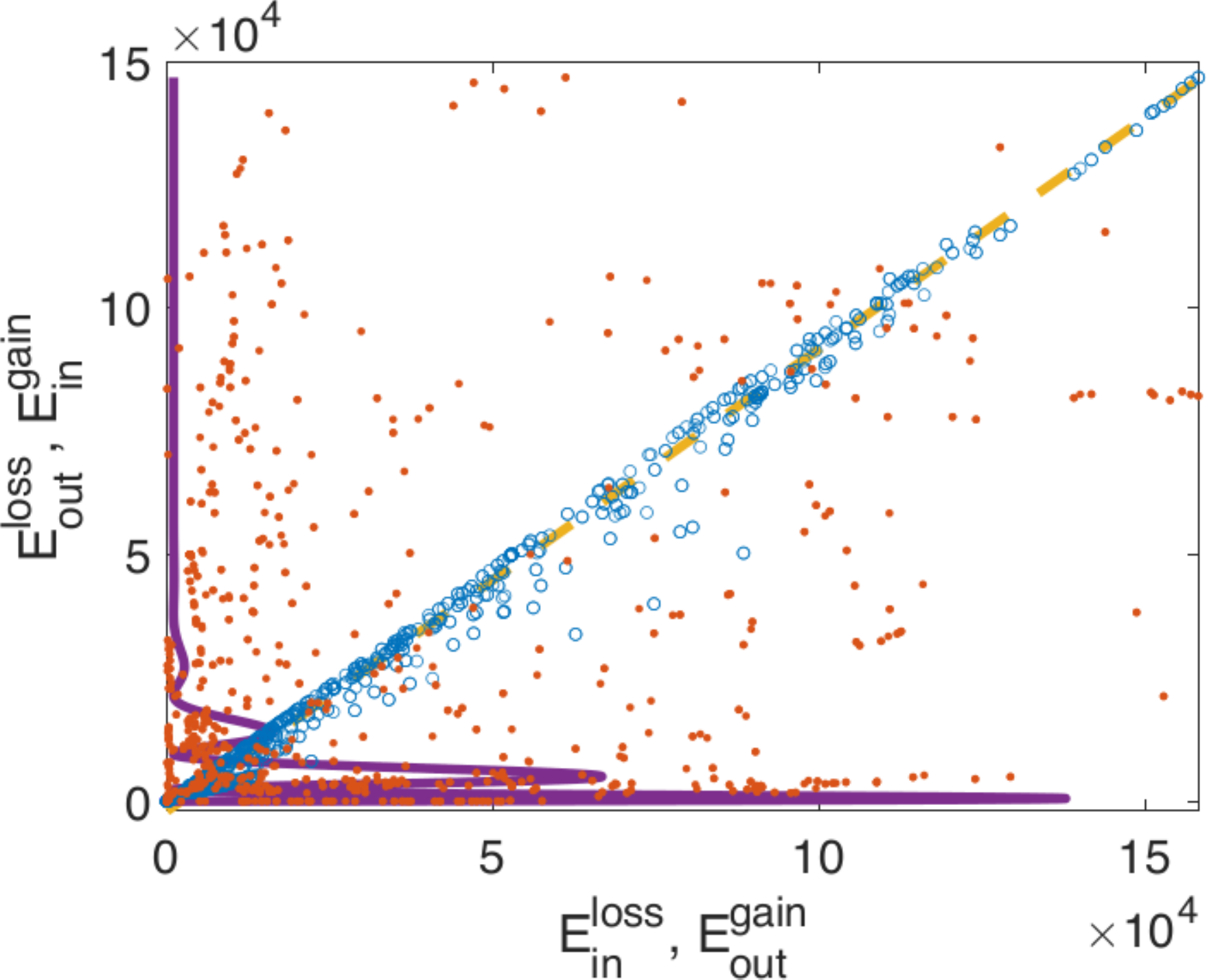}}\qquad
\stackinset{l}{}{t}{}{\textbf{(b)}}{\includegraphics[width = 0.42\textwidth]{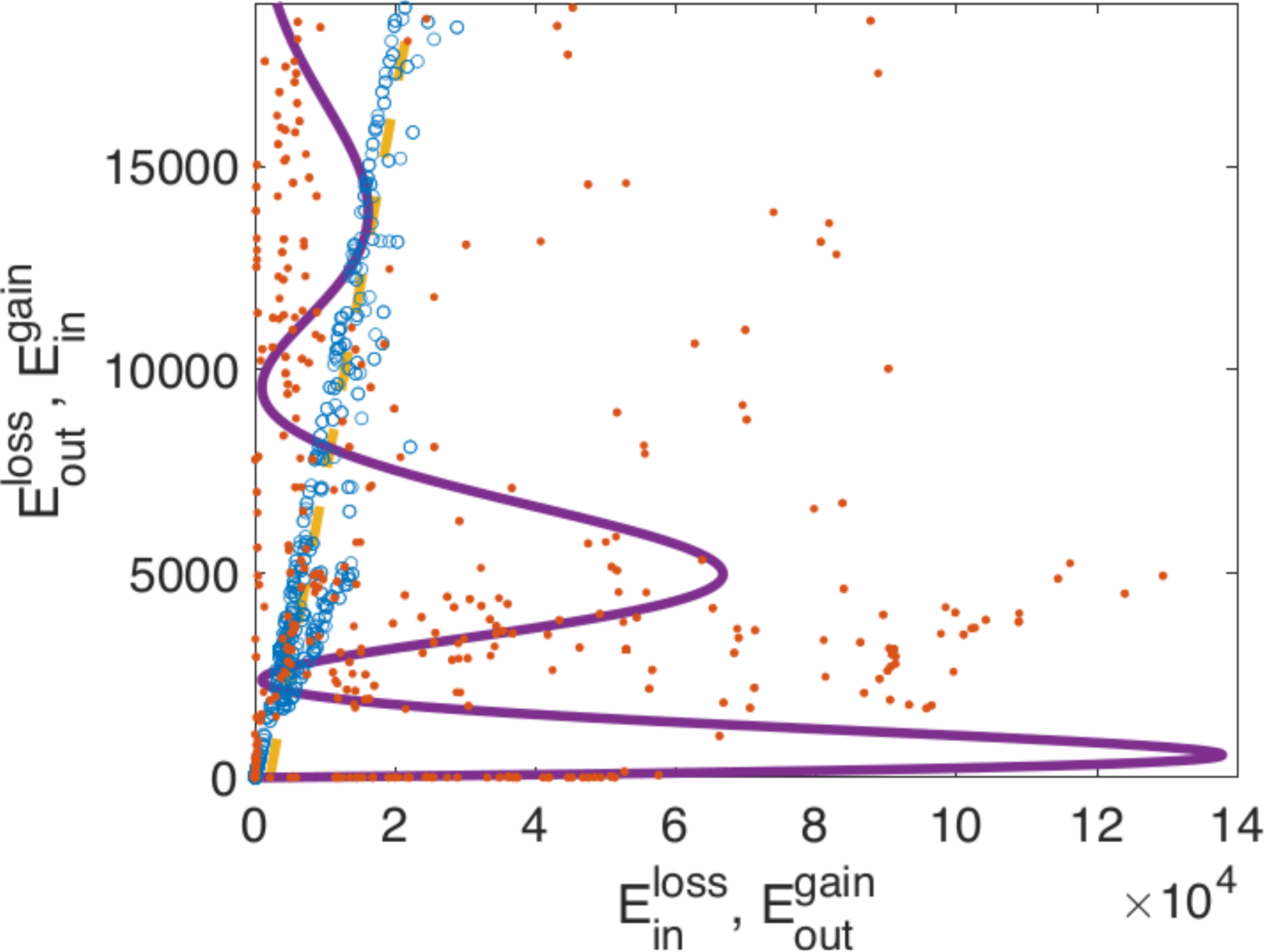}}
\caption{The solid (purple) curve fit to the closed (red) markers and the standard linear fit (dashed yellow) superimposed with the scatter plot of the gain-loss relation, where the closed (red) markers represent the relation for the gain \eqref{Eq: Energy gain H-K} and the open (blue) markers represent the relation for the loss \eqref{Eq: Energy loss H-K}.}\label{Fig: H-K_Gain-Loss}
\end{figure*}

While the averaging done in \cite{Rahman18, rahman2022walking} was through heuristic analysis of the experiments of Filoux \textit{et al.} \cite{FHV15}, the investigation in \cite{RahmanHQAAnalysis2022} allows for a more data-driven approach.  Since there are no experiments on this yet, let us use the first principles models of \eqref{Eq: Energy loss H-K} and \eqref{Eq: Energy gain H-K} as a proxy for experimental data.  If we unravel the cobweb plot of Figure \ref{Fig: HQA Period Doubling}, isolate the individual points of the cobweb, and separate gain and loss, we are left with a scatter plot as shown in Figure \ref{Fig: H-K_Gain-Loss}(a).  The scatter plot will have some distinguishing features which allow us to fit it to curves similar to Figure \ref{Fig: Kick Transition}.  If we zoom in towards the intersection points in Figure \ref{Fig: H-K_Gain-Loss}(b) we observe that it is quite similar to the averaged gain-loss curves of \cite{rahman2022walking}.

\begin{figure*}[htbp]
\centering
\stackinset{l}{}{t}{}{\textbf{(a)}}{\includegraphics[width = 0.42\textwidth]{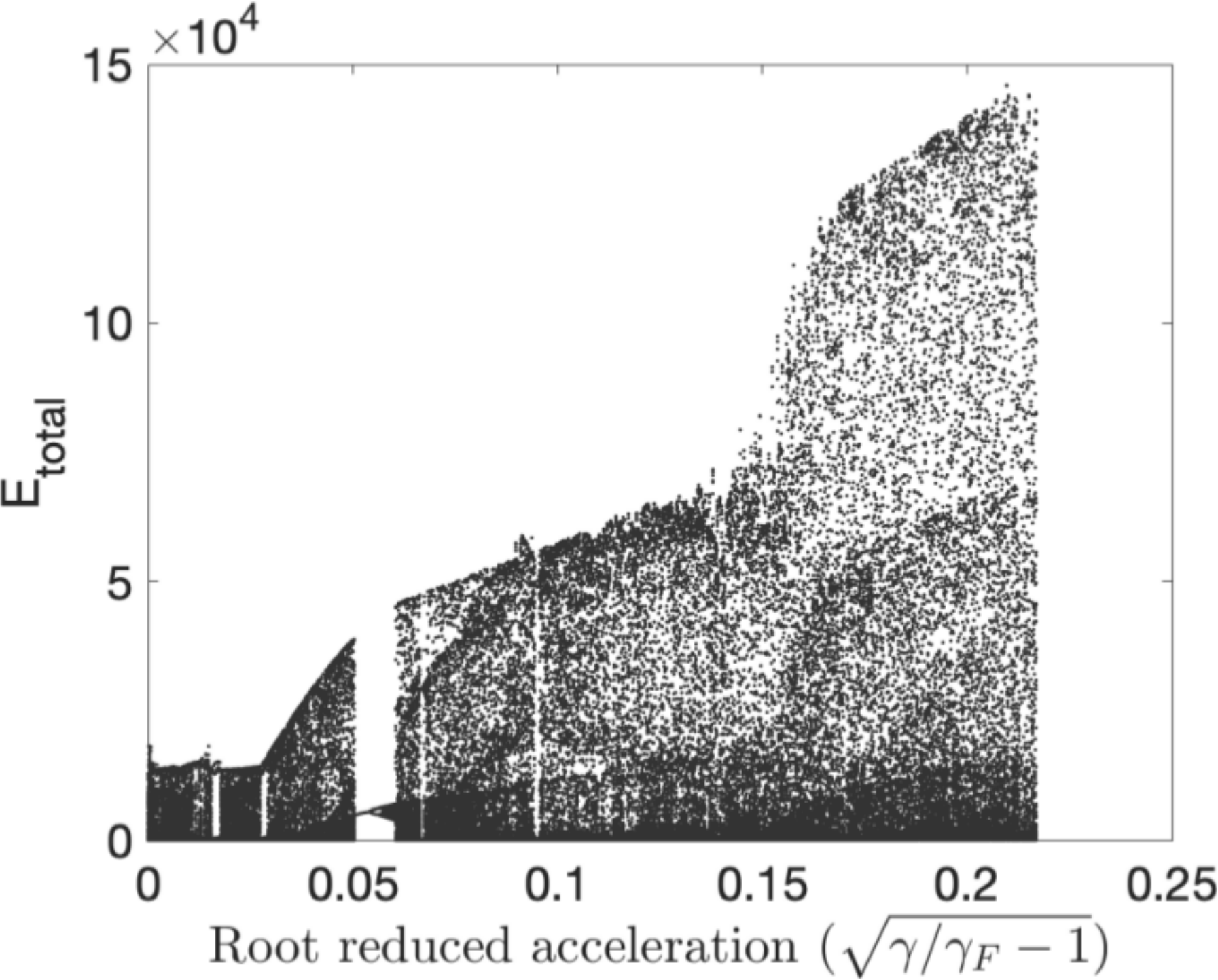}}\qquad
\stackinset{l}{}{t}{}{\textbf{(b)}}{\includegraphics[width = 0.42\textwidth]{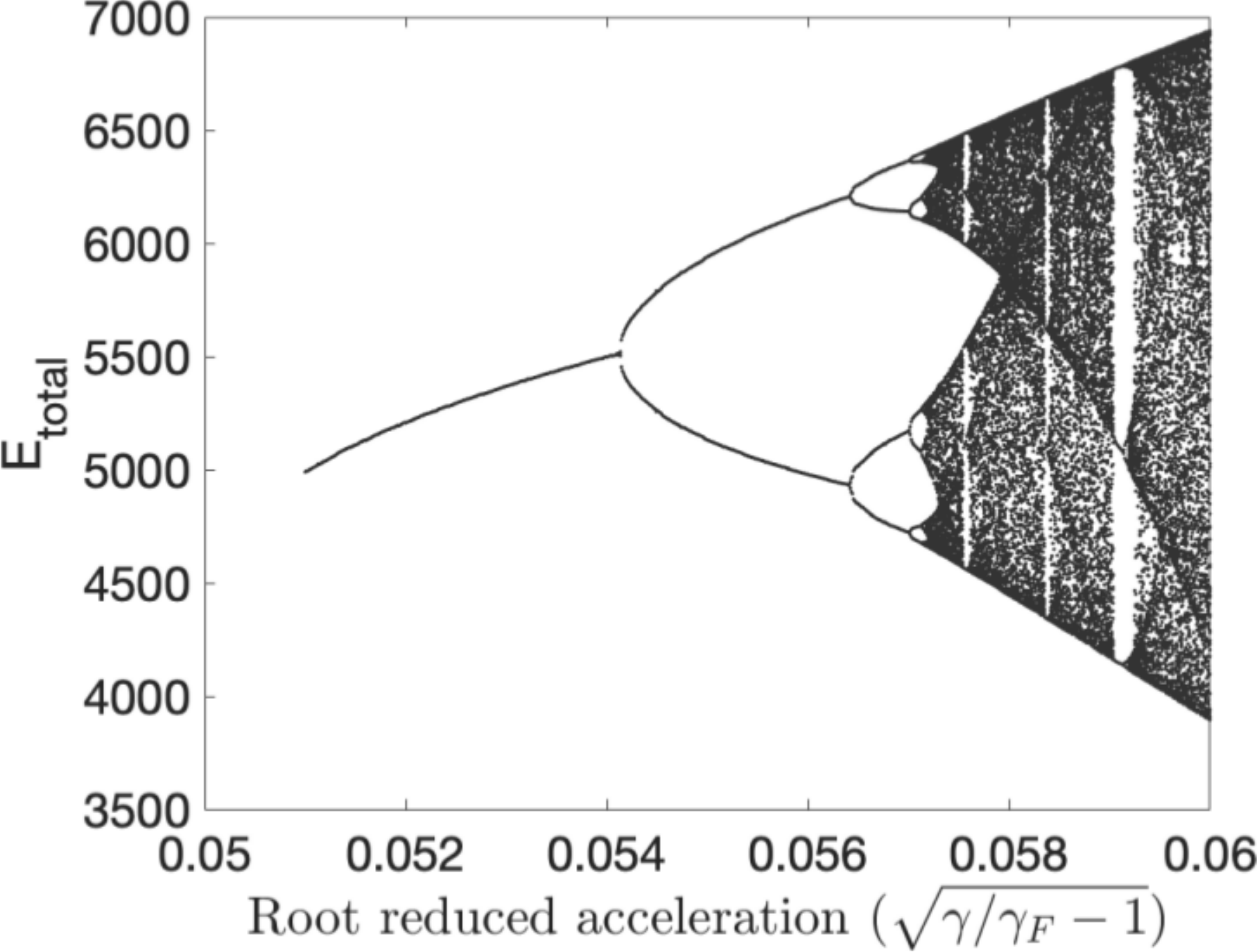}}
\caption{Bifurcation diagram for the total kinetic energy \eqref{Eq: Energy gain H-K} after one full recurrence as $\gamma$ is increased ($0 \leq \sqrt{\gamma/\gamma_F - 1} < 0.22$).  \textbf{(a)} Bifurcations in the reduced model \eqref{Eq: HQA H-K fit}.  \textbf{(b)} Period doubling bifurcation for $0.05 < \sqrt{\gamma/\gamma_F - 1} < 0.06$.}\label{Fig: HQA Fit Bifurcation}
\end{figure*}

We may now observe the bifurcation behavior of the fitted energy gain and loss curves.  First let us write the functions for the fitted curves, 
\begin{subequations}
\begin{align}
    E_\text{out}^\text{gain} &= c_1\sin^2\left(c_2\sqrt{E_\text{in}^\text{gain}}\right)e^{-c_3E_\text{in}^\text{gain}},\\
    E^\text{loss} &= \tilde{E}^\text{loss}E^\text{loss}/\tilde{E}^\text{gain};
\end{align}
\label{Eq: HQA H-K fit}
\end{subequations}
where the fitting parameters, $c_1$, $c_2$, and $c_3$ are resolved heuristically \cite{RahmanHQAAnalysis2022}.  Next we plot the full bifurcation diagram in Fig. \ref{Fig: HQA Fit Bifurcation}\textbf{(a)}.  This reduced model will inevitably miss some of the features of the full bifurcation diagram from Fig. \ref{Fig: HQA Period Doubling}, however it preserves many of the overall features.  Interestingly we also observe indication of a period doubling bifurcation in the neighborhood of the interval $0.05 \leq \sqrt{\gamma/\gamma_F - 1} \leq 0.06$ as shown in Fig. \ref{Fig: HQA Fit Bifurcation}\textbf{(b)}.  This indicates that the underlying progression of the instabilities in these types of damped-driven systems are canonical even if they are often obfuscated by noise within the system.

\subsection{Bio-locomotion}
\begin{figure}[t]
\centering
\begin{overpic}[width = 4cm]{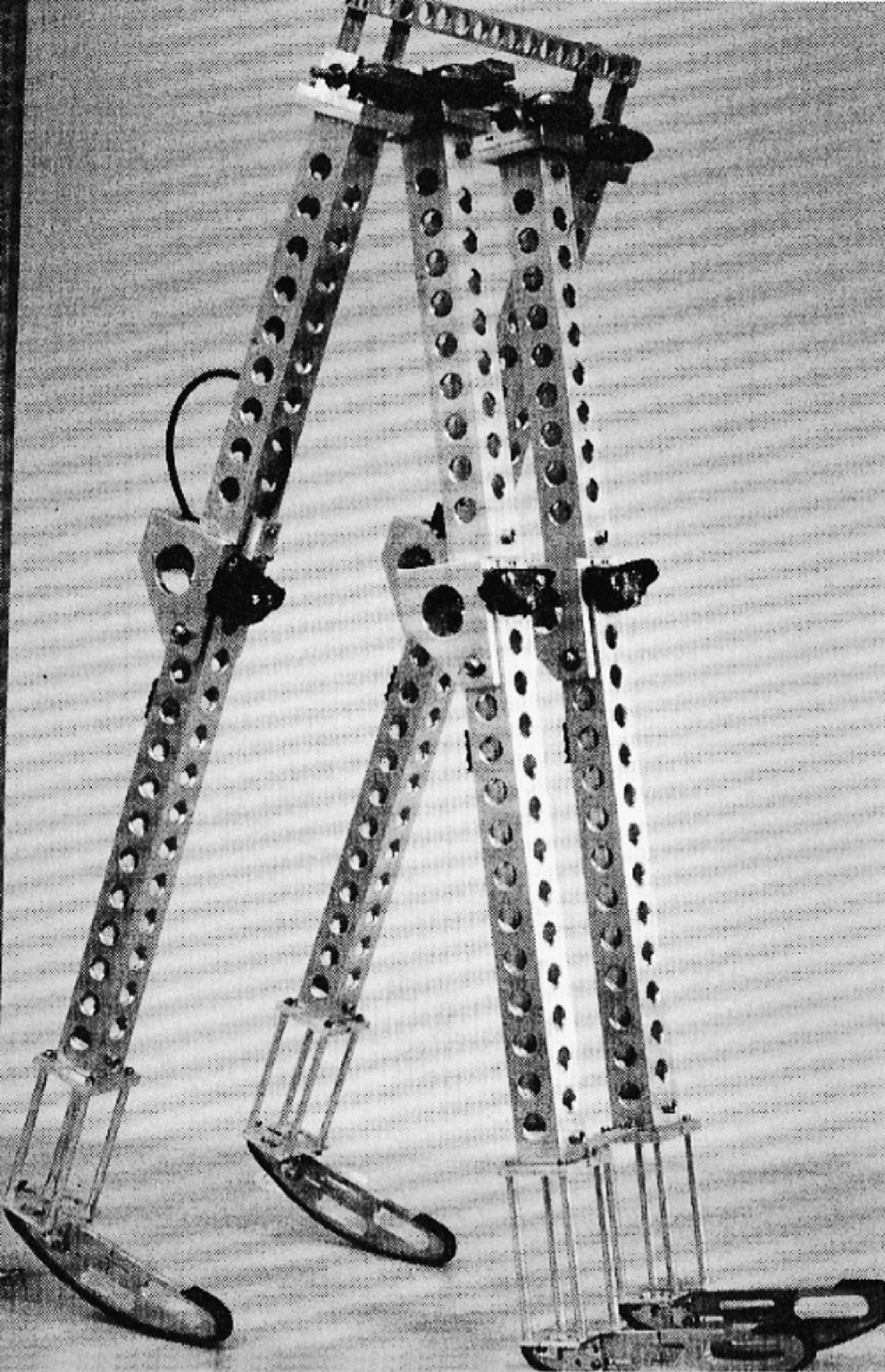}
\end{overpic}
\vspace*{-.10in}
\caption{McGeer's experimental quasi-2D passive walking robot with knees \cite{mcgeer1990passive}.  An uncontrolled and unpowered mechanical biped can produce and maintain stable steady periodic gait down a shallow slope by gravity as its only source of energy/actuation.}
\label{fig:mcgeer_robot}
\end{figure}

\begin{figure}[t]
\centering
\stackinset{l}{}{t}{}{\text{(a)}}{\includegraphics[width = 0.22\textwidth]{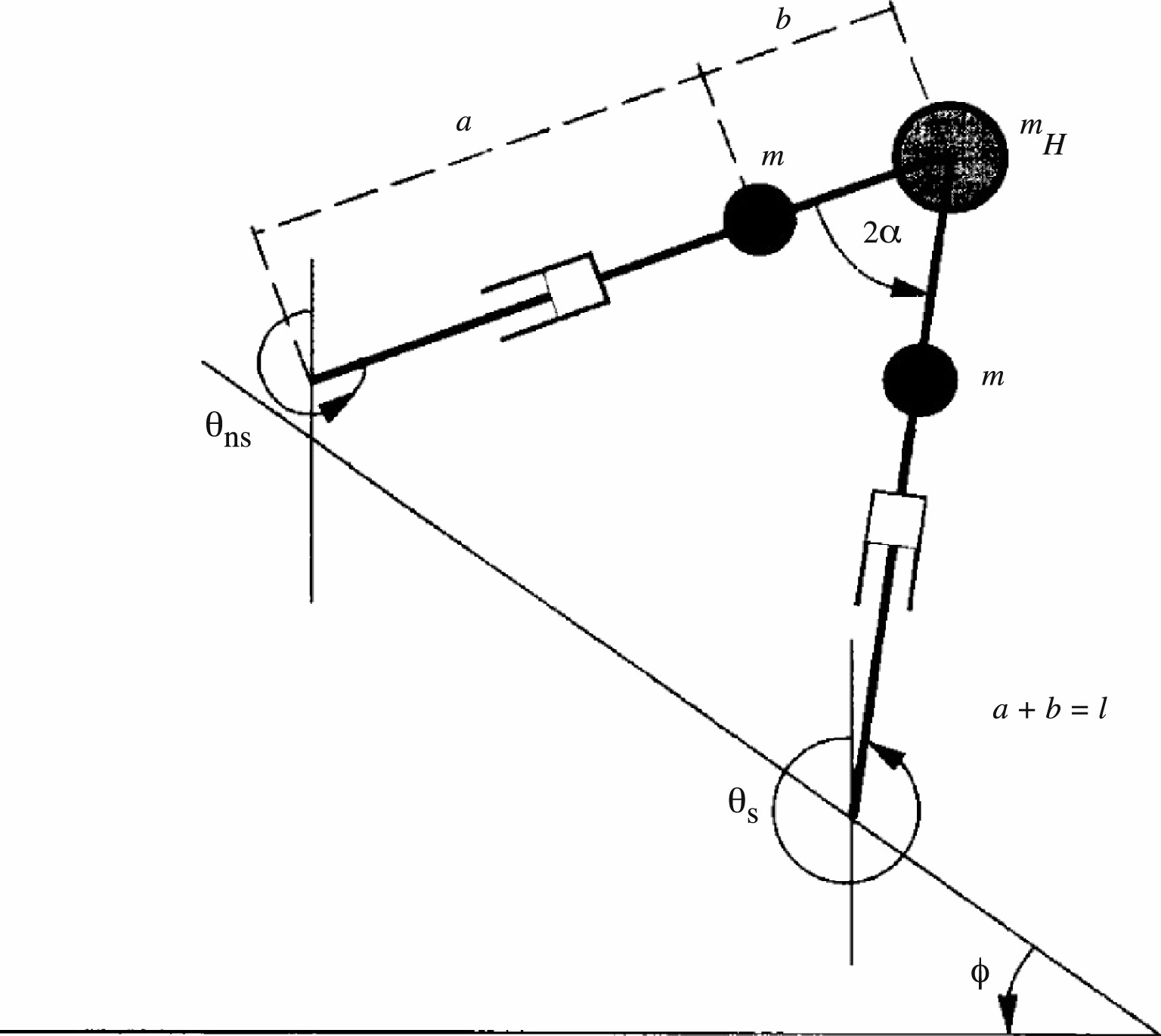}}
\stackinset{l}{}{t}{-5mm}{\text{(b)}}{\includegraphics[width = 0.22\textwidth]{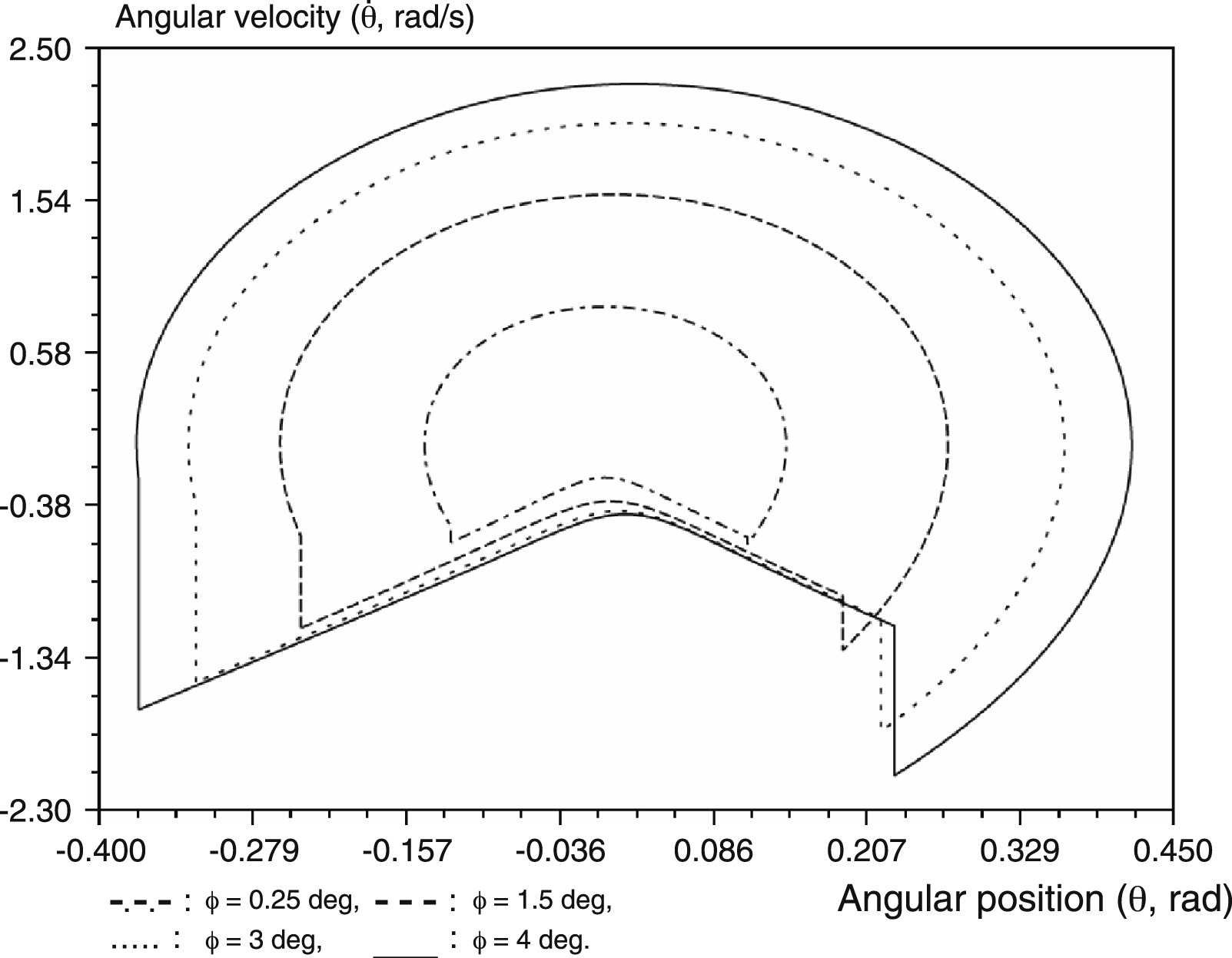}}
\caption{Compass-gait model of passive dynamic walking \cite{goswami1996compass}. (a) Compass-gait walker in swing phase descending a slope. Mass is concentrated at three points: one at the hip $m_h$ and one on each leg $m$, located distances $a$ and $b$ from the foot and hip, respectively. Both legs are identical, such that each leg measures $l = a + b$. It uses prismatic-jointed knees and massless lower legs to address foot clearance during swing. The compass-gait walker is not actuated, as it can walk by its dynamics alone via descending a shallow slope at a constant angle. Gait occurs in two phases: swing and stance. Each phase is interrupted by foot-collision with the ground, in which the swing leg makes instantaneous, inelastic, and non-sliding impact and transitions to the stance phase. (b) Limit cycle for the compass-gait walker. The steady periodic gait is associated with the limit cycle behavior of this nonlinear system. It is important to note that periodic solutions for an uncontrolled system must be defined according to orbital stability \cite{hayashi2014nonlinear}.} 
\label{Fig:compass_model_limit}
\end{figure}

Bipedal bio-locomotion, such as human walking, is considered one of the most sophisticated forms of legged motion due to its stability and efficiency while often not at static equilibrium. Indeed, the simultaneous stability and efficiency of bipedal bio-locomotion is of great interest to researchers studying and building walking machines. Researchers have hypothesized that much of the swing phase of human locomotion is passive, \textit{i.e.}, it requires no external activation nor control \cite{rose1994human}. In 1981, Mochon and McMahon introduced a mathematical model that likened the swing phase of bipedal locomotion to ballistic motion \cite{mcmahon1984mechanics}. Inspired, McGeer spent the early 1990s studying the effect of gravity-induced passive locomotion down a shallow slope in unpowered and uncontrolled bipedal robots, as depicted in Figure~\ref{fig:mcgeer_robot}, and demonstrated their stable steady periodic motion \cite{mcgeer1990passive, mcgeer1990passiveknees, mcgeer1993dynamics}. Further, he analyzed the passive dynamics using a linearized mathematical model. In this, he formalized the seminal theory of \textit{passive dynamic walking}, bipeds that walk by dynamics alone. In 1996, using the full set of nonlinear equations, Goswami and colleagues were the first to link the stable passive dynamics of a compass-gait (inverted-jointed double-pendulum) biped to chaos theory using phase-plane limit cycles and discovering period-doubling bifurcations in gait \cite{goswami1996compass}. See Figure~\ref{Fig:compass_model_limit} for an illustration. 

Shortly thereafter, Garcia and colleagues published their work on ``the simplest walking model'': a 2D straight-legged biped with point-feet \cite{garcia1998simplest}. This gait model is irreducible, a special case of the compass-gait biped robot; Garcia and colleagues also independently reported period-doubling bifurcations of gait \cite{garcia1998some, garcia1996passive, garcia1998speed, coleman1999stability}. In the years to follow, researchers studied how the effects of changing both the mechanical structure and control impacted the biped's ability to maintain stable steady periodic gait; studies included walking models with knees \cite{garcia1999stability}, a torso \cite{wisse2004passive, howell1998simple}, three dimensions \cite{adolfsson19983, dankowicz1999existence}, virtual slope \cite{howell1998simple}, non-point foot models \cite{li2012new, kumar2009simplest},  and actuation \cite{uchida2000constant}. Indeed, the phenomenon of \textit{passive dynamic walking} has been extensively evaluated mathematically \cite{kurz2005template, asano2008pseudo, garcia1998simplest}, numerically \cite{piiroinen2001normal, piiroinen2002recurrent, piiroinen2003breaking, piiroinen2005low, narukawa2005biped}, and experimentally \cite{wisse2004passive, mcgeer1990passive, collins2005efficient}, and principles elucidated from passive dynamic walking are used to inform efficient robot control, prosthetic limb design, and human gait pathology \cite{iqbal2014bifurcations}. 

The compass-gait biped is kinematically equivalent to a 2D inverted-jointed double pendulum: two knee-less legs, each with a point mass, joined at the hip, with its own point mass, as shown in Figure~\ref{Fig:compass_model_limit}(a). Despite the passive biped’s simple mechanical structure, complex walking dynamics emerge \cite{goswami1996compass, garcia1998simplest, garcia1999stability}. The compass-gait biped’s governing equations of motion are hybrid in that the nonlinear differential equations are interrupted by algebraic switching conditions at foot-collision with the ground. As the walker descends a shallow slope $\gamma$, the nonlinear dynamics are described by \cite{garcia1998simplest}:
\begin{subequations}
\begin{align}
    \ddot{\theta}(t) &- \text{sin}(\theta(t) - \gamma) = 0 \\
    \ddot{\theta}(t) &- \ddot{\phi}(t) + \text{sin}\phi(t) \left[ \dot{\theta}(t)^2 - \text{cos}(\theta(t)-\gamma) \right] = 0
\end{align}
\label{Eq:compass_hybrid_system}
\end{subequations}
When the swing leg reaches foot-collision with the ground, that leg transitions to stance phase, modeled by the algebraic switching condition:
\begin{equation}
    \phi(t)-\text{2}\theta = 0.
    \label{Eq:compass_switch} 
\end{equation}

\begin{figure}[t]
\centering
\begin{overpic}[width = 8cm]{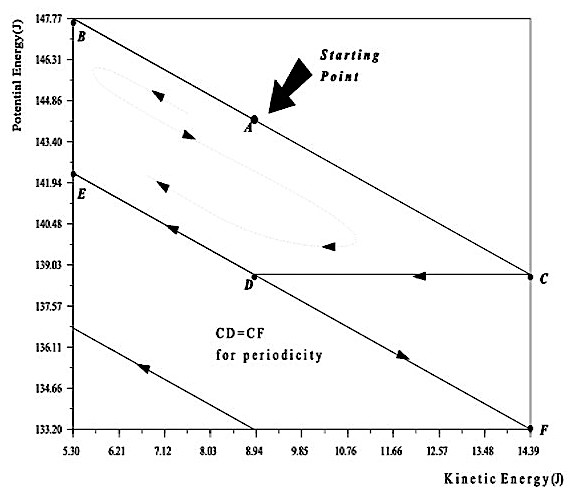}
\end{overpic}
\vspace*{-.10in}
\caption{Kinetic versus potential energy diagram of a compass-gait walker \cite{goswami1996compass}. The mechanical energy $E=K+P$ is conserved during the swing phase, corresponding to line BC, where point C is foot-collision. The instantaneous, inelastic, and non-sliding foot-collision is denoted by line CD, \textit{i.e.}, the loss of kinetic energy; the loss of potential energy is denoted by the distance AD=CF. Periodic gait is thereby CF=CD; any loss from foot-collision is recovered from gravity.}
\label{fig:PDW_energy}
\end{figure}

\begin{figure*}[t]
\centering
\stackinset{l}{}{t}{-5mm}{\text{(a)}}{\includegraphics[width = 0.38\textwidth]{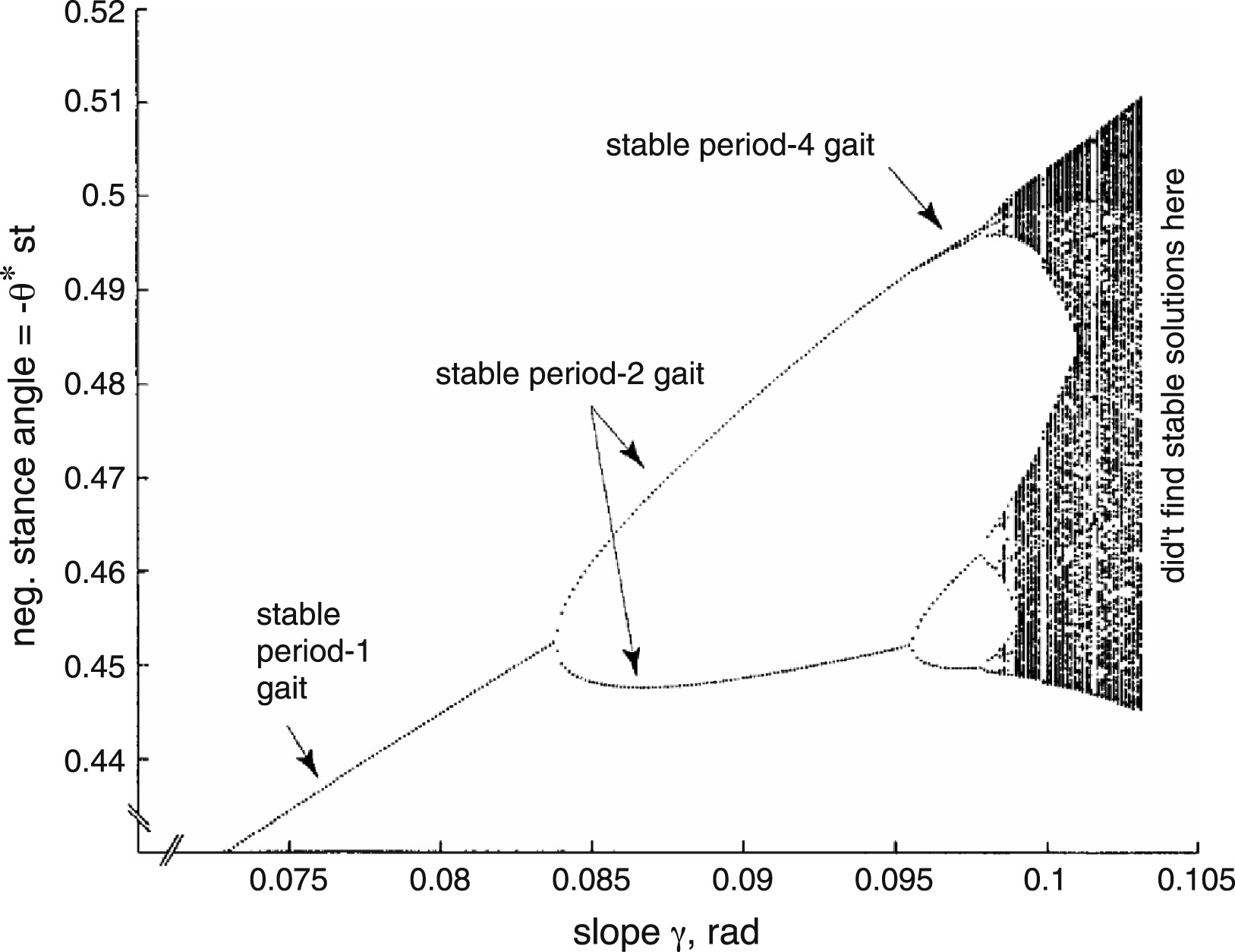}}
\stackinset{l}{}{t}{-5mm}{\text{(b)}}{\includegraphics[width = 0.44\textwidth]{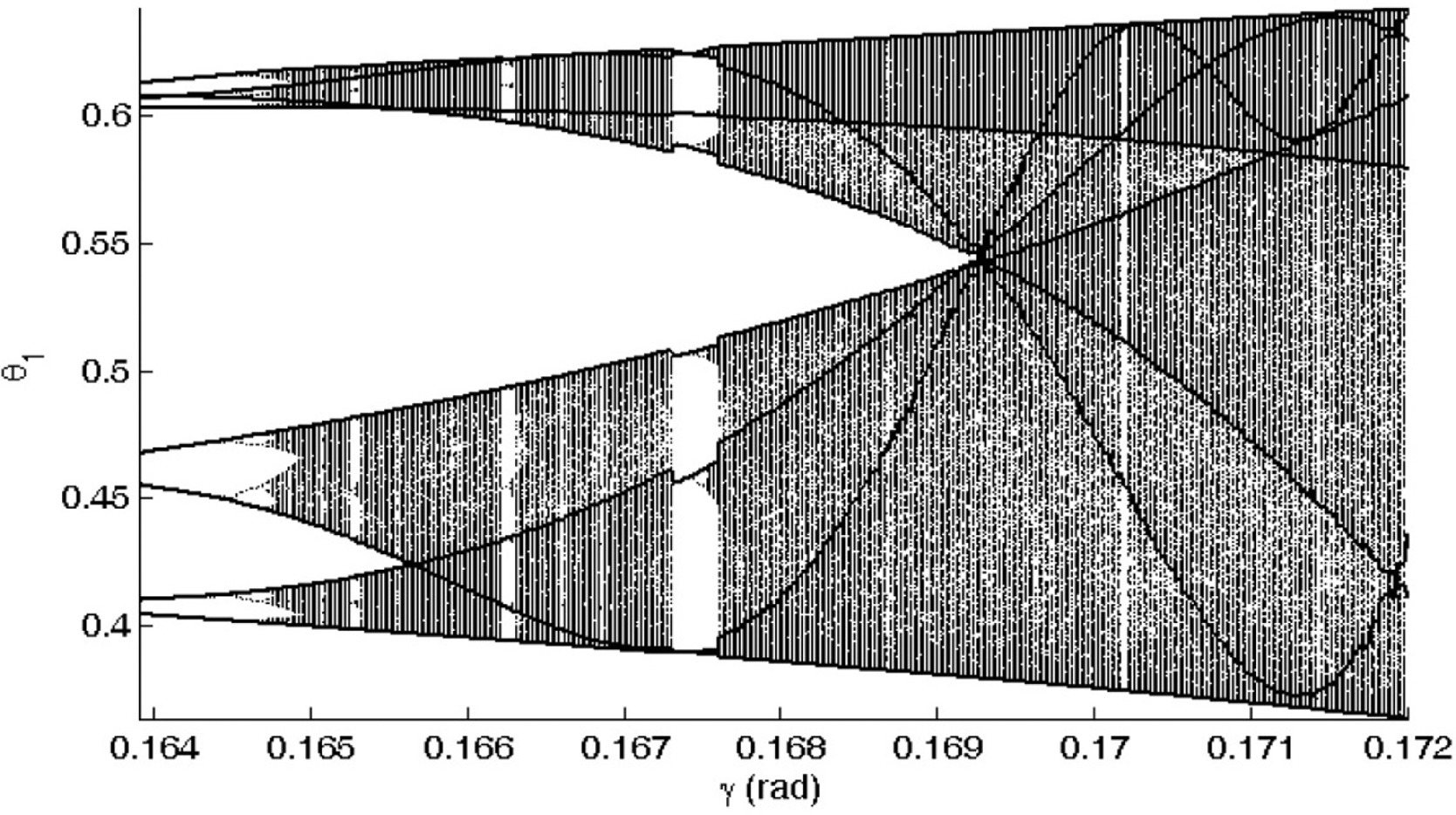}}
\caption{(a) Bifurcation diagram for Garcia’s kneed walker showing period-doubling bifurcations of stable walking. Walking was unable to be maintained at slopes steeper than 0.019 radians \cite{garcia1999stability}. (b) Zhao and colleagues elucidated the intrinsic rules underlying period-doubling bifurcations cascading to chaos in bipedal gait \cite{zhao2011analysis}.}
\label{fig:compass_bifurcation}
\end{figure*}

The biped's energy gain due to the conversion of gravitational potential energy at each step is lost on impact of foot-collision. As show in Figure~\ref{fig:PDW_energy}, its locomotion is sustained by the continual exchange of kinetic and potential energy as it descends a shallow slope. The biped’s exact gait pattern is determined by its geometric and inertial parameters, as well as ground slope. Furthermore, gait is characterized by limit cycle stability in the phase space, as is demonstrated in Figure~\ref{Fig:compass_model_limit}(b), and can exhibit both symmetric and asymmetric step lengths \cite{goswami1997limit, garcia1998simplest}. Figure~\ref{fig:compass_bifurcation} shows that passive bipedal walkers can exhibit chaotic gait through a cascade of period-doubling bifurcations \cite{goswami1996compass, garcia1998simplest}. Precisely, by increasing the ground-slope angle, a period-1 gait bifurcates to stable cycles of successively higher periods, until no two steps are the same, and the robot falls. Such instabilities associated with passive dynamic walkers arise by changing the relationship between the biped and gravity. However, the hybrid nature of the biped's differential-algebraic system made it challenging to study a passive bipedal walker's stability and symmetry with analytical methods. 
Goswasmi and colleagues \cite{goswami1996compass} extensively analyzed the energy balance and contraction of phase space volumes in steady passive gait. Zhao and colleagues studied the ground slope-angle differences between successive period-doubling bifurcations to reveal the Feigenbaum constant \eqref{Feigenbaum}, testifying to the universality of the period-doubling phenomenon observed here \cite{zhao2009improved, zhao2011analysis, wu2011restraint, zhao2012analysis}. Using chaos theory to study passive dynamic walking has invigorated bipedal robot research in recent years, for example by implementing methods of controlling chaos \cite{ott1990controlling, gritli2013chaos, suzuki2002enhancement, asano2009efficiency, osuka2002stabilization}, and is spurring on new investigation into improved analytical methods for understanding passive bipedal walkers.

\subsection{Aerodynamic Flutter}

\begin{figure}[t]
\centering
\begin{overpic}[width = 9cm]{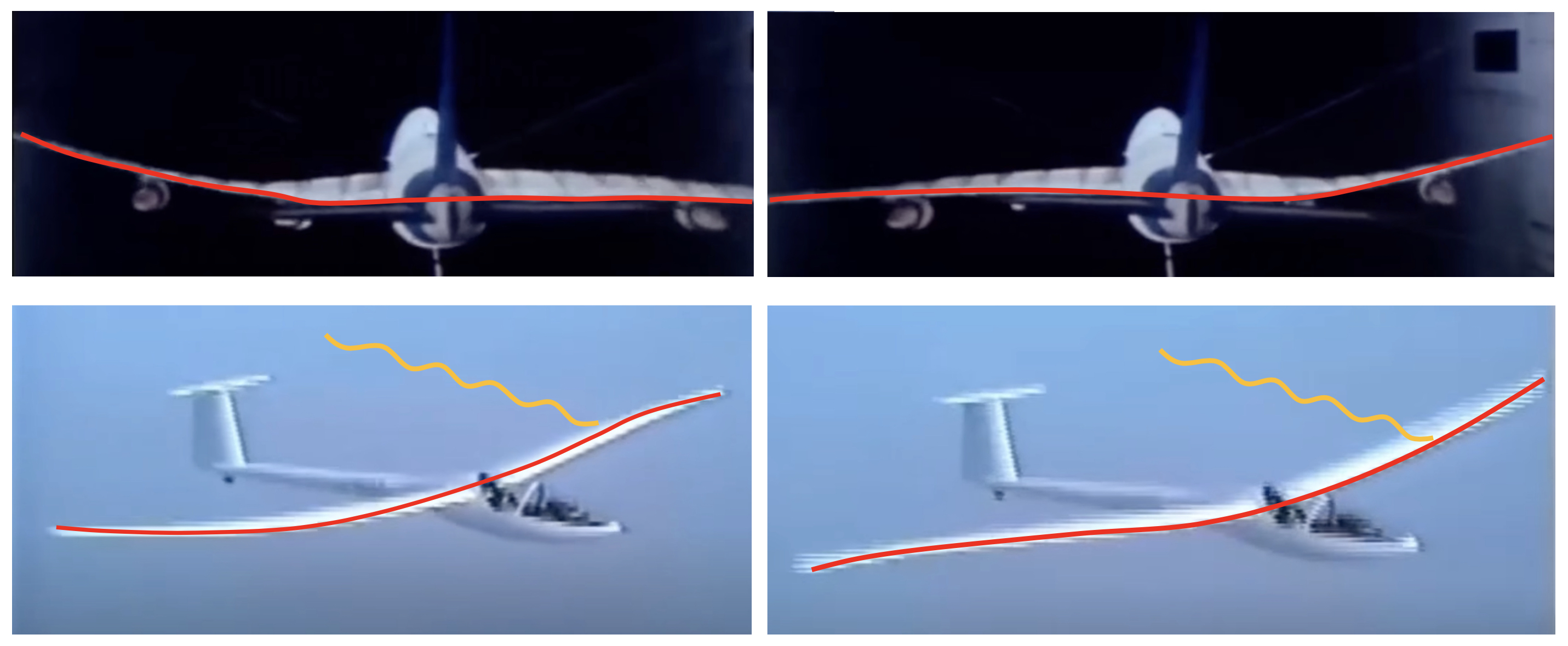}
\end{overpic}
\vspace*{-.10in}
\caption{Manifestation of flutter instability on in an airplane design in a wind tunnel and a glider in flight.  Flutter can be induced through a gust of wind which drives the system into instability.  In these snapshots, the wing profile is approximated to show the rapid (several Hertz) and large oscillations on the wing.  Structural failure is a central concern of flutter which prevents the direct observation of the cascading effects of the damp-driven instability. }
\label{fig:flutter}
\end{figure}
%

The flutter instability in aerodynamics is one of the most important to characterize given its importance in engineering design.  For flight dynamics, it is critical to circumvent this instability as it can produce structural loads and oscillations on aircraft wings that compromise safety and performance.   Indeed,  the unconstrained vibration of flutter can lead to the destruction of an aircraft.  Examples of flutter induced oscillations on airfoils leading to failure include the NASA Helios prototype~\cite{noll2004investigation}  which was a solar electric-powered flying wing concept designed to fly at high altitudes.  More famously, the Tacoma narrows bridge collapse in 1940 was induced by aeroelastic flutter~\cite{larsen2000aerodynamics}, highlighting the potential for structural failures more broadly in engineering design.  Figure~\ref{fig:flutter} highlights two examples of induced flutter on aircraft:  a model of a commercial airplane in a wind tunnel and a glider in flight.  Both produce large amplitude oscillations of several Hertz on the wings that produce significant structural stresses.

The theoretical study of flutter dates back to seminar contributions of Lord Rayleigh on the instability in jets~\cite{rayleigh1878instability}.  His theoretical approach could be used to prove that an infinite flag is always unstable.  More than that, it can be proved easily that an elastic plate of infinite dimension (both in the span- and streamwise direction) is always unstable when immersed in an axial potential flow.  Theodorsen~\cite{theodorsen1949general} extended early theories of flutter and provided a seminal theoretical study of the mechanisms of flutter and solutions for finite size airfoils.   The aerospace industry continued to develop the understanding of airfoils~\cite{kornecki1976aeroelastic,dowell1970panel}, with critical studies identifying flutter modes, their frequencies and growth rates~\cite{eloy2007flutter}.  Indeed, the fundamental observation concerning flutter is the onset of unstable modes at specific frequencies that depend on the airfoil design.

More recently, studies have considered the detailed onset of the oscillatory instability induced in mechanical systems~\cite{lee2022modelling}, fluter being a subset of this study.   In their study, the develop a hybrid modeling approach that leverages mechanistic modeling with machine learning towards modeling the behavior of a physical aeroelastic structure exhibiting limit cycle oscillations during wind tunnel tests.   In the case of fluter on airfoils, it is not clear that one can observe the onset of period doubling before structural failure.  This is in contrast to the applications of lasers, rocket engines and droplets where there is a clear route to re-distributions of energy available without a structural failure.

\subsection{Porpoising in Formula One Cars}

Theory and experiments have shown that airfoils can achieve higher lift-to-drag ratios when operating near the ground.  This phenomenon is called {\em ground effect}.  In the 2022 season, significant regulation changes in Formula One car design produced a new generation of cars that all experienced {\em porpoising}.  Porpoising is a phenomenon where above a certain speed, typically 160-180~km/hr, the car begins to bounce at a frequency of approximately 4-6~Hz.  In addition to experiencing physical pain from the bouncing, porpoising significantly compromises performance of the car.   The 2022 season championship has largely been decided by those teams who have successfully overcome the porpoising effect while maintaining performance.

\begin{figure}[t]
\centering
\begin{overpic}[width = 8cm]{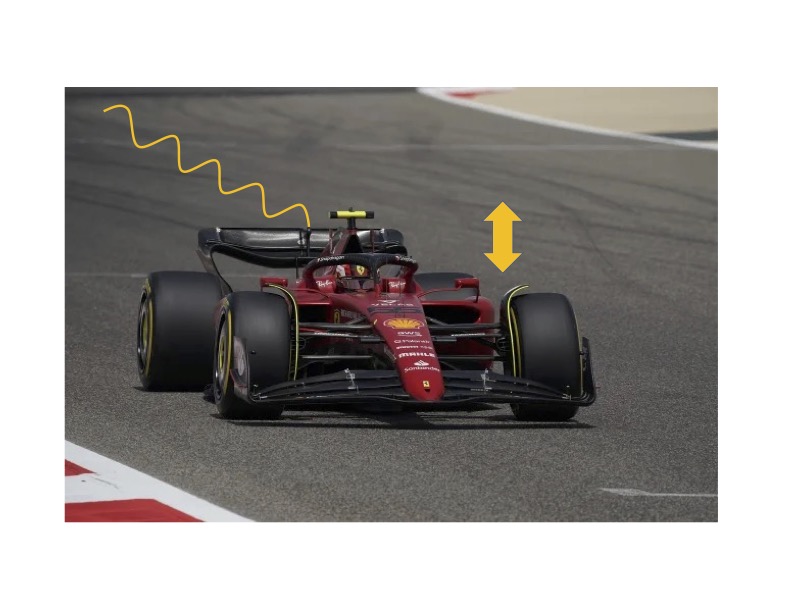}
\put(30,56){\color{white}{Frequency $4-6$ Hz}}
\end{overpic}
\vspace*{-.30in}
\caption{Porpoising dynamics in the 2022 Formula One regulation cars.  At speeds between 160-180~km/hr, the car begins to bounce at a frequency of approximately 4-6~Hz.  Over the course of the 2022 season, teams have worked at suppressing the onset of porpoising. }
\label{fig:porp1}
\end{figure}
%

Ground effects have been extensively studied in both theory and experiment.  Indeed, it is known that ground effects are used by bird to improve their efficiency in flight and gliding, this includes herring gulls~\cite{baudinette1974energy}, brown pelicans~\cite{hainsworth1988induced}, black skimmers~\cite{withers1977significance}.  Fish also use ground effects for improved efficiency in swimming, including the mandarin fish~\cite{blake1979energetics}, steelhead trout~\cite{webb1993effect} and flying fish~\cite{park2010aerodynamic}.  Our understanding of ground effects, while qualitatively known, is largely limited to steady flow, perhaps highlighting the reason that the porpoising effect was missed in the 2022 formula one regulations.  

\begin{figure}[t]
\centering
\begin{overpic}[width = 8cm]{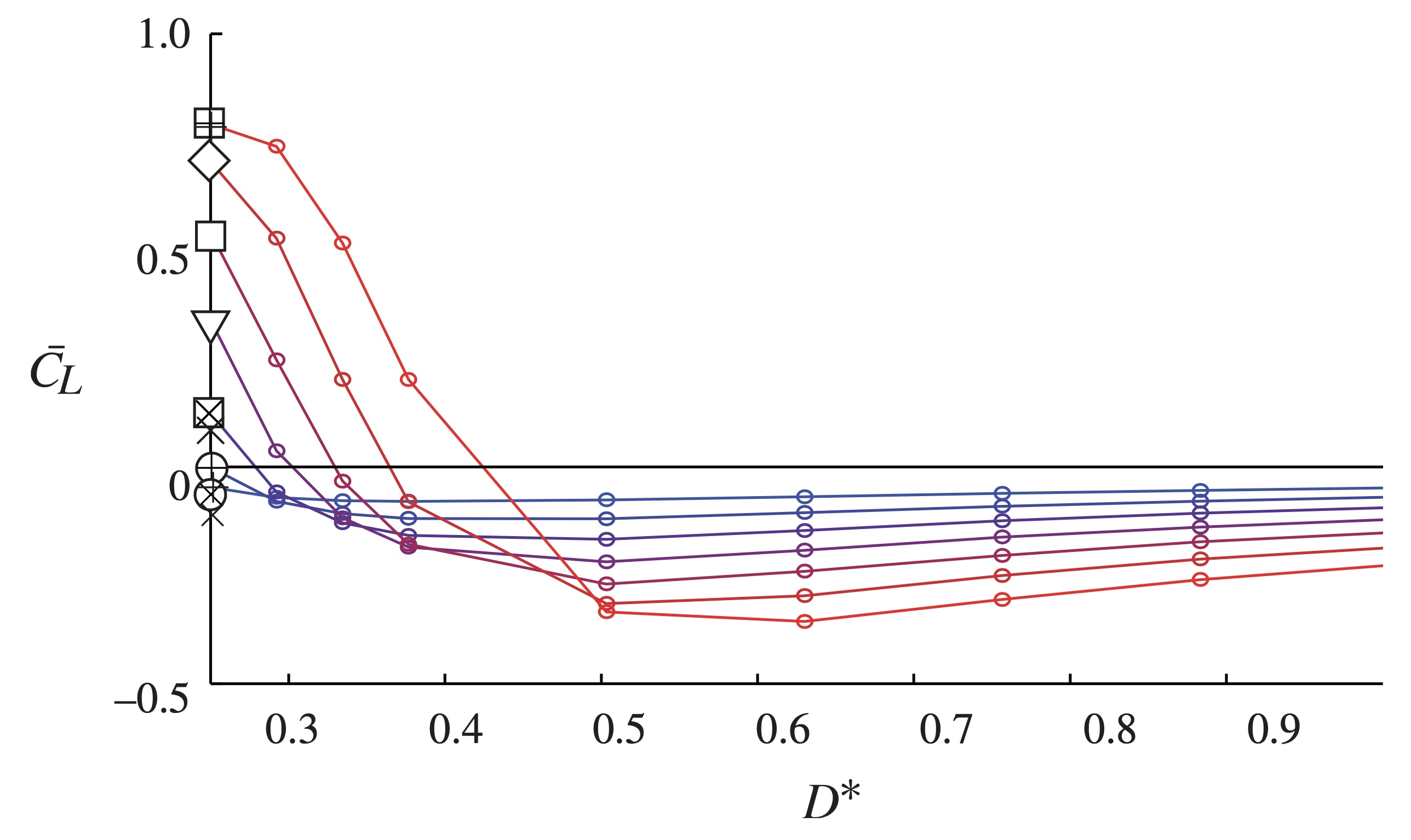}
\end{overpic}
\vspace*{-.10in}
\caption{Computation of the time averaged lift coefficient $\bar{C}_L$  as a function of the ratio of the pitching aerofoil and the chord length $D^*$ (From Quinn et al~\cite{quinn2014unsteady}).  The curves are for Strouhal number $St=0.13, 0.16, 0.19, 0.22, 0.25, 0.28$ and $0.31$ (from blue to red respectively).  These curves show qualitatively that a peak in downforce (negative lift) is achieved in the range of $D^*$ between 0.4-0.6.  This decrease in downforce as the car is closer to the ground is a primary contributor to porpoising. }
\label{fig:porp2}
\end{figure}

Tanida~\cite{tanida2001ground} was the first to study ground effects analytically by considering the linearized Euler equations of a fluttering plate in a wind tunnel.  Asymptotics were later used to predict the drag and lift for weak ground effects~\cite{iosilevskii2008asymptotic}.  Numerical studies of unsteady ground effects have previously considered Formula One car aerodynamics for an inverted car with a front wing undergoing heave oscillations~\cite{moryoseff2001computational,molina2011aerodynamics}.  More recently, experiments and computations by Quinn et al~\cite{quinn2014unsteady} have considered the ground effect for a rigid aerofoil pitching near a planar boundary.  This study is particularly valuable as it shows an important transition in the aerofoil dynamics.  Specifically, they consider the overall lift from the aerofoil as a function of the ratio of the pitching aerofoil and the chord length, denoted by parameter $D^*$ in their work.  When $D^*$ is between 0.4 and 1, the lift force pulls the aerofoil to the ground, which is the effect Formula One cars intend to leverage to produce improved downforce.  However, for $D^*$ between 0.25 and 0.4, the lift force pushes the aerofoil away from the ground.  Thus when $D^*\approx 0.4$, there is a stable equilibrium point where the time-averaged lift is zero and thrust is enhanced by approximately 40\%. 

\begin{figure}[t]
\centering
\begin{overpic}[width = 8cm]{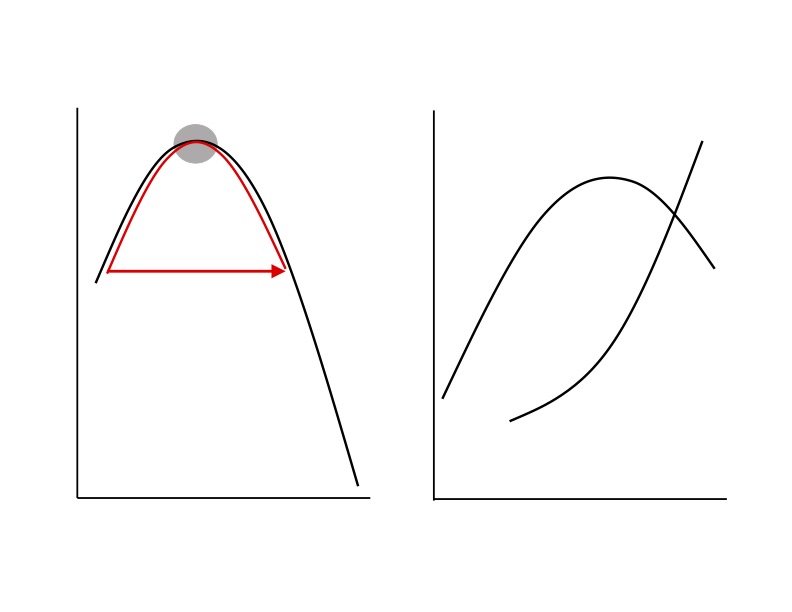}
\put(12,10){Distance to ground}
\put(5,35){\rotatebox{90}{Downforce}}
\put(3,60){(a)}
\put(48,60){(b)}
\put(78,10){$E_-$}
\put(48,45){$E_+$}
\put(78,27){gain}
\put(59,50){loss}
\end{overpic}
\vspace*{-.30in}
\caption{(a) Qualitative description (black line) of the downforce as a function of distance to the ground from computational studies of Fig.~\ref{fig:porp2}~\cite{quinn2014unsteady}.  A simple description of porpoising is given by the red trajectory:  As the speed of the car increases, the downforce increases which decreases the distance to the ground.  Once past the optimal downforce point (grey region), the downforce decreases which allows the suspension to push the car up so that the distance to the ground is increased (red trajectory).  The car is pushed down once again so as to approxinately repeat the red trajectory.  This creates the periodic motion of the porpoising.  (b) The qualitative gain and loss curves suggested by panel (a) which leads to the iterative map common of the damp-driven instability.  The decreased loss represents the losses on the suspension system which is decreased as shown by the decrease in downforce past an optimal point.  The gain is the related to the increased speed of the car.}
\label{fig:porp3}
\end{figure}
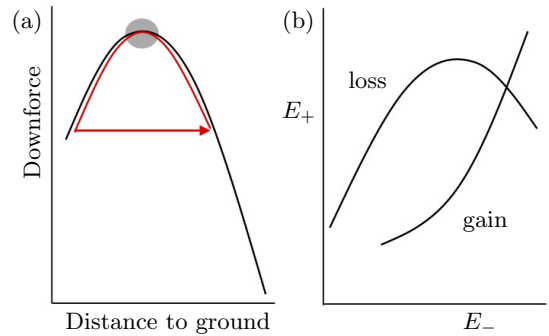

This numerical and experimental study highlights all the key features required to understand porpoising in Formula One cars.  One of the key figures of this work is reproduced here as Fig.~\ref{fig:porp2}, which shows the downforce (negative time averaged lift) of an aerofoil as a function of the distance to the ground.  The key feature of this graph is the identification of a clear optimal downforce for a given distance to the ground.  Past this point, the downforce is reduced.  In Formula One cars, the downforce interacts with the suspension, which is tasked with bringing the car back to its equilibrium ride height.  Figure~\ref{fig:porp3}(a) shows the basic dynamics of this interaction:  As the speed of the car increases, the downforce increases which decreases the distance to the ground.  Once past the optimal downforce point (grey region), the downforce decreases which allows the suspension to push the car up so that the distance to the ground is increased (red trajectory).  The car is the pushed down once again so as to approxinately repeat the red trajectory.  This creates the periodic motion of the porpoising.  From the energy perspective taken here, Fig.~\ref{fig:porp3}(b)  shows the qualitative energy balance for this ground effects--suspension interaction system.  In this case, the decreased energy represents the losses on the suspension system which is decreased as shown by the decrease in downforce past an optimal point.  The gain is the related to the increased speed of the car.  As already highlighted, the basic qualitative structure of the gain and loss curves generates the damp-driven instability cascade.  It is not clear that Formula One cars could undergo a secondary bifurcation without breaking the car.

\subsection{Chaotic logical circuits}\label{Sec: Circuits}

It has been known since the early 1800s that RLC circuits exhibit oscillatory behavior \cite{RLC_Circuit_History}, and could have complexities such as critical or overdamping depending on the strength of the resistor.  However, we cannot get chaotic behavior since the phase space is only 2-dimensional, and thus satisfies the Poincar\'{e} -- Bendixson theorem \cite{PoincareBendixson1, PoincareBendixson2, PoincareBendixson3}.  Van der Pol later developed a circuit that can exhibit chaotic behavior by inducing non-autonomous forcing through a triode coupled to the usual RLC circuit \cite{vanderpol1926, vanderpol1927}.  Then a majority of the damping comes from the resistor and the driving from the triode.

\begin{figure}[htbp]
    \centering
    \stackinset{r}{}{t}{}{\textbf{(a)}}{\includegraphics[width = 0.35\textwidth]{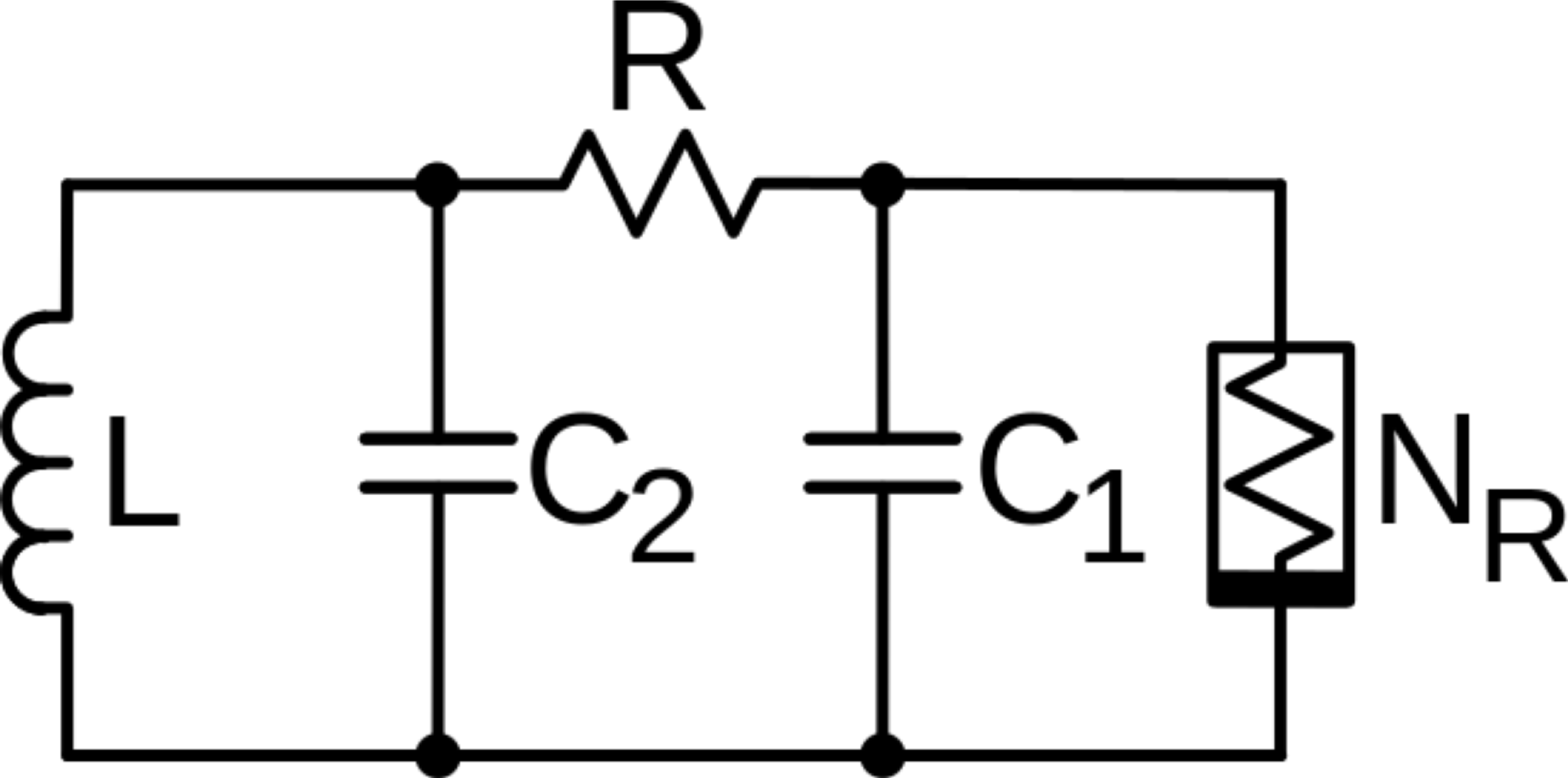}}
    \stackinset{r}{}{t}{}{\textbf{(b)}}{\includegraphics[width = 0.35\textwidth]{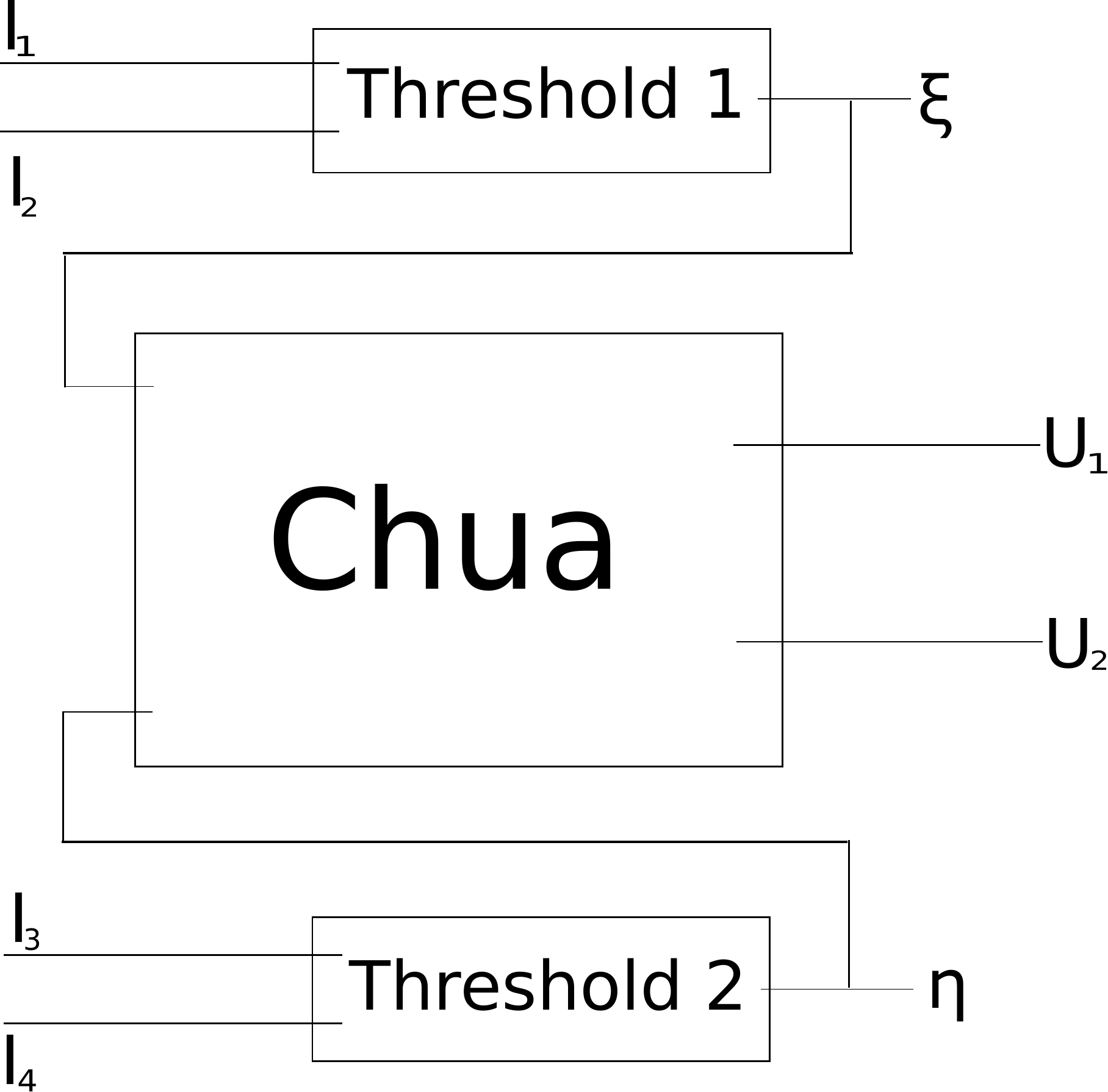}}
    \caption{\textbf{(a)} Diagram of Chua's circuit with two capacitors, $C$, an inductor, $L$, a linear resister, $R$, and a nonlinear resistor, $N_R$.  {\footnotesize (Reproduced under \emph{CC0 License};
Source: Chetvorno. \url{https://en.wikipedia.org/wiki/File:Chua's_circuit_with_Chua_diode.svg}.
Date accessed: October 11, 2017.  Date added: February 2014.)}  \textbf{(b)} Schematic of dual NOR gate via thresholding of Chua's circuit, where $I$ are the input voltages, $\xi$ are the threshold output voltages, and $U$ are the NOR output voltages.}
    \label{Fig: Circuits Schem}
\end{figure}

More recently, Chua developed a circuit that places the the components in parallel as shown in Fig. \ref{Fig: Circuits Schem}\textbf{(a)}, which enables us to directly write out a system of three ODEs via Kirchoff's laws \cite{Matsumoto84},
    \begin{subequations}
    \begin{align}
    C_1\frac{d V_{C_1}}{d t} &= G(V_{C_2} - V_{C_1}) - g(V_{C_1}),\\
    C_2\frac{d V_{C_2}}{d t} &= G(V_{C_1} - V_{C_2}) + \chi_L,\\
    L\frac{d \chi_L}{d t} &= -V_{C_2};
        \end{align}
        \label{Eq: Circuits Chua}
    \end{subequations}
rather than a second order ODE with a forcing term as in the van der Pol oscillator.  Since the Chua circuit is far less convoluted than the van der Pol oscillator, we can make it produce logical outputs by thresholding the voltage across the two capacitors \cite{Cafagna-Grassi06, RJorB18} as shown in Fig. \ref{Fig: Circuits Schem}\textbf{(b)}.  While the circuitry itself is less convoluted, the damping and driving mechanisms are now related in a more convoluted fashion.  However, if we focus only on the threshold control units (TCUs) modeled as tent-like maps \cite{RJorB18}, we observe behavior similar to that of other damped-driven systems.  

\begin{figure*}[htbp]
    \centering
    \stackinset{l}{6mm}{t}{1mm}{\textbf{(a)}}{\includegraphics[height = 0.16\textheight]{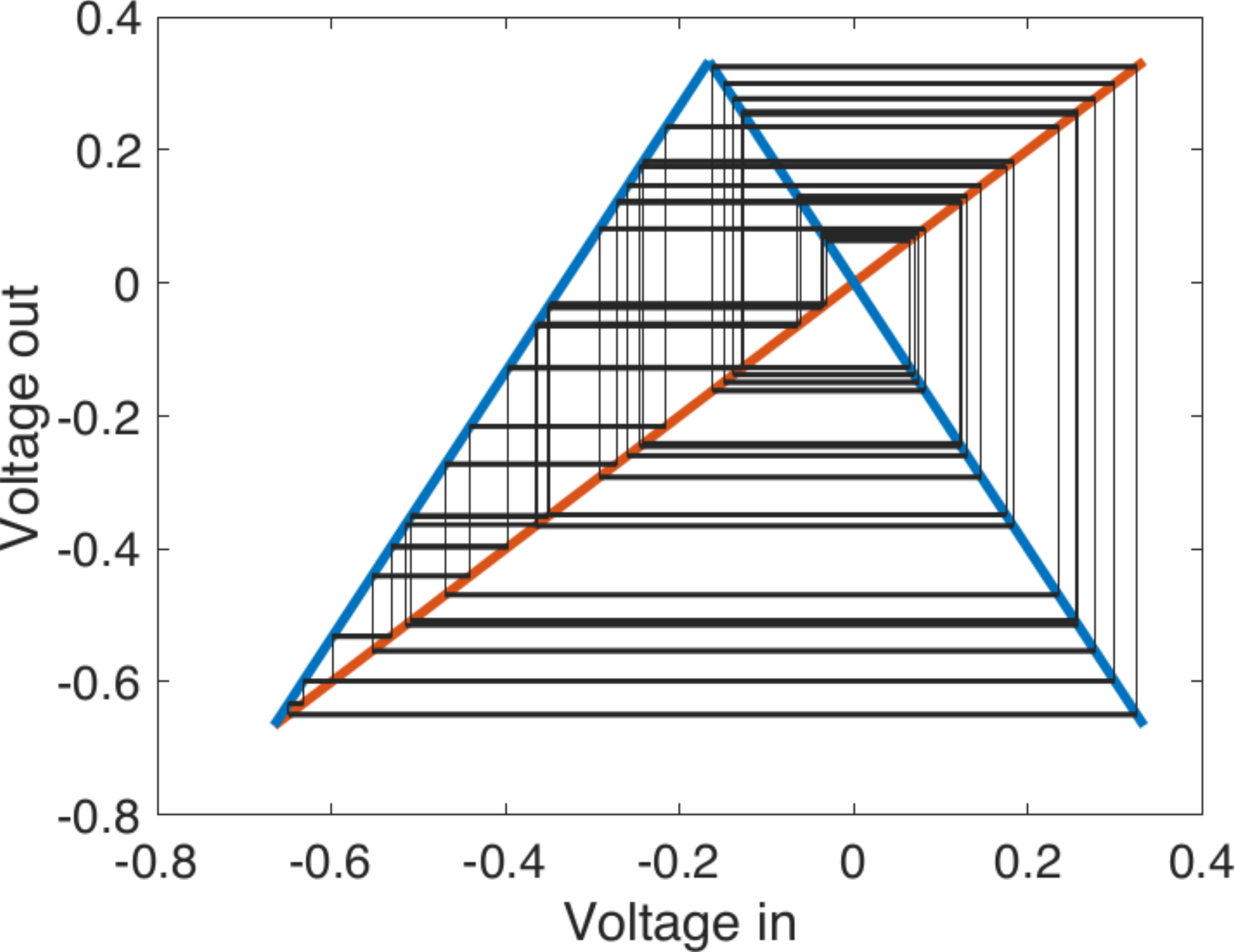}}
    \stackinset{l}{4mm}{t}{1mm}{\textbf{(b)}}{\includegraphics[height = 0.16\textheight]{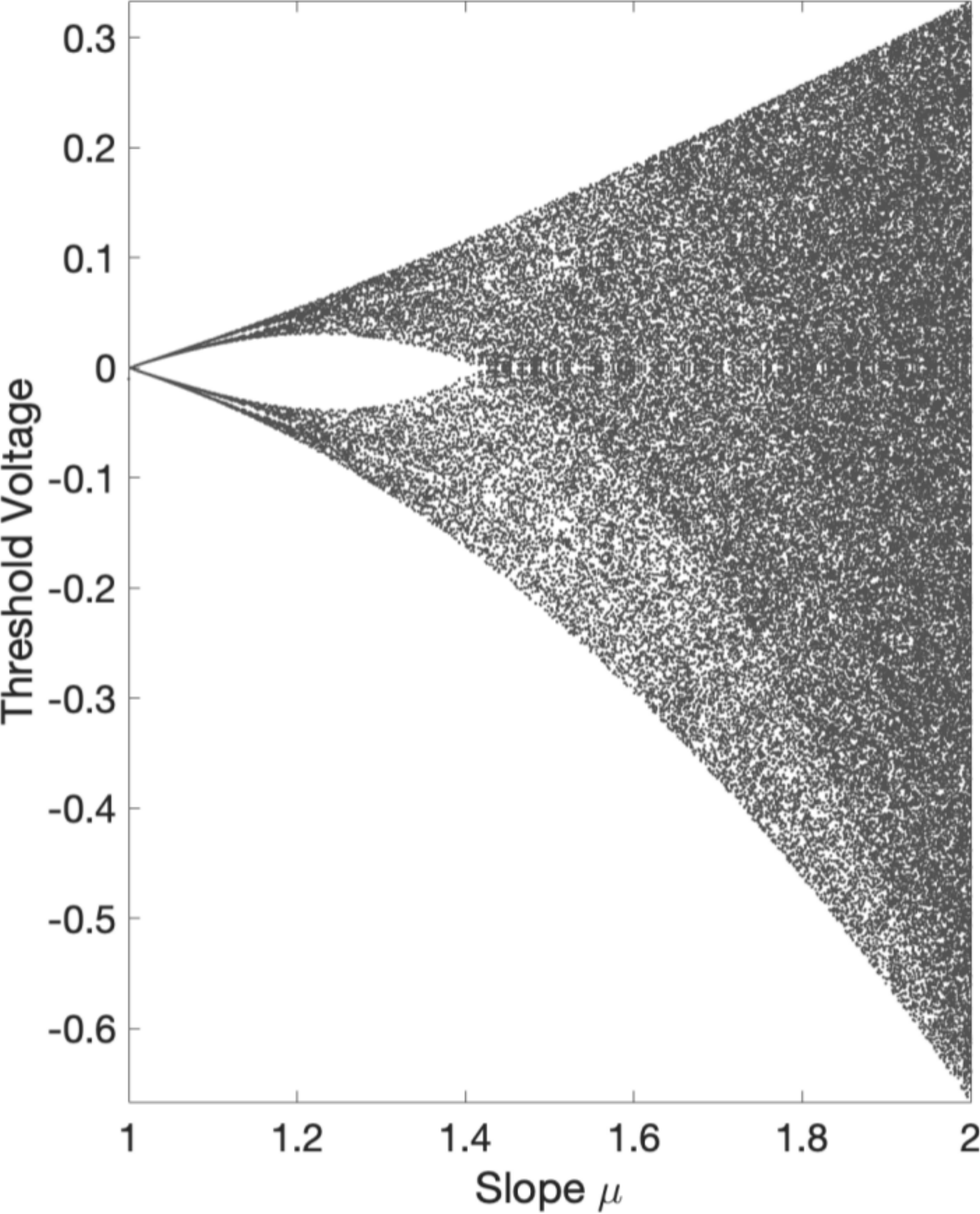}}
    \stackinset{l}{5mm}{t}{1mm}{\textbf{(c)}}{\includegraphics[height = 0.16\textheight]{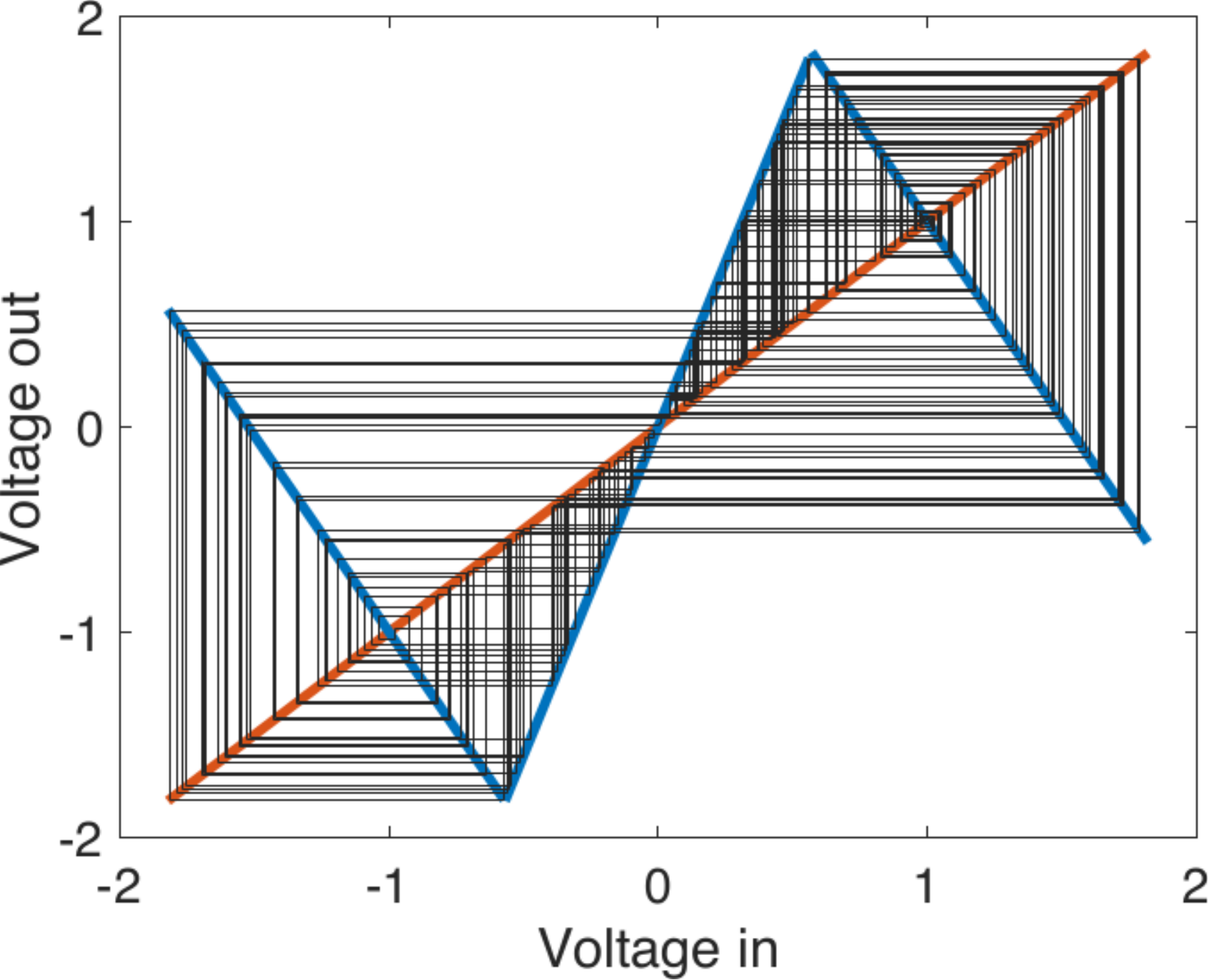}}
    \stackinset{l}{6mm}{t}{1mm}{\textbf{(d)}}{\includegraphics[height = 0.16\textheight]{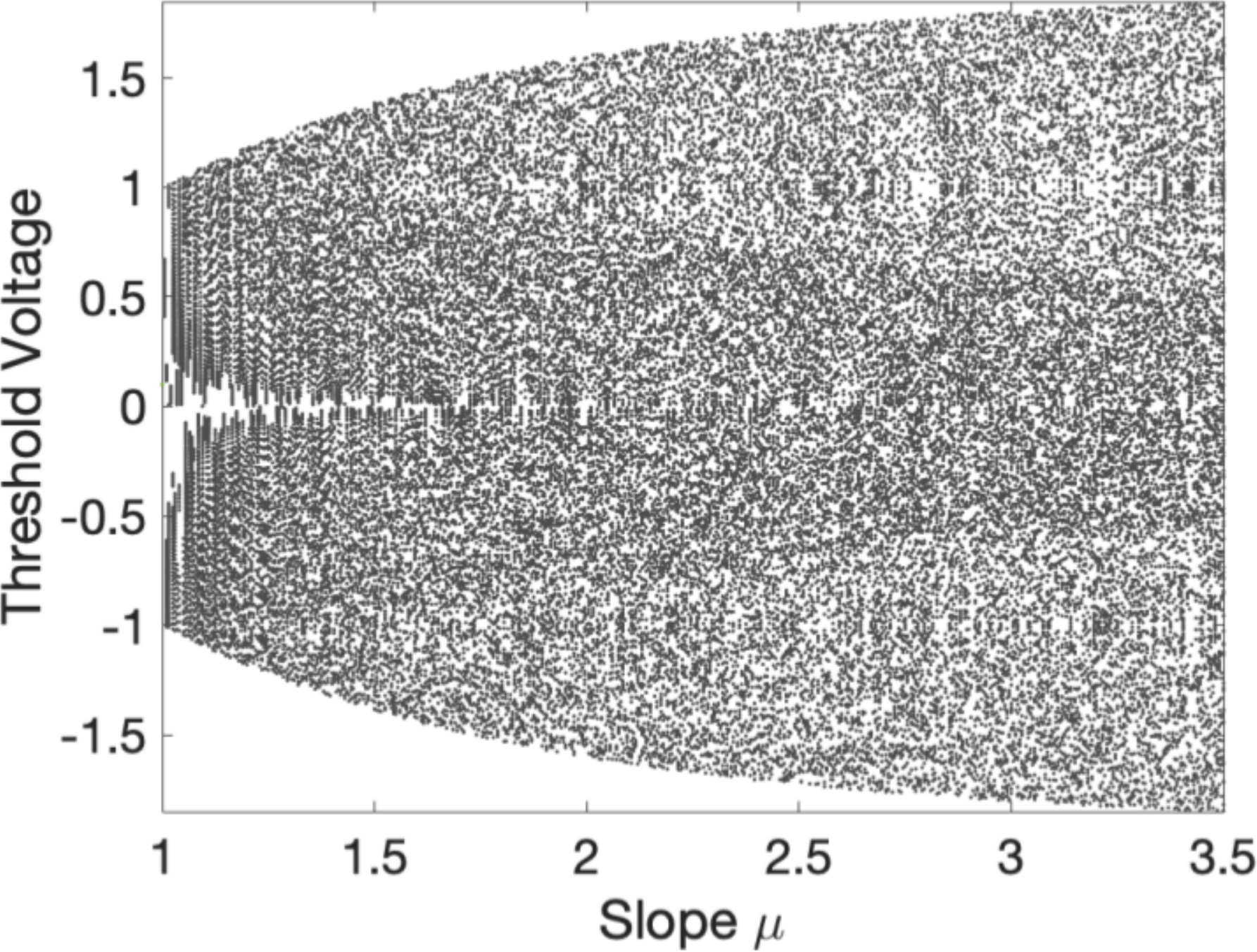}}
    \caption{Cobweb and bifurcation diagrams for the thresholding mechanisms in the dual NOR gate system.  \textbf{(a)} Cobweb plot of Threshold 1 from Fig. \ref{Fig: Circuits Schem}.  \textbf{(b)} Bifurcation diagram for Threshold 1 as the slope $\mu$ of the piecewise-linear blue curve is increased.  \textbf{(c)} Cobweb plot of Threshold 2 from Fig. \ref{Fig: Circuits Schem}.  \textbf{(d)} Bifurcation diagram for Threshold 2 as the slope $\mu$ of the piecewise-linear blue curve is increased.}
    \label{Fig: Circuits Bifurcations}
\end{figure*}

We model the TCUs as either modifications of a tent map or combinations of tent maps.  Due to the inherent asymmetry in Chua's circuit, the first TCU will behave like a modified tent map with repelling fixed points for all allowable parameter values (the slope of the pieces, $\mu$) at the origin and at a subzero voltage (Fig. \ref{Fig: Circuits Bifurcations}\textbf{(a)}), and the second TCU will behave like a combination of tent maps with a repelling fixed point for all allowable parameter values (the slope of the pieces, $\mu$) at the origin and two fixed points symmetrically off the origin (Fig. \ref{Fig: Circuits Bifurcations}\textbf{(c)}).  Interestingly, unlike many of the other systems shown in this review, the additional intersections between the gain and loss curves will not come from increasing a bifurcation parameter.  Rather, the two types of maps come from the different parts of the circuit.  Therefore we can analyze the the single and multiple nontrivial intersections separately.  For the first TCU we only have one intersection, and hence observe a period doubling bifurcation similar to that of the logistic map (Fig. \ref{Fig: Circuits Bifurcations}\textbf{(b)}).  This is to be expected as the tent map is homeomorphic to the logistic map.  On the other hand, for the second TCU we have two intersections, and hence does not experience a period doubling bifurcation.  This shows why many physical damped-driven systems can seem to bypass the period doubling stage, as observed in Sec. \ref{Sec: Walkers} for walking droplets.

\subsection{Faraday waves and Sloshing Tanks}

\begin{figure}[t]
\centering
\begin{overpic}[width = 8cm]{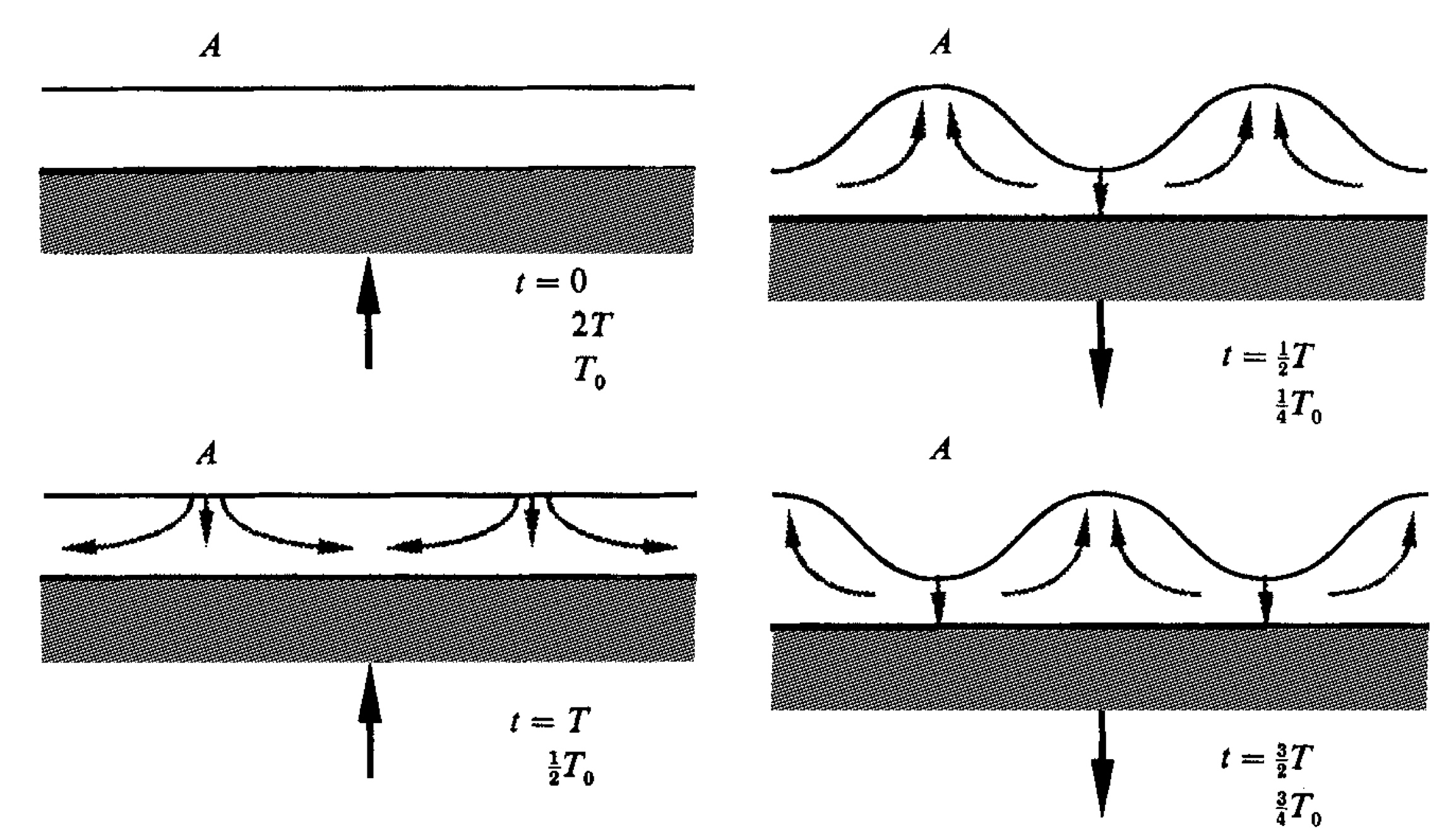}
\put(30,56){\color{white}{Frequency $4-6$ Hz}}
\end{overpic}
\vspace*{-.10in}
\caption{ As explained by Douady, "Excitation at half the excitation frequency of a fluid layer undergoing a vertical oscillation. When the vessel goes down, the fluid inertia tends to create a surface deformation, as in the Rayleigh-Taylor instability. This deformation disappears when the vessel comes back up, in a time equal to a quarter-period of the corresponding wave (T). The decay of this deformation creates a flow which induces, for the following excitation period T, the exchange of the maxima and the minima. Thus one obtains = 2T." (From Douady~\cite{douady1990experimental})  }
\label{fig:slosh1}
\end{figure}

Faraday waves refer to nonlinear standing waves, which appear on liquids enclosed by a container excited vertically with a frequency that is close to twice the natural frequency of the free surface (See Fig.~\ref{fig:slosh1} for a physical explanation of this process). This condition is referred to as parametric resonance, since the motion of the liquid free surface is due to an excitation perpendicular to the plane of the undisturbed free surface, and thus the generated waves are also known as parametric sloshing or Faraday waves. 
The Faraday instability has been extensively studied in both theory and experiments (See Ibrahim~\cite{ibrahim2015recent} and Douday~\cite{douady1990experimental} for exceptional reviews of the field).  More broadly, Faraday waves are a manifestation of pattern forming systems, for which there is an extensive body of literature across disciplines~\cite{cross1993pattern}. 

%

\begin{figure}[t]
\centering
\begin{overpic}[width = 9cm]{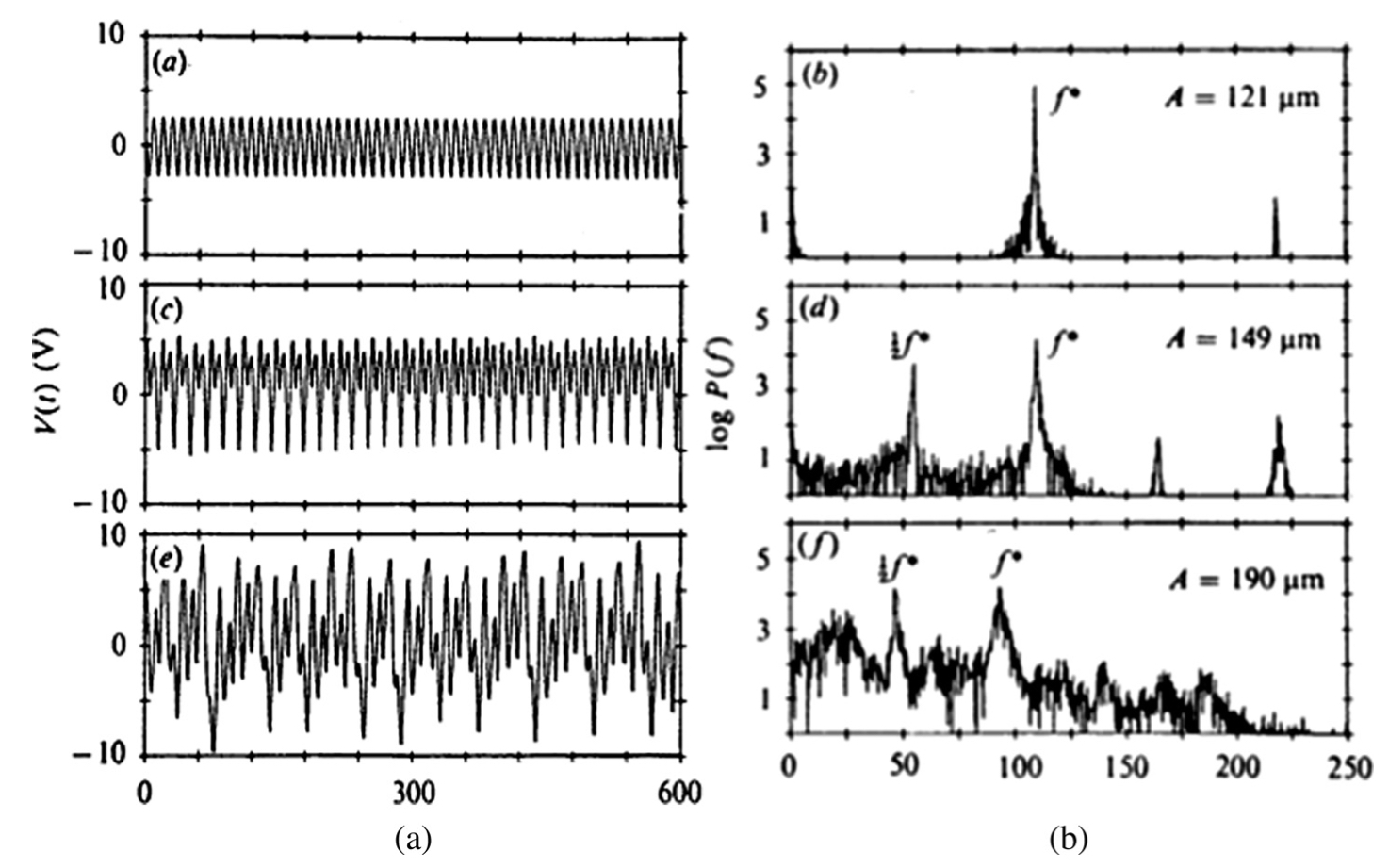}
\put(30,56){\color{white}{Frequency $4-6$ Hz}}
\end{overpic}
\vspace*{-.10in}
\caption{(left panels) Time history records showing the transition from periodic to chaotic oscillations and (right panels) corresponding power spectra under excitation frequency of 16.05 Hz and three different excitation amplitudes.  Note the period doubling from the first to second rows and the onset of chaos in the third row.  All this happens as the energy input is increased, via larger amplitude oscillations (from Ciliberto and Golub~\cite{ciliberto1984pattern}).  }
\label{fig:slosh2}
\end{figure}

The literature on Faraday waves is vast, and in this brief discussion, we do not indent to give a broad or comprehensive overview of all the experimental and theoretical studies that have been done to date.  Instead, we focus on an important recent application of the Faraday instability.  Specifically, in recent years, the Faraday instability has become important for understanding the dynamics of fuel loads on aircraft.  The slowing induced by flight vibrations induces forces on wings that should be characterized in order to understand structural mechanics~\cite{meziani2014capillary,colella2021sloshing,pizzoli2022nonlinear}.  Fuel sloshing rarely is considered  in the preliminary design and development phase of an aircraft, with its impact determining the static response and the inertial characteristics of rigid body motion and aeroelasticity of the aircraft.  Indeed,  the fluid inside the tank moves relative to the tank walls with its own dynamics that causes additional forces and moments that can eventually modify the stability margins of the structure.  This presents many engineering challenges for improving the design of more efficient aircraft and the integration of different propulsion systems for an improved environmental sustainability. This suggests the re-engineering of the position and shapes of the tanks in order to circumvent or suppress the onset of the Faraday instability and cascading period doubling to chaos.  Experiments clearly show this trend as shown in Fig.~\ref{fig:slosh2} where the driving amplitude is increased and a period doubling cascade to chaos is observed.

%




\section{Control of Damped-Driven Systems}

\begin{figure}[t]
\begin{overpic}[width=8cm]{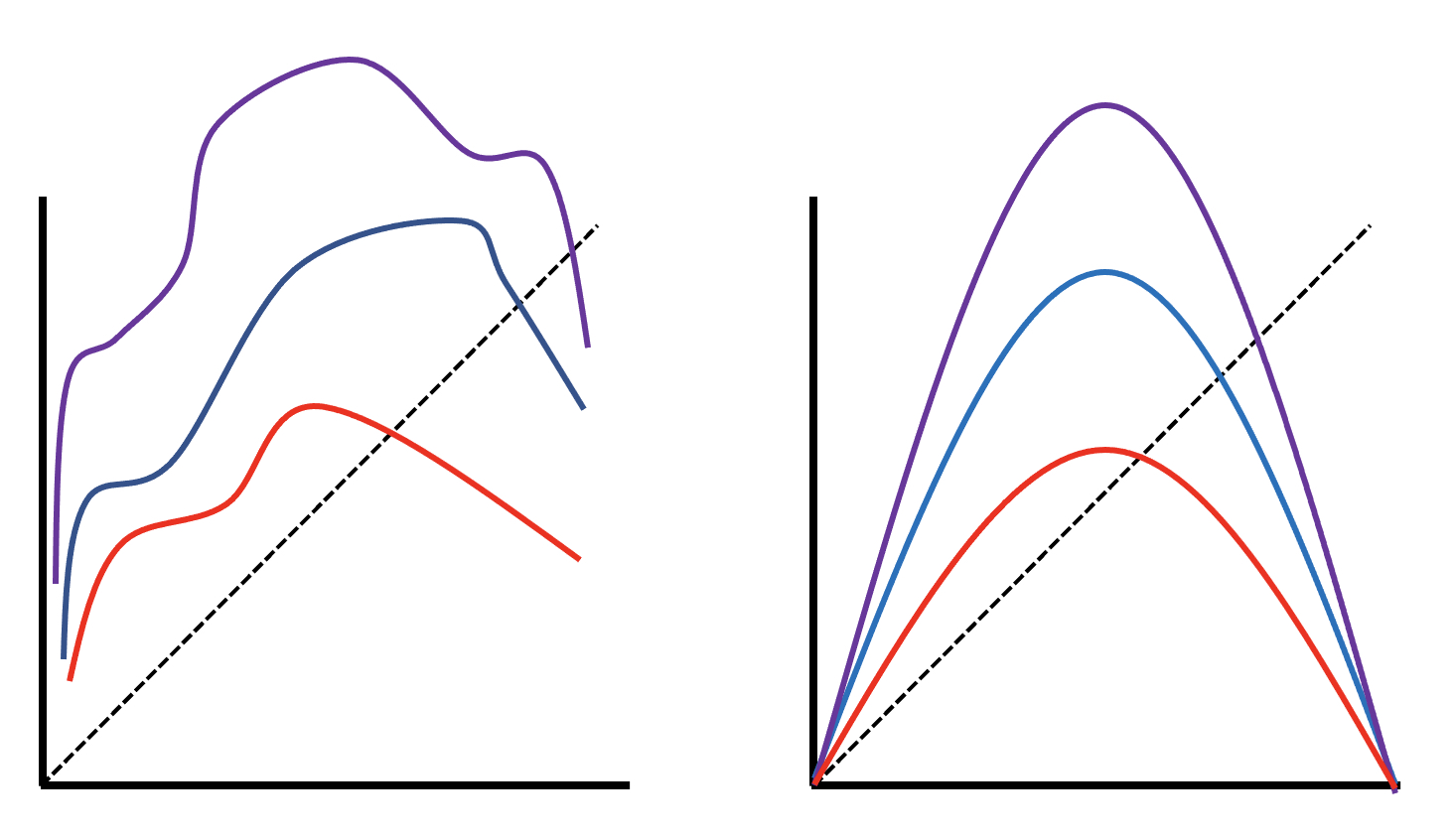}
\put(44,25){$h({\bf x}_n)$}
\put(46,20){$\longrightarrow$}
\end{overpic}
\caption{Depiction of the learned transformation of a generic combined gain-loss curve of Fig.~\ref{fig:balance}(b) to the logistic map $rh({\bf x}_n(1-h({\bf x}_n))$.  Engineering the loss curve has significant impact on the value of the bifurcation parameter $r$ when mapped to the logistic equation.\label{fig:engineer}}
\end{figure}

In almost all engineering scenarios, the onset of the period doubling cascade to chaos is undesirable. Thus one is incentivized to determine a strategy to suppress the universal damped-driven instability induced by the gain-loss dynamics.  Ultimately, to control the instability, it is necessary to engineering the loss and gain curves in order to circumvent the onset of bifurcation.

Figure~\ref{fig:engineer} depicts various combined energy mappings for a given system (See Fig.~\ref{fig:balance}).  Each curve maps to a different value of the bifurcation parameter $r$ in the logistic map.  The goal in controlling the damped-driven instability is to engineerin the combination of loss and gain to a desired value of $r$ which is below the onset of unstable dynamics.  Thus even if the underlying dynamics are exceptionally complicated, this analysis suggests one only the overall gain and loss curves need to be estimated in order to understand how to engineer a potential solution to avoiding bifurcations.  

As a specific example, recent studies have shown how to engineer the loss transmission curves in mode-locked lasers~\cite{li2011dual,fu2013high}. This can be done in order to produce much higher laser pulses without inducing the damped-driven instabilities highlighted throughout the examples. Indeed, these studies theoretically demonstrate that in a laser cavity mode-locked by a set of waveplates and passive polarizer, the energy performance can be increased by incorporating a second set of waveplates and polarizer in the cavity. The two nonlinear transmission functions acting in combination can be engineered so as to suppress the multi-pulsing instability responsible for limiting the single pulse per round trip energy in a myriad of mode-locked cavities. In a single parameter sweep, the energy is demonstrated to double. It is anticipated that further engineering and optimization of the transmission functions by tuning the eight waveplates, fiber birefringence, two polarizers and two lengths of transmission fiber can lead to further power increases. Moreover, the analysis suggests a general design and engineering principle that can potentially realize the goal of making fiber based lasers directly competitive with solid state devices. The technique is feasible and easy to implement without requiring a new cavity design paradigm.

\section{Conclusions and Outlook}

As illustrated through the numerous examples herein, damped-driven systems are abundant in physics and engineering.  Despite the significant diversity in underlying physical processes governing the dynamics, the various systems considered all exhibit the same underlying instability:  a period doubling cascade to chaos that can be mapped to the logistic equation.  By considering the energy balance generated by the gain and loss curves, the overall dynamics can be characterized by 
a Verhulst diagram (cobweb plot).  Thus irrespective of the physics and its complexities, this simple geometrical description dictates the universal set of logistic map instabilities that arise in complex damped-driven systems.

The geometrical description of the universal behavior manifested can be understood with a minimal set of assumptions.  Assuming the
energy increases monotonically as a function of increasing gain, and the losses become increasingly larger with increasing energy, allows for the construction of the iterative mapping characterizing the gain and loss dynamics.  The intersection of the gain and loss curves define an energy balanced solution, and the construction of an iterative map between the loss and gain curves is shown to be a unimodal map. As detailed in Section~\ref{sec:Logistic}, unimodal maps are homeomorphic to the logistic map, which helps us to understand its resulting period doubling cascade to chaos.

The characterization of the underlying instability has significant implications for engineering design.  Specifically, instead of engineering at the level of the complex physical processes, one should instead consider the potential for engineering the overall gain and loss curves since the manner of their intersection ultimately determines the onset of the logistic map instability as a function of the parameter $r$ in \eqref{eq:logistic}.  We have shown that deep learning methods can warp the gain/loss curves into the logistic map, showing that this mapping is indeed a underlying normal form for damped-driven dynamics.



\section*{Acknowledgments}
We acknowledge support from the National Science Foundation AI Institute in Dynamic Systems (grant number 2112085).  JNK further acknowledges support from the Air Force Office of Scientific Research (FA9550-19-1-0011).

\bibliographystyle{unsrt}
\bibliography{DampedDrivenNotes, Bouncing_droplets, Biolocomotion, rde}

\end{document}